\documentclass[numberedappendix,letterpaper]{emulateapj}
\usepackage{amstext}
\usepackage{apjfonts}
\usepackage{amsmath}
\usepackage{amssymb}
\usepackage[colorlinks=true,linkcolor=blue,citecolor=blue]{hyperref}
\usepackage{natbib}

\begin{document}

\newcommand{\beq}{\begin{equation}}
\newcommand{\eeq}{\end{equation}}
\newcommand{\beqar}{\begin{eqnarray}}
\newcommand{\eeqar}{\end{eqnarray}}
\newcommand{\e}{\epsilon}
\newcommand{\rCI}{r_{\rm CI}}
\newcommand{\rmin}{r_{\rm min}}
\newcommand{\mtot}{m_{\rm tot} }
\newcommand{\vc}{v_{\rm c}}
\newcommand{\acc}{{\bf a}}

\title{The Formation of Eccentric Compact Binary Inspirals and the Role of Gravitational Wave Emission in Binary-Single Stellar Encounters} 

\author{Johan Samsing$^{1}$, Morgan MacLeod$^{2}$, Enrico Ramirez-Ruiz$^{2}$} 
\altaffiltext{1}{Dark Cosmology Centre, Niels Bohr Institute, University of Copenhagen, 
Juliane Maries Vej 30, 2100 Copenhagen, Denmark}
\altaffiltext{2}{Department of Astronomy and
  Astrophysics, University of California, Santa Cruz, CA
  95064}
   
\begin{abstract} 
The inspiral and merger of eccentric binaries leads to gravitational waveforms distinct from those generated by circularly merging binaries. Dynamical environments can assemble binaries with high eccentricity and peak frequencies  within the {\it LIGO} band. In this paper, we study binary-single stellar  scatterings occurring in dense stellar systems as a source of  eccentrically-inspiraling binaries. 
Many interactions between compact binaries and single objects are characterized by chaotic resonances in which the binary-single system undergoes many exchanges before reaching a final state. 
During these chaotic resonances, a pair of objects has a non-negligible probability of experiencing a very close passage. Significant orbital energy and angular momentum are carried away from the system by gravitational wave (GW) radiation in these close passages and in some cases this implies an inspiral time shorter than the orbital period of the bound third body.  We derive the cross section for such dynamical inspiral outcomes through analytical arguments and through numerical scattering experiments including GW losses.  We show that the cross section for dynamical inspirals grows with increasing target binary semi-major axis, $a$, and that for equal-mass binaries it scales as $a^{2/7}$.  Thus, we expect wide target  binaries to predominantly contribute to the production of these relativistic outcomes. We estimate that eccentric inspirals account for approximately one percent of dynamically assembled non-eccentric merging binaries. While these events are rare, we show that binary-single scatterings are  a more effective  formation channel  than single-single captures for the production of eccentrically-inspiraling binaries, even given modest binary fractions. 
\end{abstract}

\maketitle

\section{Introduction}
The density of stars in galactic nuclei and in the centers of some globular clusters can be more than a
million times higher than that in our solar neighborhood \citep{Lightman:1978go}. In such cases, 
a primordial binary will undergo a close encounter with at least one other star with high
probability within its lifetime \citep[e.g.][]{2007HiA....14..215R}. 
It is in these environments, called
dense stellar systems, that binary populations will no longer be truly primordial as their stellar composition, eccentricity, and period distributions will be largely
determined by past interactions with other stars \citep[e.g.][]{1991ASPC...13..324M, 1992PASP..104..981H, 2003astro.ph.12497I, 2005ASPC..328..231I, Hopman:2006ha, 2006MNRAS.372.1043I, 2008ApJ...673L..25F, Ivanova:2008jx, 2010ApJ...717..948I}. This transformation of binary systems was envisioned by \citet{Hills:1976vi}, who suggested that exchanging neutron stars into     
preexisting binaries might be a natural way to form X-ray binaries as
byproducts. 

Dynamical friction
causes the heaviest stars and primordial binaries to concentrate towards the cluster's core \citep{1997A&ARv...8....1M, 2002ApJ...570..171F, 2009ApJ...707.1533F}.  Since the heaviest stars tend to be left in the binary following such three-body encounters (this can be understood as consequence of the tendency toward energy equipartition, in which the lighter star would have the highest velocity in the final state), binaries are quite effective at soaking up heavy stars such as neutron stars and heavy white dwarfs \citep{Hills:1980bm,Sigurdsson:1993jz, Sigurdsson:1995gh, Heggie:1996fza}, even if none of them originally had a companion. 

After such an exchange, the binary will not only be slightly  wider but also heavier, which will result in gravitational focusing being more effective. The binary's cross section for encounters will thus be larger than  before the exchange. For this reason, a binary likely to undergo one exchange over some time period is likely to have several more encounters coming rapidly after the first exchange \citep{Sigurdsson:1993jz}. The tendency to exchange the heaviest compact  stars also has the consequence that the rates of ejection of binaries involved in three-body exchanges are less than those predicted  by models in which all stars have equal masses. The recoil speeds of the light, single stars are consequently larger.  

A large fraction of the encounters where the field star approaches within approximately a binary semi-major axis (SMA), $a_0$, of the binary center of mass result in resonant interactions, in which the three stars wander for a long time on chaotic orbits and approach each other repeatedly \citep{Heggie:1975uy, Hut:1993gs}.  During these chaotic encounters, the stars have many opportunities for close encounters. If the stars are compact,  angular momentum loss due to gravitational
radiation may become a noticeable effect during  close passages  \citep{Peters:1964bc}, and could cause the two stars to be driven together. 
It is the interplay between binaries  and compact objects in such dense environments and their ability to manufacture eccentric merging binaries in three-body exchanges that forms the main topic of this work.  

Our main goal  in this paper  is to study how the inclusion of gravitational wave (GW) losses modifies the compact binary  outcomes that originate from three-body scatterings, in particular during resonant interactions. The inclusion of GW losses into the binary-single dynamical system, we argue, introduces a new potential outcome in which a pair of objects may dynamically inspiral and merge while the three-body system is still in resonance. These outcomes are rare, and they are typically only realized during resonant interactions. Chaotic, resonant orbits augment the probability of very close passages when compared to direct interactions, and they can produce systems with correspondingly short GW inspiral time. \citet{2006ApJ...640..156G} first explored the cross section for these inspiral outcomes in the context of IMBH formation and growth. A surprising result of \citeauthor{2006ApJ...640..156G}'s simulations is that the cross section for inspiral outcomes {\it increases} with increasing binary SMA. This is perhaps counterintuitive because one might expect that the cross section for relativistic outcomes would be largest in very tight binaries. However, we will show that this is a natural consequence of resonant binary-single interactions, and that the scaling with binary SMA can be analytically derived.  

In this paper, we explore the cross section for dynamical inspiral outcomes during binary-single interactions through numerical experiments and analytic calculations. In Section \ref{sec:binary_single_encounters}, we review some of the dynamical properties and outcomes of binary-single interactions. In order to build intuition for how the inclusion of GW losses modifies binary-single interaction dynamics, in Section \ref{sec:Newtonian} we summarize the results binary-single scatterings with point masses in Newtonian gravity. Readers familiar with previous work in binary-single dynamics may wish to skip to Section \ref{sec:GW_losses}, in which we describe the inclusion of post-Newtonian (PN) corrections to the binary-single system equation of motion. Section \ref{sec:Inspirals} describes the formation of dynamical inspirals from resonant interactions between hard binaries and single objects. We explain the origin of these inspirals through numerical scattering experiments, and use our results to motivate an analytic derivation of the scaling of the inspiral cross section with binary SMA.  In Section \ref{sec:ecc_insp_in_LIGOband}, we show that dynamical inspirals give rise to inspirals that pass with high eccentricity through the {\it LIGO}\footnote{http://www.ligo.caltech.edu/} band. We compare this process to eccentric inspirals arising from single-single interactions and show that the cross section is greatly enhanced in binary-single interactions. In Section \ref{sec:discussion}, we extend our calculations to consider binaries containing white dwarfs, we discuss binary lifetimes and the role of GW emission, and we estimate whether the products of binary-single interactions are ejected or retained in their host stellar system. Finally, we estimate the rates of eccentric inspirals given typical globular cluster core properties.

\section{Binary-Single Encounters}\label{sec:binary_single_encounters}

Binary-single stellar encounters in dense stellar systems may be broadly divided into a few well-defined  categories. In the majority of encounters, the incoming object passes on a hyperbolic trajectory relative to the binary at a distance large compared to the binary separation \citep{Heggie:1975uy}. The passage time is greater than the binary's orbital period and the binary is subjected to a weak perturbation (WP). A strong perturbation (SP) is possible \citep{Heggie:1975uy} when the incoming object approaches the binary on a hyperbolic trajectory that happens to pass at a distance comparable to the binary SMA. In this case, the interaction time is less than  or similar to the binary's orbital period. 

The accumulation of WPs and SPs across the lifetime of a binary in a dense stellar system modifies the expected eccentricity and SMA distributions as compared to more isolated binaries. To quantify this effect, one must rely on integrations of the coevolution of binaries and their parent clusters over the cluster's relaxation time \citep[e.g.][]{Aarseth:1975kf,Hills:1975io,Hills:1975jk,Heggie:1975uy,Lightman:1978go,1986ApJ...306..552M,Baumgardt:2002eb,2003ApJ...593..772F,2005MNRAS.358..572I,2007ApJ...658.1047F,2009ApJ...707.1533F}.

\subsection{Close Interactions and Their Cross Section}\label{sec:Close Interactions and Their Cross Section}

\begin{figure}[tbp]
\centering
\includegraphics[width=1\columnwidth]{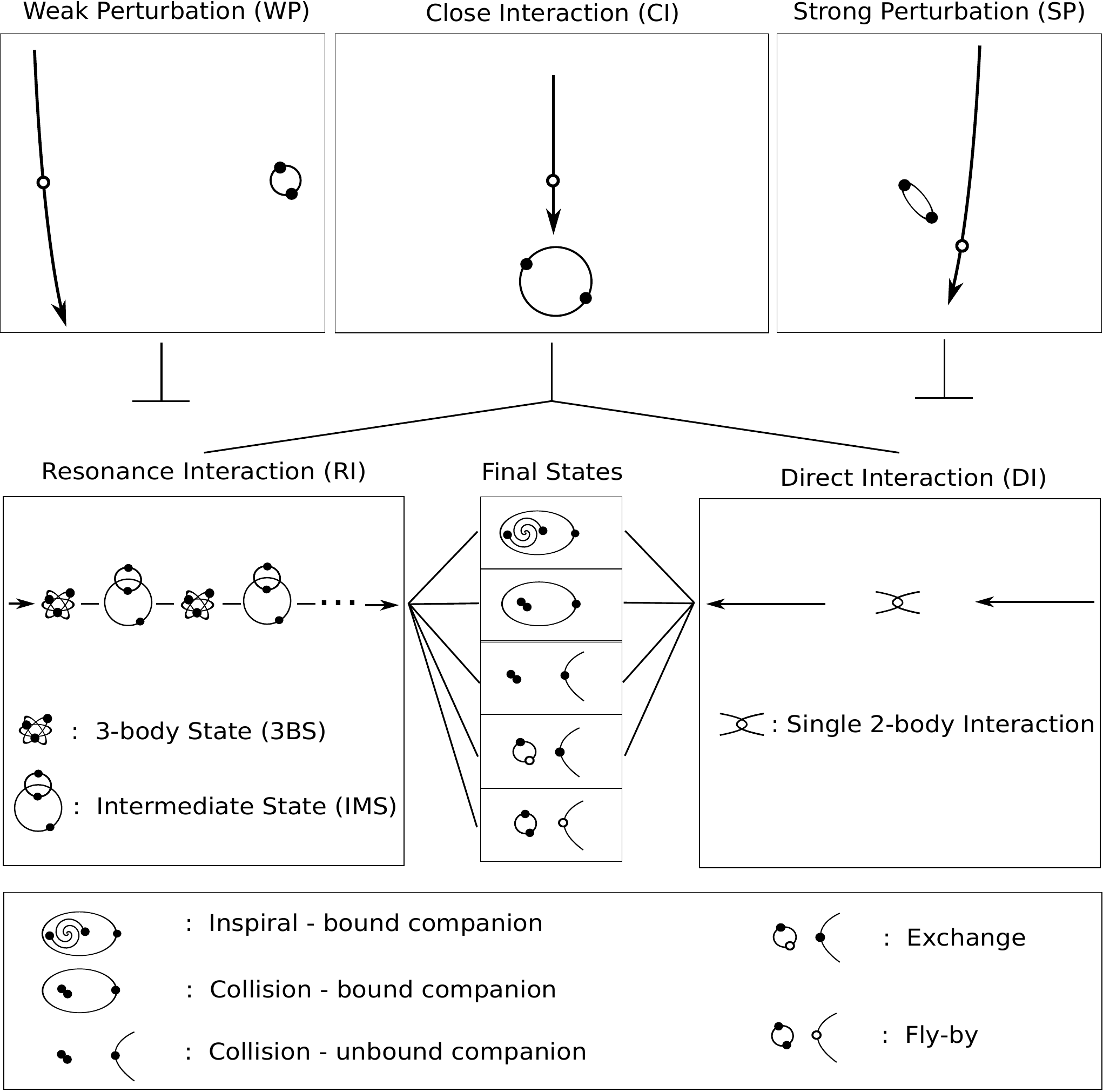}
\caption{Schematic illustration of binary-single interactions and their final states.
The {\it top panel} shows three
different types of interactions. The {\it top left panel} shows a weak perturbation (WP) where the single encounter
is only weakly perturbing the binary, but over several orbital periods. The {\it top right panel} shows a short but strongly
perturbing encounter (SP). A close interaction (CI)  is shown in the {\it middle panel}.
The evolution of the system from this CI channel can further be divided into the two interaction channels: direct interaction (DI) and
resonant interaction (RI). These are illustrated in the {\it middle panel}. The RI channel can be decomposed into
intermediate binary-single states (IMS), where an intermediate binary is formed with a bound companion.
Several IMS are created and destroyed in the chaotic RI before a final state is reached. The RI erases
any information of initial conditions. 
The DI channel is on the other hand very fast  and, as a result, the endstate depends sensitively  on the initial state.
Which channel dominates depends particularly on the mass ratio between the objects and the velocity of the incoming object.
The set of endstates from both the RI and the DI interactions are
listed in the {\it middle panel} where the individual interaction diagrams are defined in the {\it bottom panel}.
There is, in general, a similar final state scheme for each permutation of the objects.}
\label{fig:binsin interactions}
\end{figure}

A close interaction (CI), by contrast, occurs when the incoming object passes within a sphere of influence marked
by the binary's separation.  In these cases, the gravitational interaction between all three bodies may be of similar strength, and the outcomes are chaotic. In this work, we will focus on CIs and the dramatic role they play in reshaping binaries. Figure \ref{fig:binsin interactions} shows a schematic overview of the different  interactions and their expected outcomes.

We define a CI to have occurred when the third body passes within a distance $\rCI$ from the binary center of mass. We choose $\rCI$ as the distance from the center of mass to the lighter object in the binary,
\begin{equation}\label{eq:rCI}
\rCI = \frac{m_{2}}{m_1 + m_2 }a_{0},
\end{equation} 
where $1,2$ are the binary members in order of ascending mass $(m_2 > m_1)$, $3$ is the incoming object, and $m_{1}+m_{2}$ is the mass of the target binary.
This value is always between $a_{0}/2$ (if $m_1=m_2$) and $a_{0}$ (if $m_2\gg m_1$).

Whether a CI will occur is analytically predictable given the impact parameter, $b$, and velocity, $v_{\infty}$, of the third body relative to the target binary. 
At large separations between the binary and the incoming object, the fact that the binary is composed of two objects is unimportant and thus the encounter can realistically be treated as the interaction between two point masses: the binary with total mass $m_{\rm bin} = m_1 + m_2$ and the incoming object with mass $m_3$. In this case, a given distance of closest approach between the incoming single and the center-of-mass of the binary, $\rmin$, corresponds directly to an impact parameter, $b$, defined at infinity \citep{Sigurdsson:1993jz},
\beq
b = \rmin  \sqrt{1+\frac{2G \mtot}{  \rmin  v_{\infty}^{2}  }},
\eeq
where $v_\infty$ is the initial relative velocity at infinity of the binary center of mass and the single object, and $m_{\rm tot} = m_{\rm bin} + m_3$. The second term in this expression corresponds to the gravitational focusing of trajectories from an initially large impact parameter to a closer pericenter distance.
Because the argument of the square root is always larger than unity, $b$ is always greater than $\rmin$.

If we now consider the interactions with a closest approach less than the sphere of the binary, $\rCI$, then we see that all encounters with impact parameter less than the corresponding $b_{\rm CI}=b(\rCI)$ will
have $r_{\rm min}<\rCI$.
Therefore, all encounters coming from within the area $\sigma_{\rm CI} = {\pi}b_{\rm CI}^{2}$ will lead to an interaction
with $r_{\rm min} \leq \rCI$. This area $\sigma_{\rm CI}$ is defined as the \emph{cross section} for a close interaction. Given the definition of $b$ above, this may be written 
\begin{equation}
\begin{split}
\sigma_{\rm CI}  = {\pi}b_{\rm CI}^{2}  = {\pi} \rCI^{2}\left(1+\frac{2G \mtot}{  \rCI v_{\infty}^{2}  }\right).
\end{split}
\label{eq:def_of_crosssection}
\end{equation}
Whether the first (geometric) or second (gravitational focus) term in parenthesis dominates depends on the relative binding energy of the binary and the kinetic energy of the incoming object. 

Given a distribution of single stars, the CI cross section, $\sigma_{\rm CI}$, gives an estimate of how often such interactions can occur. 
As $\sigma_{\rm CI}$ increases,
the more encounters will be focused into the binary system. In a stellar system with an isotropic stellar density, $n$, and typical relative velocity, $v_\infty$, this rate of CIs per binary may be approximated as
\beq
\Gamma_{\rm CI} \simeq n \sigma_{\rm CI} v_{\infty}.
\eeq
Thus, given a stellar distribution, the cross section is the only factor that determines the relative rates of different processes. For this reason, significant  effort will be invested  in deriving the cross sections of the  various outcomes of CIs as fractions of the total CI cross section. 
In the following section, we explore the role of the relative energy of the binary and the single object in shaping binary-single interactions. 

\begin{figure*}[tbp]
\centering
\includegraphics[width=0.33\textwidth]{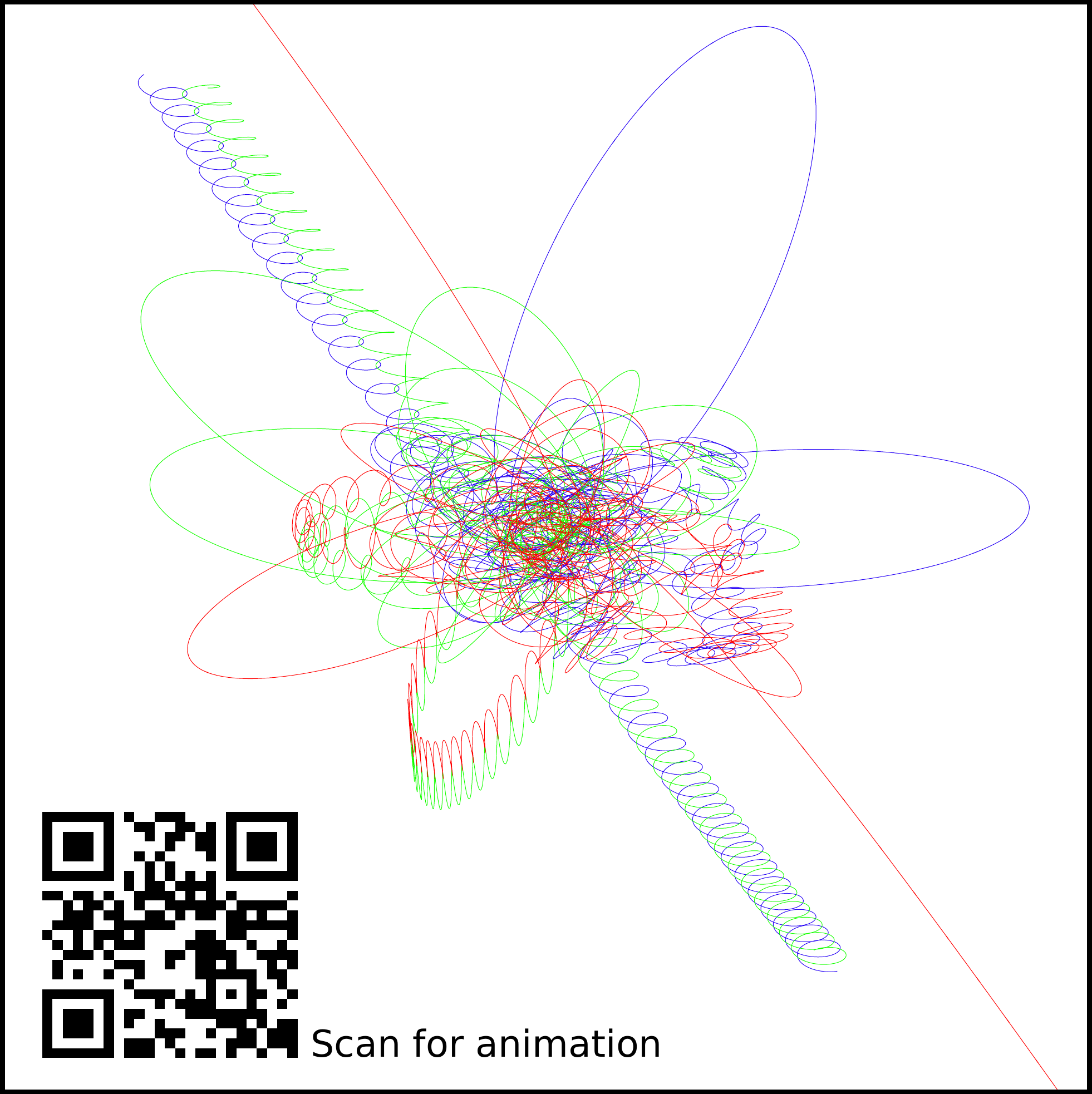}
\includegraphics[width=0.33\textwidth]{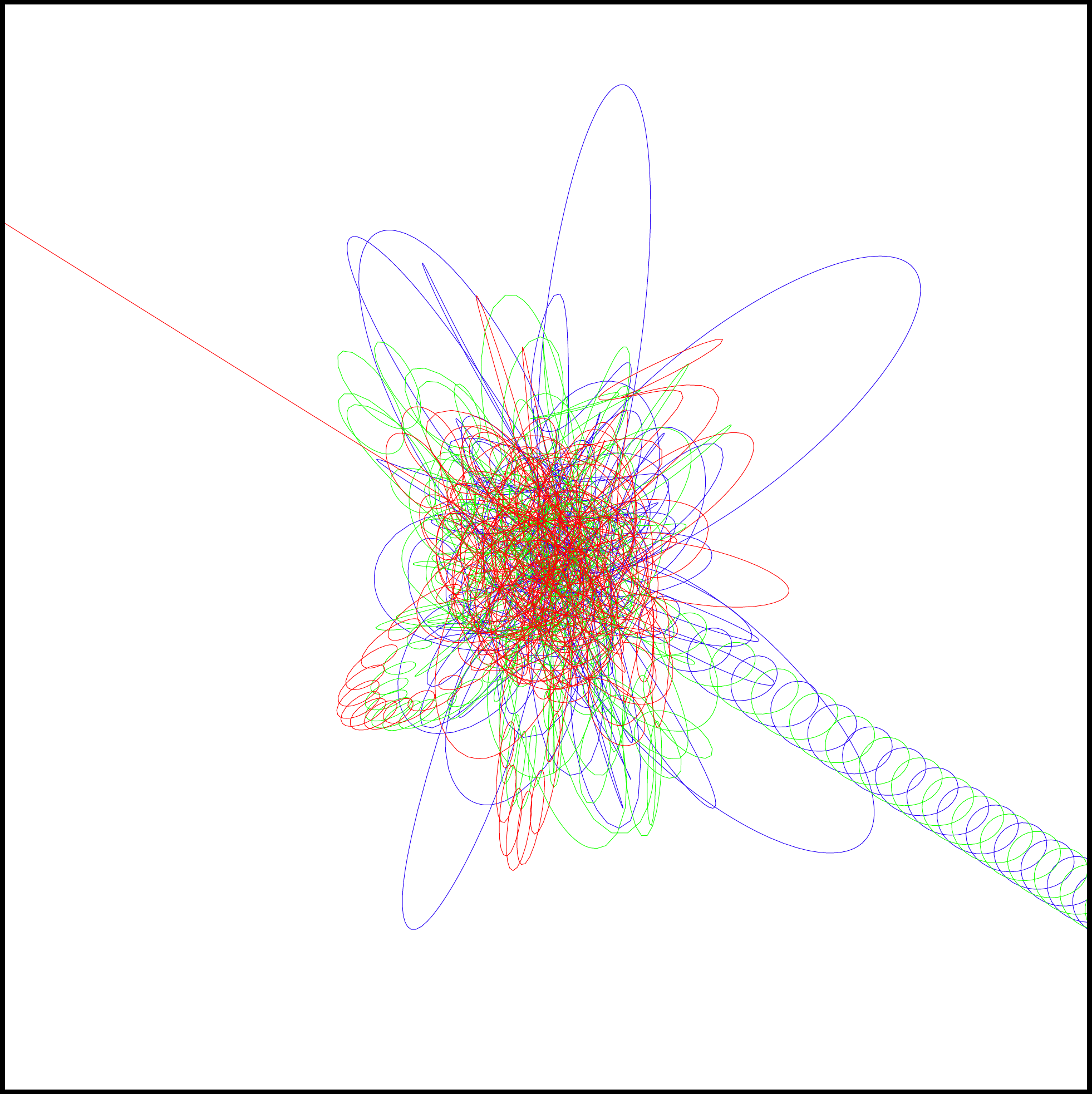}
\includegraphics[width=0.33\textwidth]{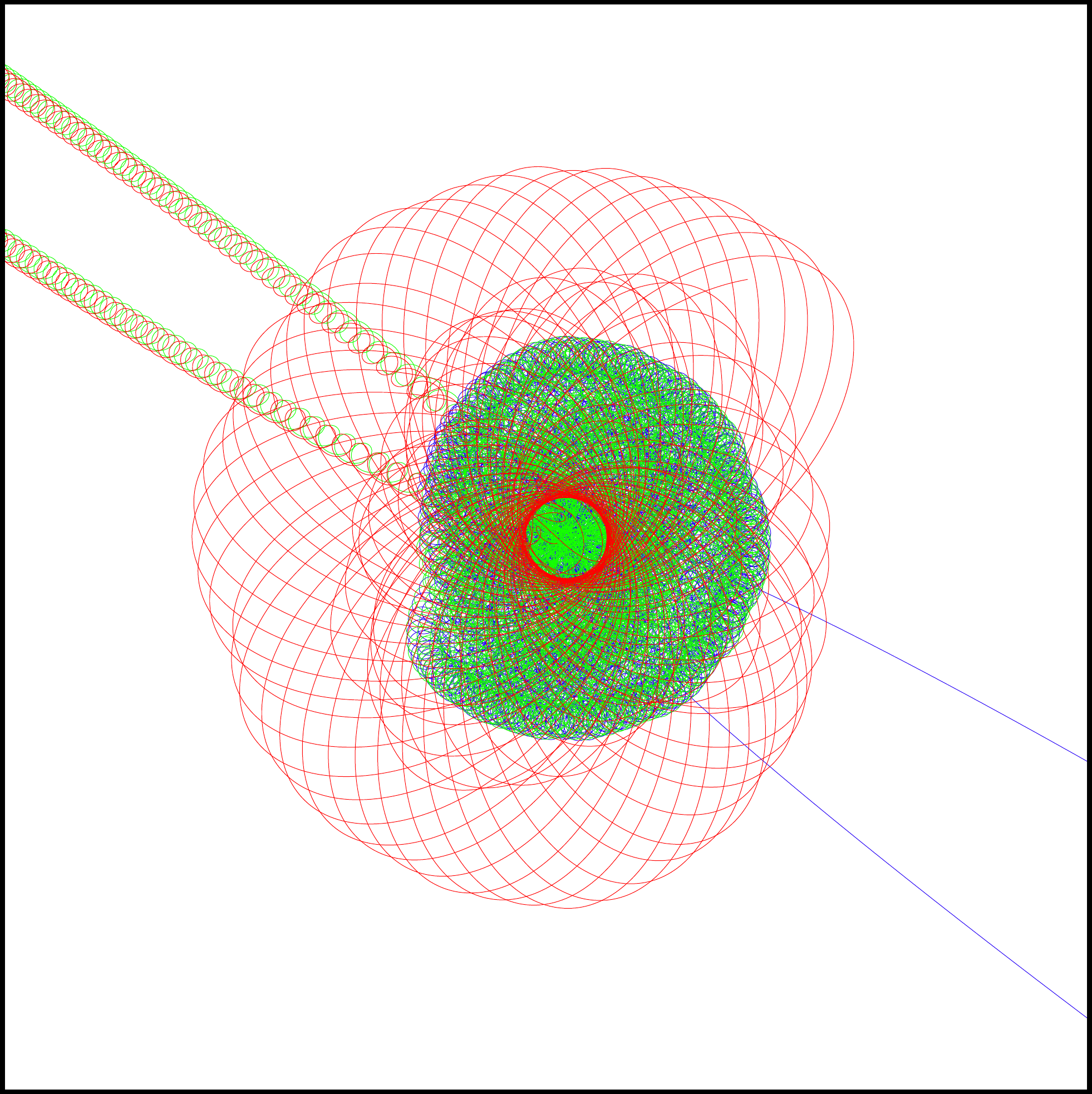}
\caption{Examples of binary-single resonance interactions between equal mass black holes.
\textit{Left:} The close interaction forms a few intermediate binary states (red-green)  with a bound companion (blue). We denote these intermediate binary-single states (IMS).
Scan the {\tt QR code}\footnote{or visit {\tt http://youtu.be/ipPniBvZvxY}}
to see an
animation of this interaction.
The last frame in the animation is  shown in the plot.
\textit{Middle}: This interaction have the same total energy as in the {\it left} panel
but the energy is distributed  differently in the system. 
\textit{Right}: A relative rare interaction  class  is displayed in this panel, with a system  composed of a binary (blue and green) that remains  bound to a  companion (red)  for many orbits. The final state is a collision.
The interactions shown in the {\it left} and {\it middle} panels are generally refer to as {\it democratic resonances} where the right panel  shows an example of an 
{\it hierarchical resonance}.
The many IMS created during  these resonant interactions aid  the formation of  eccentric binaries with  short inspiral times. Binaries that inspiral and merge due to GW radiation during a resonance interaction are called {\it inspirals}. An example is shown in Figure \ref{fig:example_inspiral_fromsim}.}
\label{fig:scatterex123}
\end{figure*}

\subsection{Hard and Soft Target Binaries}

The relative velocity of the binary and the single object, $v_\infty$, as compared to the characteristic velocity of a binary, $\vc$, determines the outcomes that are possible in a binary-single interaction.  A binary's characteristic velocity is defined as \citep{Hut:1983js}
\begin{equation}
\begin{split}
\vc^2	& = G \frac{ m_{1}m_{2}(m_{1}+m_{2}+m_{3})}{m_{3}(m_{1}+m_{2})}\frac{1}{a_{0}}.
\end{split}
\label{eq:formulae for v_c}
\end{equation}
This velocity is written such that if the relative velocity at infinity is larger than $\vc$ ($v_{\infty}>\vc$),
then the total energy of the three-body system is positive \citep{Heggie:1975uy}. 

A binary with $v_\infty > \vc$ is described as a soft binary (SB) relative to its environment. 
The cross section for close interaction, equation \eqref{eq:def_of_crosssection}, is well approximated by the binary's geometrical cross section, $\pi \rCI^2$.
Because the velocity at infinity is greater than the binary's orbital velocity, the binary appears nearly static during the interaction. 
The resultant  encounters can thus be viewed mainly as  two-body interactions that are well described by  {\it impulsive} approximations \citep{Heggie:1975uy,Hut:1983by}.
Additionally, with $v_{\infty}>\vc$ the incoming body carries a  large amount of energy when  compared to the binary's binding energy.
That excess of energy can effectively be utilized to split the binary \citep{Heggie:1975uy}.

Hard binaries  (HB) are characterized by $v_{\infty}<\vc$. In this case, the cross section for CI is dominated by the gravitational focus term, and 
\beq
\sigma_{\rm CI}  \simeq \frac{2\pi G \mtot \rCI}{ v_\infty^2 }
\label{eq:CC_CI_gravfoc_dominates}
\eeq
Thus, in this limit, $\sigma_{\rm CI} \propto a_0 / v_\infty^2 $.
Further, the energy carried from the encounter into the system is relatively small and a temporary bound triple state can be formed \citep{Hut:1983by}.

In dense stellar systems, the HB limit is typically the relevant limit for the steady-state binary population. 
Equation \eqref{eq:formulae for v_c} can be re-written for equal mass encounters as 
\beq\label{eq:vc_numbers} 
\vc \approx 36.5 \left( m /M_\odot   \right)^{1/2} \left( a_0 / {\rm AU}   \right)^{-1/2} \text{km s}^{-1}.
\eeq
Values for $v_{\infty}$ are in the 10-50 km s$^{-1}$ range for galactic GCs \citep{Lightman:1978go}. Thus any binaries with SMA smaller than $\approx 1$ AU will be in the HB limit. In clusters, HBs tend to be the ones that survive as encounters tend to split soft binaries \citep{Heggie:1975uy}. Further, based on a statistical trend toward energy equipartition \citep{Heggie:1975uy,Hills:1975jk}, hard binaries tend to become harder (as energy is transferred from the binary to put the single on an unbound orbit) while soft binaries get  softened or disrupted (as the incoming single star pumps energy into  the system before leaving). This natural selection makes a hard binary population even harder and causes a soft binary population to evaporate. 

Binary-single CIs involving HBs may be decomposed into direct interactions (DIs) and resonant interactions (RIs). DIs are brief, two-body interactions which occur when the incoming body passes very close to only one of the binary members. In these cases, the interaction is brief and the initial conditions with which the single object entered the binary are key in determining the outcome. By contrast, a RI is comprised of many intermediate exchanges of binary and single star hierarchy. We denote these temporary triple-object states, comprised of a binary and a bound single, as intermediate states (IMSs).  The IMS decomposition is illustrated in Figure \ref{fig:binsin interactions}.

The number of resonances  a system undergoes during a RI depends
on the mass ratio of the interacting objects and is maximized for equal mass objects  \citep{Sigurdsson:1993jz}. 
In the equal mass case, these RIs can have lifetimes extending from one to several hundred times the orbital period of the initial target binary.  If one of the objects is lighter compared to the others, this object is likely to be dynamically kicked out, leaving the heavier objects behind as a binary \citep{Sigurdsson:1993jz}. An illustration of the possible orbital morphologies of RIs is shown in Figure  \ref{fig:scatterex123}.  Examples of both democratic (similar pairwise binding energy) and
hierarchical (disparate pairwise binding energies) resonances can be clearly seen in  Figure \ref{fig:scatterex123}.

Compared to the entire duration of a RI, the lifetimes of individual IMSs are relatively short. This implies that a single RI encompasses many IMS exchanges in which close encounters  occur and the binary-single system is transformed. The IMSs themselves are unstable because they are disrupted every time the current bound single object makes a close passage.  Over the course of several such IMS changes (through  three-body interaction {\it knots}), the triple system evolves chaotically, loosing memory of the initial conditions with which the single object first entered the binary \citep{Heggie:1975uy}. Rare outcomes may be achieved with higher likelihood in RIs for the simple reason that the single object makes many randomized close passages through the binary system. This is particularly significant when GW radiation is included into the three-body equation of motion because there is a non-negligible  probability that a very close (and thus highly dissipative) passage will take place.

\begin{figure}
\centering
\includegraphics[width=1\columnwidth]{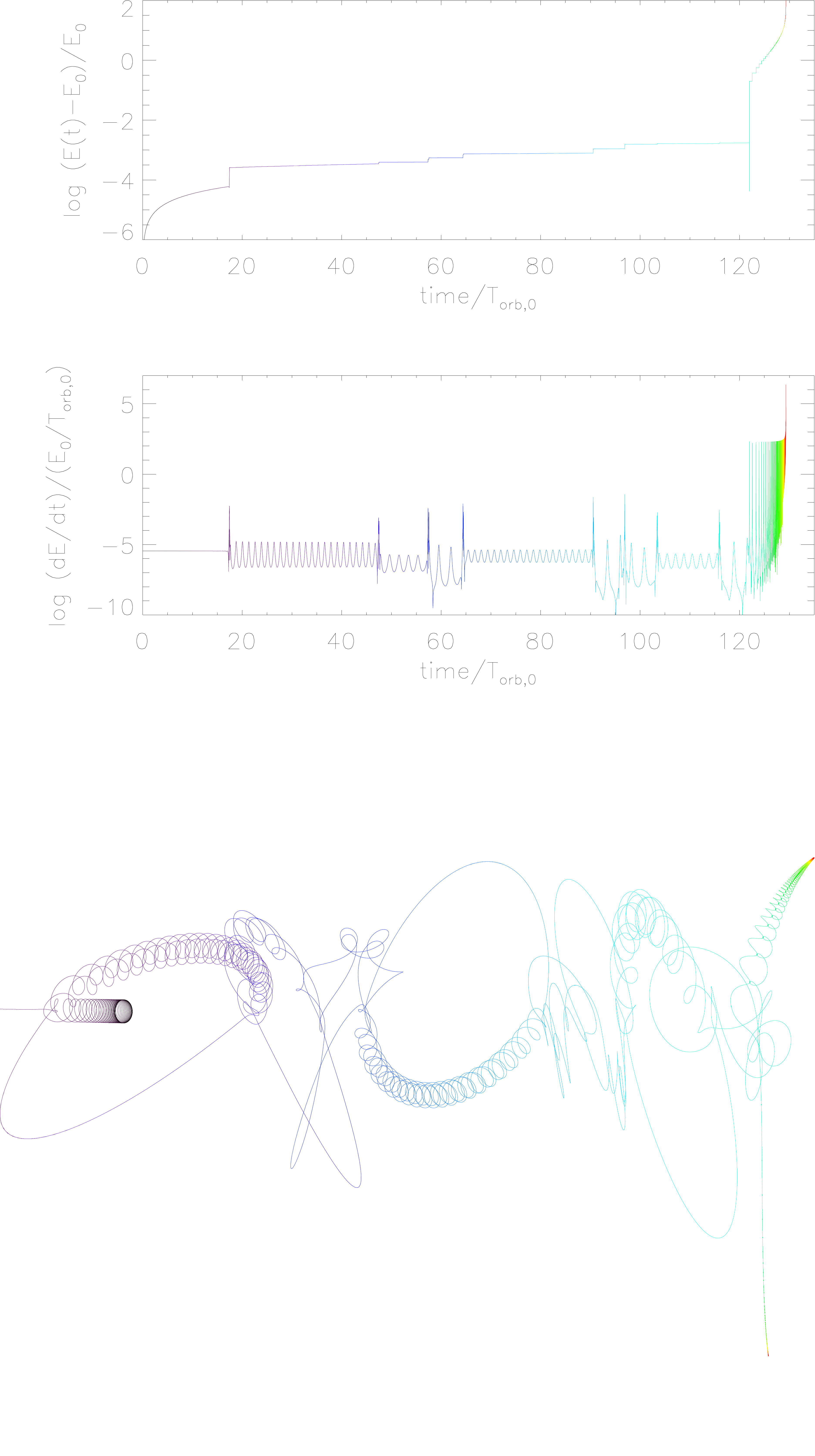}
\caption{Example of a binary-single interaction which ends as an {\it inspiral}. These are the new general relativistic (GR) endstates that  are the main
focus of our work. {\it Inspirals} are IMS-formed binaries that  merge due to GW radiation during the resonance interaction,
i.e. while the single object is still bound. The color across all plots denotes time.
\textit{Top}: Fraction between the energy loss of the system after a time $t$ and the initial energy of the system $E_{0}$.
Any deviations from zero are due to 
energy radiated away by GWs.
\textit{Middle}: Energy loss $dE/dt$ as a function of time. The oscillating form of $dE/dt$ arises because the
system evolves between multiple  IMS.
\textit{Bottom}: The ensuing  binary-single trajectories. The interaction starts at the left where the binary interacts
with the  incoming object. The final state seen at the far right is an IMS binary that inspirals due to GW radiation while the third object is
still bound. The final inspiral is as expected characterized by a
large and rapid increase in GW losses. These inspirals can by observed with {\it LIGO} and as we will show later are
likely to be highly eccentric at the time of observation, which  makes them particularly interesting.  }
\label{fig:example_inspiral_fromsim}
\end{figure}

\subsection{Outcomes of Close Interactions}

In the previous section we have described how CIs arise in binary-single star encounters and how their likelihood can be quantified by their cross section, $\sigma_{\rm CI}$. During a CI, the system is in a three-body state, but no three-body state is stable \citep{Hut:1993gs} and the
system will thus  invariably evolve (through the DI or the RI channel) into one out of the several possible final-states (or \emph{outcomes}) as illustrated in Figure \ref{fig:binsin interactions}.  In general, there is a given  cross-section for each of these possible outcomes to occur. In Section \ref{sec2numerical}, we describe how we compute these outcome cross sections statistically based on the fraction of binary-single scatterings that can generate a given outcome. 
In the two sections below we describe in detail the particular final outcomes expected from  CI interactions.

\subsubsection{Outcomes from Newtonian Gravity}\label{sec:Outcomes from Newtonian Gravity}

In Newtonian gravity, a binary-single interaction can result in a binary with an unbound companion, a collision, or three unbound objects. The cases in which a binary is left behind may be further subdivided based on the properties of the surviving binary \citep{Heggie:1975uy,Hut:1983js}. If the binary is composed of the original two objects (1,2) then we refer to the encounter as a {\it fly-by} even though the endstate binary may be the result of a more complex interaction than the fly-by label suggests.  If instead the binary is composed of one of the original binary members and the third body, we denote the encounter an {\it exchange}. In this case, the binary may either be (1,3) or (2,3). An outcome in which all three members are mutually unbound is possible when the total system energy is positive, $v_\infty > \vc$. This outcome is denoted as an {\it ionization}.  Collisions are possible at all values of $v_\infty$, but they are most likely to occur at negative total binding energies where the gravitational focus cross sections of the individual objects are larger.

\subsubsection{Inspiraling Binaries due to GW Emission}

With GW emission included in the three-body equation of motion, a new outcome is possible: {\it dynamical inspirals}. Inspirals are characterized by the gravitational radiation driven inspiral of an IMS binary, while the third object is bound to the binary. Inspirals are particularly likely to occur during RIs. The magnitude of GW emission depends strongly on the distance of closest approach between two objects \citep[e.g.][]{Peters:1964bc}.  In relatively widely separated binaries, inspirals do not result from tightly bound circular orbits, but rather they are the product of orbits of very high eccentricity in which the objects experience close pericenter passages that generate significant GW emission and thus substantially reduce their orbital energy and angular momentum. High eccentricity orbits are most readily achieved in the chaotic environment of RIs, where despite the $e=0$ initial conditions we impose on the binaries, the angular momenta of the three bodies is randomized and approaches an isotropic distribution with increasing number of passages. 

Figure \ref{fig:example_inspiral_fromsim} shows an inspiral from one of our simulations. The binary-single interaction happens at the left of the plot and then propagates towards the right, terminating  with the inspiral. One important feature of this interaction is that the bulk of the energy losses occur in three body knots, where the relative orbital angular momenta of the bodies is randomized, and the objects undergo very close pericenter passages, which in turn  give rise to the spikes seen in the energy loss rate. 
Inspirals are of particular interest, as we will show in this paper, because they  occur more frequently in widely separated target binaries, and they give rise to eccentric compact object mergers. 

\subsection{Numerical Approach}\label{sec2numerical}

Here  we  study the outcomes of binary-single interactions and their associated cross sections
by performing   large sets of numerical  scattering experiments. To this end, we have developed a new N-body code to integrate the equation of motion of the three bodies using a fourth order Hermite integration scheme.
The equation of motion including the effect from GW emission is discussed in Section \ref{sec:addingGR}. For
a full description of the code and the exact state classification criteria employed,  the reader is referred  to the Appendix.\footnote{In the Appendix we also directly test the code against the \citet{Peters:1964bc} analytic solution for binaries inspiraling due to GW emission.} For each scattering experiment the target binary was randomly orientated
in phase and orbital plane orientation. 

We estimate the cross section numerically for a given outcome type $O_{i}$
by performing $N_{\rm tot}$ binary-single interactions with isotropic sampling across a disc at infinity with radius $b$.
If the total number of outcomes of type $O_{i}$ from that scattering set is denoted by $N_{i}$, then the corresponding 
cross section for outcome $O_{i}$ can be estimated by 
\beq\label{numerical_cross_section}
\sigma_{i}= \frac{N_{i}}{N_{\rm tot}}{\pi}b^{2}
\eeq
 with a corresponding error given by 
\beq
\Delta{\sigma_{i}}= \frac{\sqrt{N_{i}}}{N_{\rm tot}}{\pi}b^{2}.
\eeq
This, in turn, implies a rate of a given outcome $O_i$,
\beq
\Gamma_i \simeq n \sigma_i v_\infty
\eeq
expected from a distribution of single objects with number density $n$ and typical relative velocity $v_\infty$. Thus, the rate of outcomes of type $O_i$ compared to the rate of CIs is defined by the ratio of their cross sections, $\Gamma_i / \Gamma_{\rm CI} = \sigma_i / \sigma_{\rm CI}$.

%
%
\section{Newtonian Point-Particle Limit}\label{sec:Newtonian} 

To build intuition and to provide a direct link to previous studies in Newtonian gravity, we will first describe the most salient features of binary-single encounters of point masses in Newtonian gravity. These interactions
and their final states, or outcomes, are well studied numerically and theoretically, especially in the pioneering series
of work by \citet{Hut:1983js,Hut:1983by,Hut:1993gs,Heggie:1993hi,Goodman:1993gg,1996ApJ...467..348M,Heggie:1996fz}. More recent work by \citet{Fregeau:2004fj} and \citet{2007ApJ...658.1047F}
have extended such studies to calculate  the probability for collisions, and the coevolution of binaries and their host clusters.

When the three objects are equal point-masses, the outcome of an interaction will always be
either a {\it fly-by},  an {\it exchange} or an  {\it ionization}. These outcomes were described in Section \ref{sec:Outcomes from Newtonian Gravity}.
In this Section, we calculate their associated cross section over a broad range of encounter velocities $v_{\infty}/v_{c}$ using  a series of numerical scattering experiments.
In our equal mass case, 
\begin{equation}
{v_\infty \over v_{\rm c}} = v_\infty\sqrt{2 a_0  \over 3 m},
\end{equation}
thus any defining characteristics  of the system can be   rescaled using this ratio.
We perform a total of $8\times10^5$ binary-single scatterings divided into $40$ sets each with $2\times10^4$ interactions. 
For each scattering experiment, the target binary is randomly
orientated in phase and orbital plane.
The velocities of the encounters for the 40 sets are equally spaced in $\text{log} (v_{\infty}/\vc )$ from
$0.01$ to $8$. The maximum impact parameter, $b_{\rm max}$, is kept fixed for all
scatterings at $5a_{0}$. In this setup, outcomes from all the three interaction channels WP, SP and CI will  occur depending on ${v_\infty/v_{\rm c}}$.
Our numerical approach is closely related to the one used in \cite{Hut:1983js}.
We also refer the reader to the Appendix for further details on our numerical approach.

\begin{figure}
\includegraphics[width=1\columnwidth]{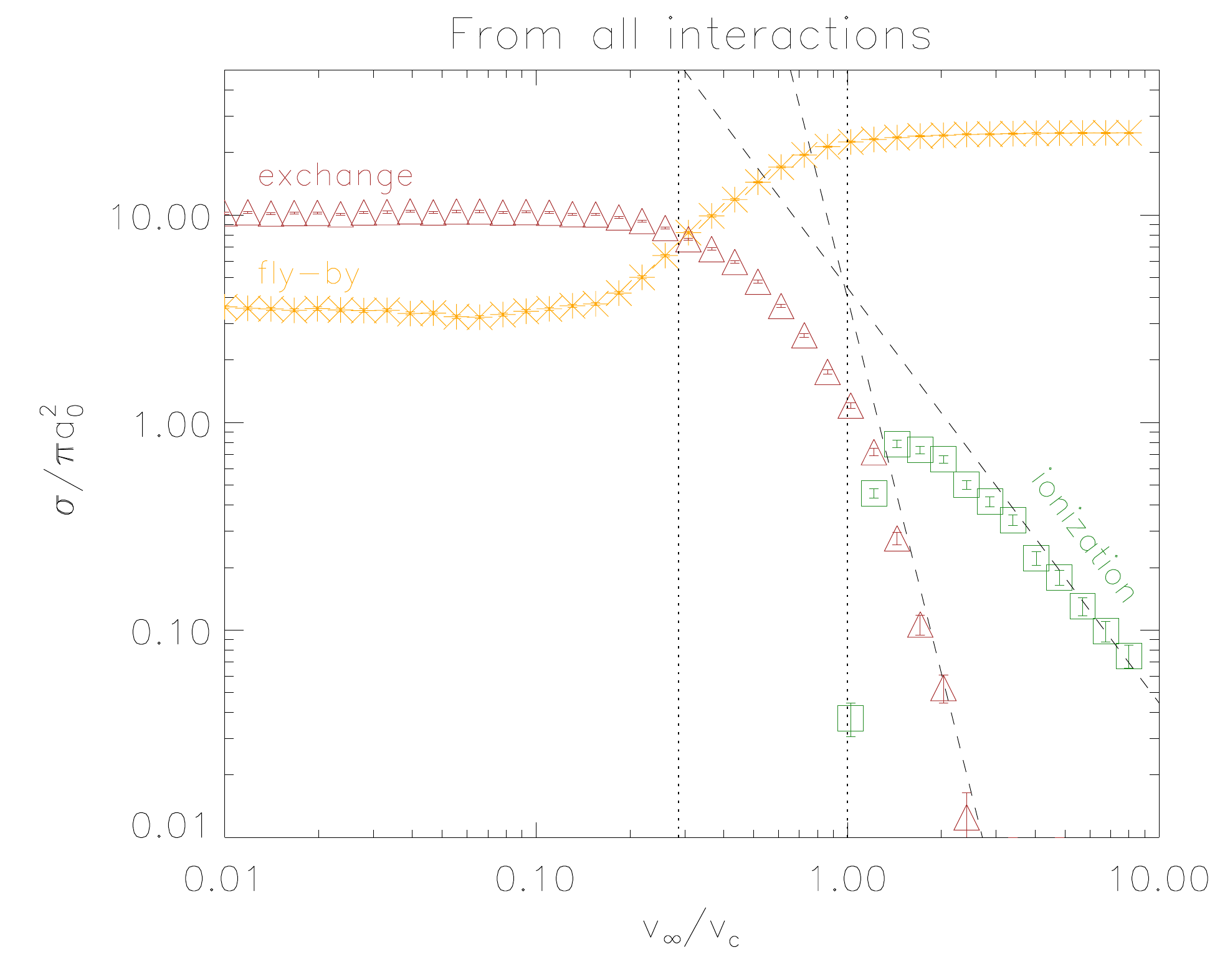}
\includegraphics[width=1\columnwidth]{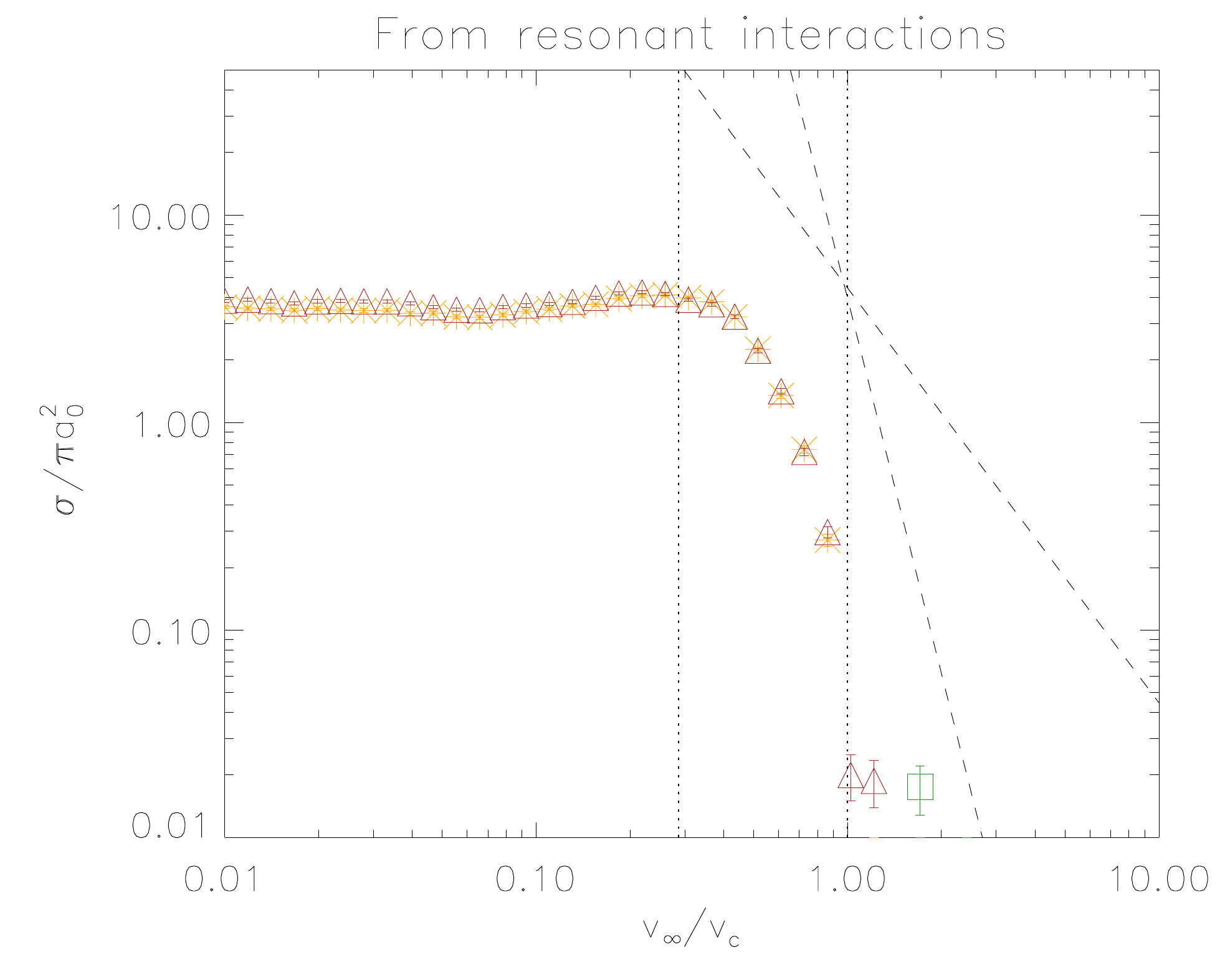}
\caption{Integrated cross sections for the classical outcomes: {\it exchange} (brown triangles), {\it ionization} (green squares)
and {\it fly-by} (orange stars) as a function of $v_{\infty}/v_{\rm c}$, where $v_{\infty}$ is the relative velocity of the incoming object at infinity
and $v_{\rm c}$ the characteristic velocity given by equation \eqref{eq:formulae for v_c}. 
The dashed lines show analytical approximations to the {\it exchange} (equation \ref{eq:exchange_crosssec}) and {\it ionization} (equation \ref{eq:ionize_crosssec}) cross
sections. 
The vertical dotted lines
indicate two characteristic velocities, the gravitational focusing velocity $v_{\rm foc}/v_{\rm c} (b_{\rm max}= 5a_0)\approx 0.28$ and
the velocity that divides the system
into having total positive or negative energy, $v_{\infty}=v_{\rm c}$.
\textit{Top}: Cross sections calculated from all interactions including RIs and DIs.
\textit{Bottom:} Cross sections only including endstates coming from  RI encounters. This channel erases any information about initial
conditions and all the three objects have  thus equal probability  to be kicked out. As a result, the {\it fly-by} and {\it exchange} cross sections are 
identical. Because  a {\it fly-by} can not result from a DI, the {\it exchange}
and {\it fly-by} cross sections are separated   in the {\it top} panel. As can be clearly seen, the cross section for a RI  is independent of $v_{\infty}$ as
long as $v_{\infty}<v_{\rm foc}$. Each plot is based on a total of $8\times10^5$ scatterings.}
\label{fig:piethutplot_ngr}
\end{figure}

Figure \ref{fig:piethutplot_ngr} shows the results from our scattering experiments. Both panels show
the cross sections for exchange, fly-by, and ionization outcomes as a function of $v_{\infty}/v_{\rm c}$.
The upper panel includes outcomes from all interactions including DIs and RIs, while the lower panel shows the outcomes coming from the RIs only.
In what follows, we detail the outcomes and their dependence with ${v_\infty/v_{\rm c}}$.

\subsection{Low Velocity $(v_{\infty}/ \vc \ll1)$}\label{sec:new_low_vel_sec}

At low velocities, gravitational focus leads to all interactions happening via the CI channel.
Therefore, all final state outcomes will be a result from either the DI or the RI channel.
Since the total energy of the three-body system is initially negative and no bound triple state can form a stable final state \citep{Hut:1993gs}, the
only possible outcome is a binary (carrying the negative energy part in form of binding energy) and
a single unbound object. Depending on which two objects that form the binary the outcome will either be labeled as an exchange or a fly-by.

Within the CI channel the probability for a given outcome depends on whether the binary has experienced a RI or a DI.
If the outcome is a result of the RI channel, then any permutation of the three objects in the final state is equally likely
since the RI erases any memory of the binary's initial configuration. As a result, the exchange and fly-by outcomes have the same cross section
when the system has evolved through a RI.
This can be seen in the lower panel of Figure \ref{fig:piethutplot_ngr}.

For interactions passing through the DI channel, fly-bys have a negligible probability to occur. The reason is that a DI
is characterized by having only a single interaction that in the majority of cases leads to an exchange between the incoming object and one of the binary members.
A typical fly-by involves at least two closest IMS pairs leading these interactions to be classified as arising from the RI channel. 
This leads to the cross section difference between exchange and fly-by when all interactions are included as seen in the upper panel of Figure \ref{fig:piethutplot_ngr}.

The critical velocity that defines the transition to all interactions happening through the CI channel, $v_{\rm foc}$, is found from equation \eqref{eq:def_of_crosssection},
\begin{equation}
{v_{\rm foc} \over \vc} = \sqrt{2}\left({a_0 \over b_{\rm max}}\right),
\label{eq:foceq}
\end{equation}
which, in our numerical setup  with $b_{\rm max} = 5a_0$, gives $v_{\rm foc}/v_{\rm c} = 0.28$. This critical velocity transition is illustrated
with a vertical dotted line in Figure \ref{fig:piethutplot_ngr}. It is clear in the lower panel in Figure \ref{fig:piethutplot_ngr} that
this line accurately separates the plot into two regimes. The cross sections are approximately flat to the left of this line, when $v_{\infty}<\vc$.
This tells us that the relative numbers of RIs and DIs are nearly constant, and as a result, independent of the exact impact parameter and encounter velocity as long as
the interaction is a CI.

\subsection{Intermediate Velocity ($v_{\infty} / \vc \approx1$)}

At intermediate velocities, the resultant encounters are a mixture of CIs, SPs and WPs, and the velocity
dependence shapes the resultant cross-sections. CIs can still occur at intermediate velocities, but their probability   decreases as $\sigma_{\rm CI} \propto {(v_{\infty}/v_{\rm c})^{-2}}$, as given by equation \eqref{eq:def_of_crosssection}.
This scaling solely determines the shape of the exchange cross section in this regime, since exchanges only can happen via a CI. This is seen in Figure \ref{fig:piethutplot_ngr} where
the exchange cross section is observed to clearly transition from being flat at low velocities to decreasing as $\propto {(v_{\infty}/v_{\rm c})^{-2}}$
at  intermediate velocities.

WPs and SPs happen with increasing frequency as the velocity is increased since more encounters pass by the binary instead of making a CI.
These perturbative encounters necessarily result in a fly-by classification since the encounter never comes close enough to make an exchange,
thus leading to a velocity dependent increase in the associated cross section.

\subsection{High Velocity ($v_{\infty}/ \vc >1$)}

In high velocity interactions ($v_{\infty} > \vc$), the total energy of the three-body system is positive
and ionization becomes a possible outcome. Ionization occurs  when all three objects are unbound with respect to each other.
This outcome dominates over the exchange
outcome in this high velocity regime as seen in the upper panel in Figure \ref{fig:piethutplot_ngr}.

Because of the high velocity, CIs are rare. The CI cross section is determined by the geometrical term in equation
\eqref{eq:def_of_crosssection}. 
Since the geometrical term only depends on the size of the target binary,
 the occurrence  of a CI is independent of velocity. 
 By contrast to the intermediate velocity range, the observed steep decrease in both the exchange and ionization cross sections as the velocity increases is  a result of properties of the interactions themselves, rather than a varying number of CIs.

As observed in the lower panel of Figure \ref{fig:piethutplot_ngr}, RIs  do not occur at high velocity. 
All outcomes from the CI channel are, therefore, only arising from the DI channel.
The main reason for this is that the incoming object  enters the binary with such a high velocity that the pair appears to be approximately stationary.
The majority of interactions between the single and the binary will therefore be a DI between the incoming single  and its nearest binary object.
The problem therefore reduces to a two-body interaction between the encounter and one of the binary members. This setup has an analytical solution and cross sections for exchange and ionization can be analytically estimated in this so called impulsive regime.
This was first done by \cite{Hut:1983by} who calculated in this high velocity regime the exchange cross section
\begin{equation}
\sigma_{\rm ex} = \frac{320}{81}\frac{{\pi}a_{0}^{2}}{v_{\infty}^{6}}, \label{eq:exchange_crosssec}
\end{equation}
and the ionization cross section
\begin{equation}
\sigma_{\rm ion} = \frac{40}{9}\frac{{\pi}a_{0}^{2}}{v_{\infty}^{2}}. \label{eq:ionize_crosssec}
\end{equation}
These scalings are also shown in Figure \ref{fig:piethutplot_ngr}. 
The similarity of this three-body scattering problem to atomic physics can be seen by comparing the exchange scenario,
in the limit where one of the binary members are very light, with 
electron capture (or charge transfer) in heavy nucleus interactions \citep{Shakeshaft:1979gz}.

%
%
\section{Gravitation Wave Losses and Three Body Dynamics}\label{sec:GW_losses}

In this section, we describe how  general relativity (GR) corrections are included  into the equation of motion in our three-body integration code, and highlight the dynamical consequences of these loss terms. 

\subsection{Adding General Relativistic Corrections}\label{sec:addingGR}

In this work, we include the energy and angular momentum losses by GW radiation using the PN formalism \citep{Blanchet:2006kp}.
In this formalism, the acceleration experienced by an object of mass $m_1$ due to the gravitational force from a second object of mass $m_2$ is expanded in series as
\begin{equation}
\acc=\acc_{0} + c^{-2}\acc_{2} + c^{-4}\acc_{4}+c^{-5}\acc_{5}+\mathcal{O}(c^{-6}).
\label{eq:pn_overview}
\end{equation}
The standard Newtonian force per unit mass, $\acc_{0}$ is
\begin{equation}
\acc_{0} = - \frac{G m_2}{r_{12}^2} {\bf \hat r}_{12},
\end{equation}
where the separation vector is ${\bf r}_{12}  = {\bf r}_1 - {\bf r}_2$, its magnitude is $r_{12} = |{\bf r}_{12}|$, and its direction is ${\bf \hat r}_{12} = {\bf r}_{12}/r_{12}$. The terms $\acc_{2}$ and $\acc_{4}$ account for the periastron shift. The leading order term that represents the radiation of energy and momentum from the system, $\acc_{5}$, is also known as the 2.5PN term.  This term
takes the following form
\begin{equation}
\begin{split}
\acc_{5} =	&  \frac{4}{5}\frac{G^{2}m_{1}m_{2}}{r_{12}^{3}}    \left[  \left ( \frac{2 Gm_1}{r_{12}}  - \frac{8 Gm_2}{r_{12}} - v_{12}^2 \right )  {\bf v}_{12} \right. \\
  		& \left. + \ ( {\bf \hat r}_{12} \cdot  {\bf v}_{12}  )\left ( \frac{52 Gm_2}{3 r_{12}} - \frac{6 Gm_1}{r_{12}} + 3v_{12}^2   \right )   {\bf \hat r}_{12}  \right],
\end{split}
\label{eq:25PNterm}
\end{equation}
where the relative velocity scalar, $v_{12}$, and vector, ${\bf v}_{12}$, are defined following the same conventions as in \citet{Blanchet:2006kp}. 
We use the modified acceleration $\acc = \acc_{0}+c^{-5}\acc_{5}$ in our numerical treatment instead of the Newtonian $\acc_{0}$.
A fundamental difference between the purely Newtonian acceleration and the 2.5PN acceleration is that $\acc_5$ depends not only on the separation between the objects but also on their relative velocity.

The energy and angular momentum losses through the 2.5PN term should coincide with those calculated using
the quadripolar formalism for two bodies. To  this end, the orbit-averaged equations for the time dependent evolution of SMA, $a$, and eccentricity, $e$, of a two-body system emitting GWs derived by  \citet{Peters:1964bc}  have provided a useful test framework  to many authors,
\begin{equation}
\frac{da}{dt} = -\frac{64}{5}\frac{G^{3}m_{1}m_{2}(m_{1}+m_{2})}{c^{5}a^{3}(1-e^{2})^{7/2}}\left(1+\frac{74}{24}e^{2}+\frac{37}{96}e^{4} \right),
\label{eq:peterssoldadt}
\end{equation}
and
\begin{equation}
\frac{da}{de} = \frac{12}{19}\frac{a}{e}\frac{\left[1+(73/24e^{2})+(37/96)e^{4}\right]}{(1-e^{2})\left[1+(121/304)e^{2}\right]}.
\label{eq:peterssoldade}
\end{equation}
By including the comparable 2.5PN terms directly in our three-body integration of the equation of motion we can capture losses in three-body interaction knots as well as reproduce equations \eqref{eq:peterssoldadt} and \eqref{eq:peterssoldade} in the case where the system develops strong hierarchy and two bodies evolve following the secular evolution described by \citet{Peters:1964bc}. In the Appendix, we show comparisons between the orbit-averaged equations \eqref{eq:peterssoldadt} and \eqref{eq:peterssoldade}  and a direct numerical integration in our code. 

With the inclusion of losses to GW radiation, binaries have  a finite lifetime.  If, for example, we consider a binary with objects of equal mass, $m$, and a circular orbit with initial SMA $a_{0}$, equation \eqref{eq:peterssoldadt}
reduces to the form $da/dt\propto{(m/a)^{3}}$ with the solution 
\beq\label{tlifee0}
t_{\rm life}(a_{0}) = 1.6\times{10^{17}}\left(\frac{a_{0}}{\text{au}}\right)^{4} \left(\frac{m}{ M_\odot }\right)^{-3}\text{ yr}.
\eeq
Here $t_{\rm life}$ is the GW inspiral time, or the time it takes for the initial binary to evolve from $a=a_{0}$ to ${a=0}$.
The dependence on the SMA to the fourth power makes the lifetime very sensitive to small changes in $a_{0}$.
In the other limit, where the initial eccentricity $e_{0}$ is not far from unity,
the inspiral time is 
\beq\label{tlifee1}
t_{\rm life}(a_{0},e_{0}) \simeq{t_{\rm life}(a_{0})} \frac{768}{425}\left(1-e_{0}^{2}\right)^{7/2}.
\eeq
The lifetime of a very eccentric binary  is shorter than that of a binary in a circular orbit with similar SMA  because as
the eccentricity increases the pericenter distance, which is given by $r_{\rm min}=(1-e)a$, decreases.
This results in a higher GW flux  every pericenter passage, which in turn decreases the lifetime
and gradually circularizes the orbit of the binary.

An analytical solution for the coupled evolution in $a$ and $e$
also exists \citep{Peters:1964bc}
\begin{equation}
a(e) = \frac{c_{0}e^{12/19}}{1-e^{2}}\left(1+\frac{121}{304}e^{2}\right)^{870/2299},
\label{eq:a_evolve_from_e_peters}
\end{equation}
where $c_0$ is a constant with dimensions of length, set according to the initial conditions $(a,e)$ of the binary system.
From this expression we see that in the high eccentricity limit, where $e\approx1$, the SMA scales as $a(e) \propto (1-e)^{-1}$. 
As a result, the orbital SMA (and thus also the orbital energy) must change by many orders of magnitude before the eccentricity becomes significantly less than unity. 
Inspiraling binaries thus only become approximately circular during the last phases of their inspiral. 

\subsection{Significance of PN corrections}

The binary's compactness determines many of the important dynamical properties of the system, especially the importance of PN corrections and collisions. A dimensionless compactness can be defined as  \citep{Blanchet:2006kp}
\beq\label{compactness}
\gamma = \frac{G m}{r c^2} . 
\eeq
Using $\gamma$, we can write the acceleration,  $\acc = \acc_{0}+c^{-5}\acc_{5}$, in terms of the dimensionless radius and mass, $\tilde{r}=r/r_{\rm u}$ and $\tilde{m}=m/m_{\rm u}$. In these units, the acceleration is  $\tilde{\acc}= \acc/({Gm_{\rm u}}/{r_{\rm u}^2})$ and we have
\begin{equation}
\tilde{\acc}_{\rm tot} = \tilde{\acc}_{0}(\tilde{m},\tilde{r}) + \gamma^{5/2} \tilde{\acc}_5(\tilde{m},\tilde{r},\tilde{v}).
\end{equation}
For systems that are strongly relativistic, the SMA $a_0\approx G m /c^2$ and, as a result,  PN corrections become very  important. For weakly PN systems, $a_0\gg G m /c^2$ and the compactness of the orbit provides an estimate for  the importance of the PN  corrections to the equation of motion of a circular, $e\approx0$, orbit.
However, a key point that we emphasize in this work is that measuring the strength of the PN corrections only in terms of the 
compactness of the initial binary orbit  can be misleading. 
In chaotic three-body interactions, the eccentric orbits and close passages that arise make it possible for strong PN corrections to be realized even in systems with initially wide SMA. As we will discuss later, the initial compactness of the binary system still determines the probability that a  very strong encounter will occur.

Close approaches in eccentric orbits lead to strong PN corrections to the equation of motion. They also may lead to direct collisions.
The maximal strength of PN corrections to the acceleration is therefore set by the physical size and mass of the objects, rather than by the initial SMA of their orbits.
This can be quantified by calculating the compactness $\gamma$ for the interacting objects themselves using their mass and radius. 
For example, if the objects are black holes, their compactness $\gamma \sim 1$, and PN corrections can therefore reach their maximal strength.
If the constituent objects are not black holes, then $\gamma < 1$, and the magnitude of the maximal PN corrections for that three-body system is reduced.
Neutron stars have typical dimensionless compactness of $\gamma \approx 0.2$, while a $0.6\ M_\odot$ white dwarf is characterized by a $\gamma \approx 10^{-4}$.
Interacting WDs will therefore in general collide before PN corrections become strong.

If a system of N interacting objects is only composed of BHs, then the dynamics of the system becomes scale free \citep[e.g.][]{Shapiro:1983wz, 2006ApJ...640..156G}.
The reason is that the equation of motion scales with the masses of the BHs, as do the BH gravitational radii. 
For example, for a binary-single interaction involving  three equal mass BHs, the expected dynamics 
for a system with $a_{0}=10^{-3}$ AU and $m_{\rm BH}=1\ M_{\odot}$ will be equivalent to that of  a system with $a_{0}=10^{-1}$ AU and 
$m_{\rm BH}=10^{2}\ M_{\odot}$. This  allows us to identify dynamically similar systems that occur in different astrophysical contexts.
If the N interacting objects are not BHs, then the system looses its scale-free behavior as the object radius no longer scales with mass.
Neutron stars, for example, exhibit relatively constant radius across their observed mass range \citep{2010ApJ...722...33S}, while white dwarfs have an inverse mass radius relationship $R_{\rm WD} \propto m_{\rm WD}^{-1/3}$.

\subsection{Energy Losses}

\begin{figure}[tbp]
\centering
\includegraphics[width=\columnwidth]{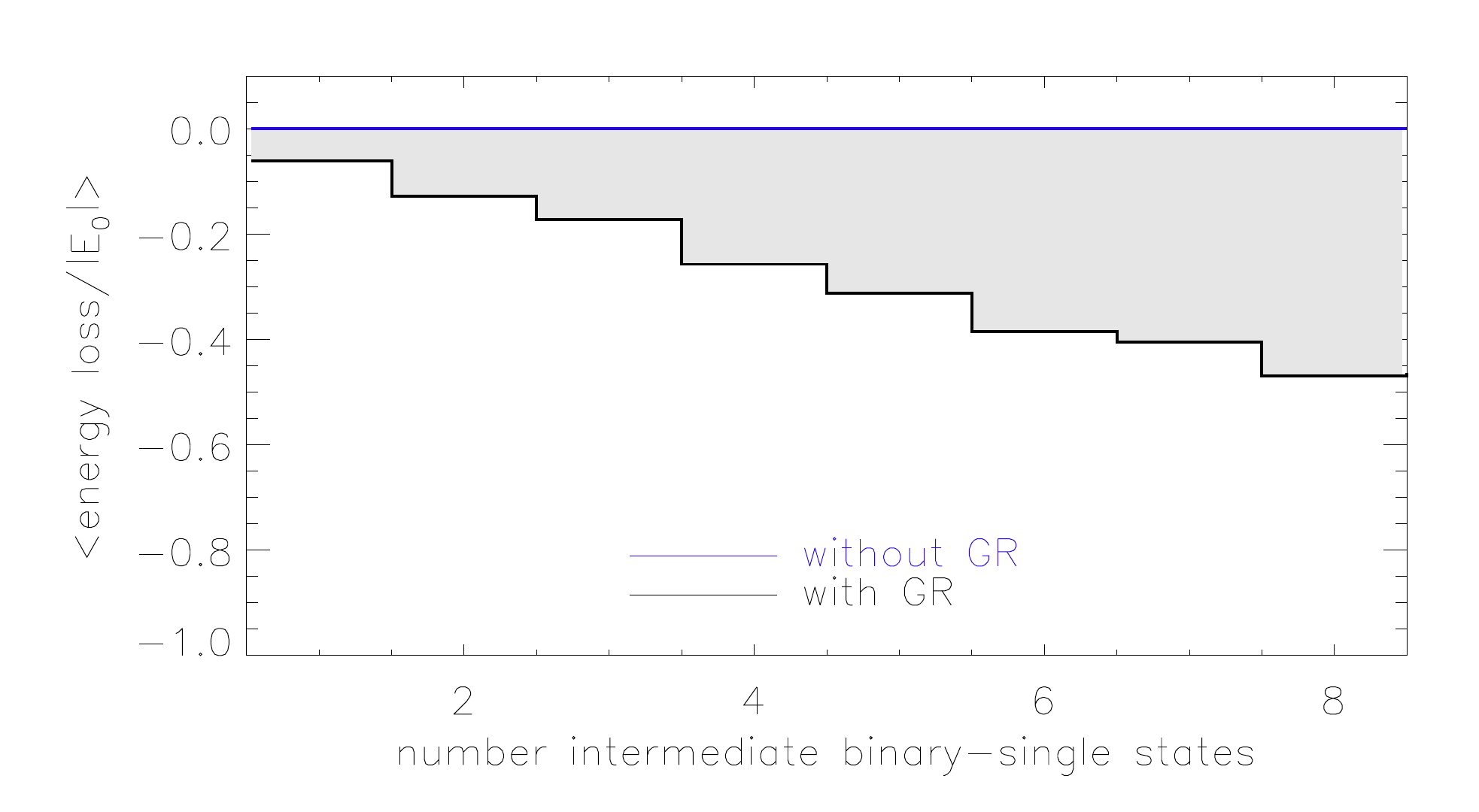}
\includegraphics[width=\columnwidth]{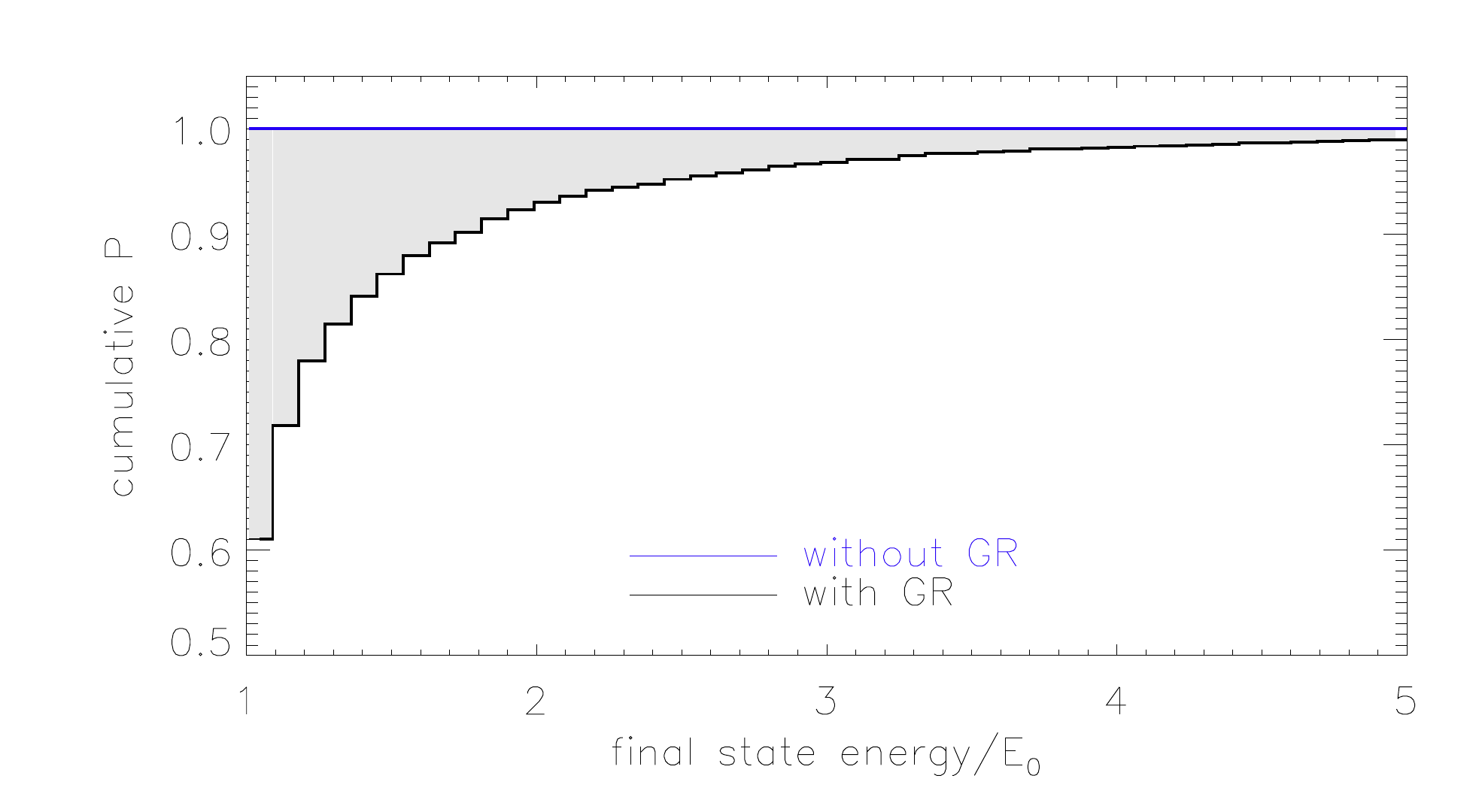}
\caption{GW energy loss in binary-single interactions between equal mass BHs. The panels show an extreme HB case with 
$a_{0}=10^{-5}\ {\rm AU}$ and $m_{\rm BH}=1\ M_{\odot}$. The black lines indicate scatterings where GR is included in the
simulation and the blue lines indicate those for which GR is  not  included.
Softer binaries will have energy losses within the grey shaded region and quickly end up near the blue line, which indicates no energy loss.
Both panels include only states from the RI channel with a finale state where the single object is unbound.
{\it Top:}  Average energy change scaled by the initial energy $E_{0}$ after a certain  number of intermediate binary-single states. Fractional energy losses of the order of  $\sim10\%$ can be achieved just after the second
instance  a new binary-single state is produced. The average energy loss increases with the number of identified IMS, indicating that 
energy has being extracted from the system.
{\it Bottom:} Cumulative distribution for the fractional energy difference between the total final state energy and the initial energy.
The figures are based on $2\times10^4$ binary-single interactions with $v_{\infty}\ll v_{\rm c}$.}
\label{fig:deltaEpumps}
\end{figure}

\begin{figure}[tbp]
\centering
\includegraphics[width=\columnwidth]{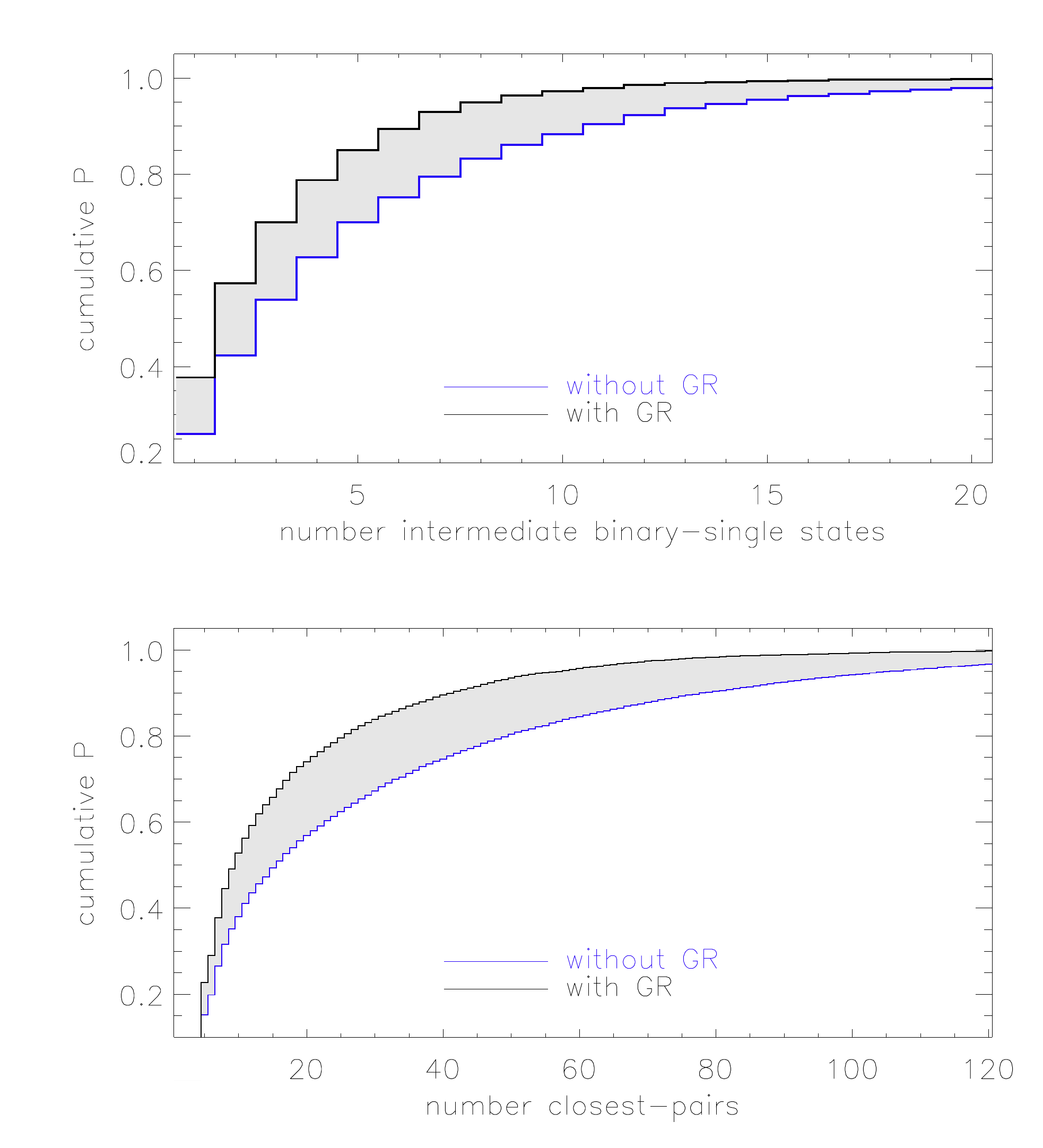}
\caption{Number of three-body interactions between equal mass BHs arising from binary-single scatterings.
Both panels include only states from the RI channel.
The target binary is chosen to be initially very hard with $a_{0}=10^{-5}\ {\rm AU}$ and $m_{\rm BH}=1\ M_{\odot}$.
The black lines indicate scatterings where GR corrections have been added while blue lines show experiments with no  GR corrections included. 
The two plots differ in  the way the number of interactions are counted.
{\it Top:} Number of times an intermediate binary-single state (IMS) is observed to occur  during a resonant interaction.
{\it Bottom:} Number of times a new closest pair has been identified during the resonant interaction. A high number of close-pairs
indicates highly chaotic motion during the encounter (see Figure \ref{fig:binsin interactions}) which occurs  between each IMS.}
\label{fig:PDF_nrinteractions}
\end{figure}

The effects of GW energy  loss  can be most easily seen by examining equation \eqref{eq:25PNterm}  in the context of a circular binary of equal mass objects. In that case, ${\bf \hat r}\cdot{\bf v}_{12}=0$, leaving only the first term in equation \eqref{eq:25PNterm}. For equal mass objects, the term in parenthesis in equation \eqref{eq:25PNterm}  evaluates to a negative number and the direction of $\acc_5$ is determined by  $-{\bf v}_{12}$, directly against the motion of the two bodies. As a result, the orbiting objects essentially experience a drag force
\begin{equation}
F_{\rm 2.5 PN} = \frac{32\sqrt{2}}{5}\frac{G^{7/2}}{c^{5}}\left(\frac{m}{r}\right)^{9/2}
\label{eq:circularbin_acc25PN}
\end{equation}
This follows directly from equation \eqref{eq:25PNterm} by substituting $v=\sqrt{2Gm/r}$.  The energy leaving the system per unit time can be easily calculated by using
${\Delta}E_{\rm orb}=\text{force}{\times}\text{distance}=F_{\rm 2.5 PN}2{\pi}r$, from which it follows that
\begin{equation}
\frac{dE}{dt} \simeq \frac{{\Delta}E_{\rm orb}}{T_{\rm orb}}=-\frac{64}{5}\frac{G^{4}}{c^{5}}\left(\frac{m}{r}\right)^{5}
\label{eq:dEdtPN}
\end{equation}
where $T_{\rm orb} = 2\pi (2Gm/r^3)^{-1/2}$ is the orbital period. 
One should notice that the distance $r$ is changing as a function of time 
with a rate that can be calculated by using the Newtonian relation $dE/dr=-Gm^2/2r^2$.

The above formalism  can be extended to a binary-single interaction. 
The distribution of  GW energy radiated during a resonant encounter is shown in Figure \ref{fig:deltaEpumps}.
The upper panel shows how energy from the system is depleted as new intermediate binary-single state are created.
The fractional  energy loss is relatively small, especially for binaries with large SMA, but at each encounter the binaries are effectively hardened and the relative likelihood for the system to undergo a collision or a merger  is increased. The lower panel shows the
cumulative distribution of the fractional energy loss between the initial state and the  final state for the same set of interactions.
Figure \ref{fig:PDF_nrinteractions} shows the corresponding cumulative distributions of the number of IMS ({\it top panel})
and the number close-pairs ({\it bottom panel}) in a binary-single interaction. The number of close-pairs is greater than the number of IMS since
it also includes all close passings that can occur within a single state (see Figure \ref{fig:binsin interactions}). 
For the set of scatterings ending with an unbound companion (exchange or fly-by)
the number of three-body interactions are reduced when GR is included. For example, without GR $20\%$ of all scatterings
shown in Figure \ref{fig:PDF_nrinteractions} have more than $50$ close interactions, but only about $ 25$ when GR is included. 
The reason is simply that the possibility of  the system  inspiraling  when GR is included, truncates the chain of resonance interactions.
 
\section{The Formation of Dynamical Inspirals}\label{sec:Inspirals}

With the inclusion of energy and angular momentum losses from GW emission a new class of dynamical outcomes appears, which we denote here as  {\it inspirals}.
These are interactions in which two of the objects inspiral and merge while all three objects are still in a bound three body state; that is,  before one of the  classical outcomes of exchange, flyby or ionization is achieved. An example of an inspiral end state   is  shown in Figure  \ref{fig:example_inspiral_fromsim}.

\begin{figure}
\centering
\includegraphics[width=1\columnwidth]{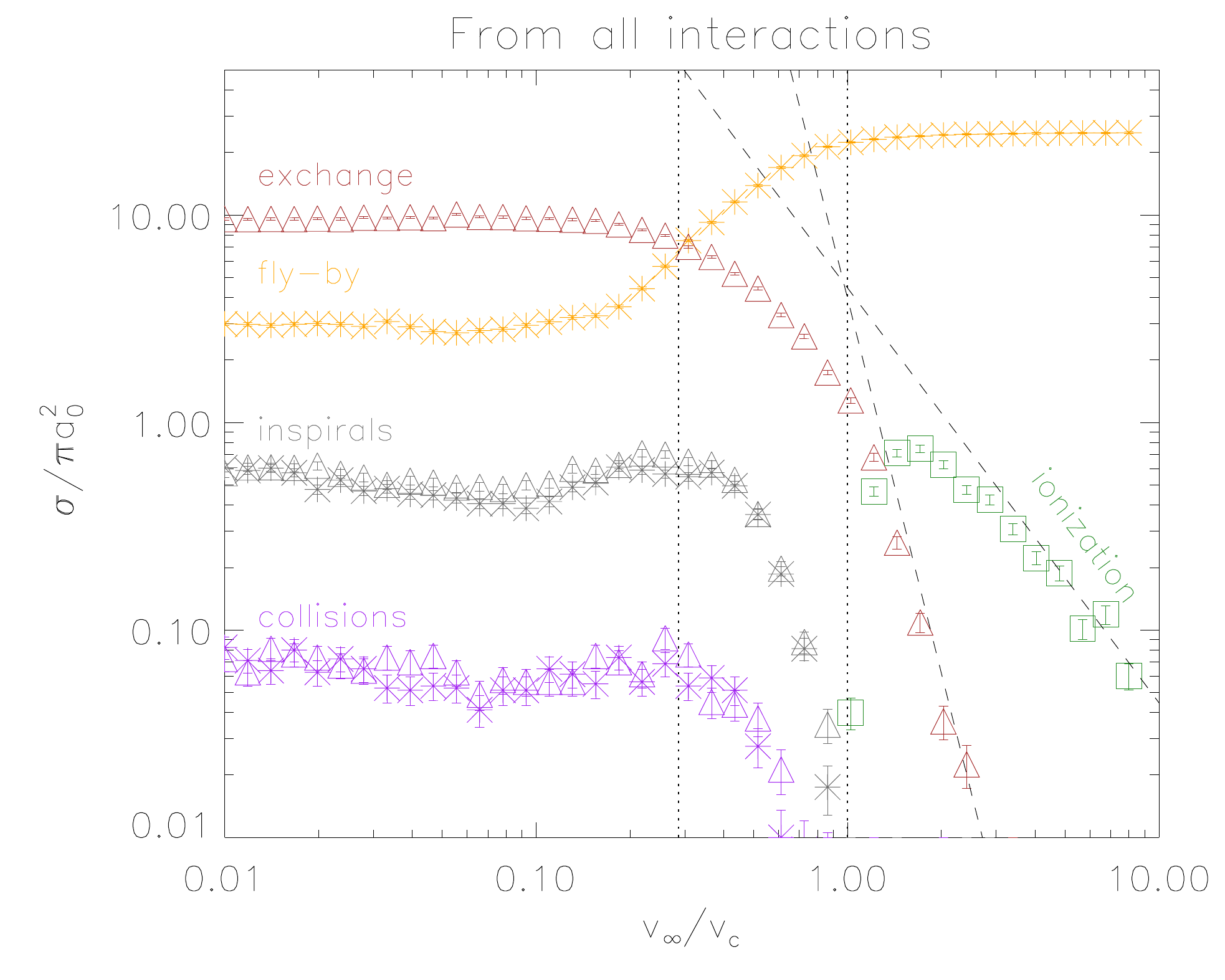}
\includegraphics[width=1\columnwidth]{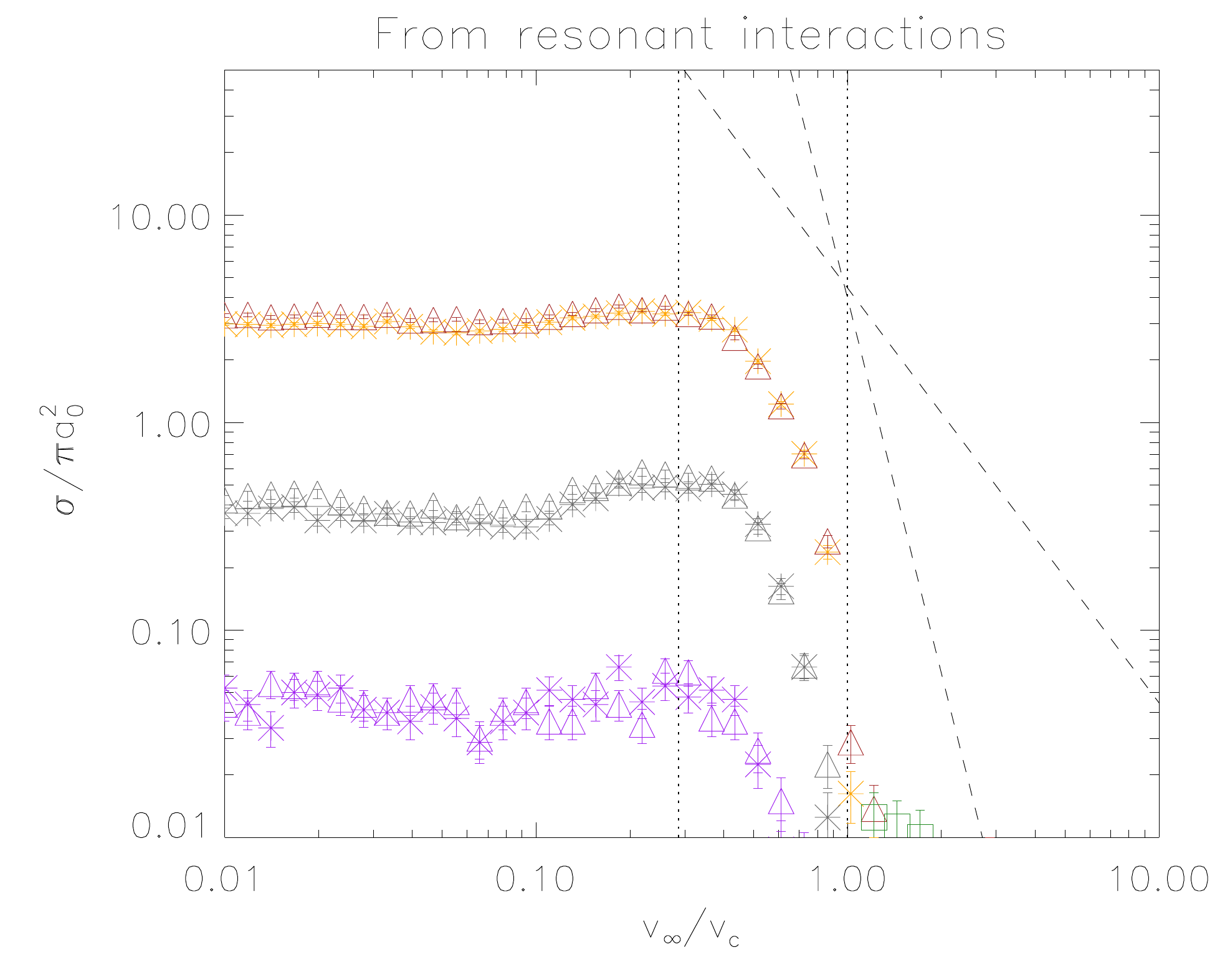}
\caption{Integrated outcome cross sections from binary-single interactions between equal mass BHs including 2.5PN corrections.
Similar to Figure \ref{fig:piethutplot_ngr} but now including collisions (purple) and {\it inspirals}  (grey).
The initial SMA of the target binary is $a_{0}=10^{-4}\ {\rm AU}$ and $m_{\rm BH}=1\ M_{\odot}$.
{\it Top:} All interactions including RIs and DIs.
{\it Bottom:} Outcomes arising  from the RI channel only.
The new outcome from including GR is a population of objects  that gravitationally inspiral
during the interaction. We denote such  endstates as {\it inspirals}.
The {\it inspiral} cross section  flattens out below the gravitational focusing
velocity, $v_{\rm foc}$, which implies  that these end states  are not
sensitive to the exact value of the impact parameter, $b$, and velocity, $v_{\infty}$, as long as the interaction is a CI.}
\label{fig:piethutplot_wgr}
\end{figure}

In order to understand how the inclusion of GR corrections changes the binary-single outcome  landscape,  
we recompute the Newtonian scattering experiments  shown in Figure \ref{fig:piethutplot_ngr}  with the  addition of  the 2.5PN term in the equation of motion. Our results are illustrated in
Figure \ref{fig:piethutplot_wgr}.
The revised cross sections  include inspirals and collisions between solar mass black holes with an initial binary SMA of $10^{-4}$ AU. 
The top panel shows the resultant  cross sections from all interaction channels including DIs and RIs while the bottom panel includes only endstates
arising from the RI channel. By comparing the two panels one can conclude that inspirals (and collisions)
are dominated by the RI channel, an observation that  will become  useful when we derive the  analytical treatment  for inspiral occurrence in Section \ref{subset:ccins}.

Another important point  is that the cross section for inspirals is approximately flat when $v_{\infty}<v_{\rm foc}$. 
This implies that the probability for an inspiral to occur  is not sensitive to the exact value of the impact parameter, $b$, or velocity, $v_\infty$, as long as the single object experiences
a CI with the binary.
The lack of a dependence on the initial conditions arises because nearly all inspirals are generated from RIs (for which memory of the initial conditions is rapidly lost through ensuing  resonances) and because the fraction of RIs and DIs is
approximately  constant for $v_\infty \ll \vc$ (see Section \ref{sec:new_low_vel_sec}).
This observation makes it possible to write the probability for an outcome to be an inspiral given the interaction is a CI as 
\beq
P_{\rm insp} \equiv N_{\rm insp}/N_{\rm CI},
\label{eq:definition_of_Pinsp}
\eeq
and the corresponding inspiral cross section  as
\begin{equation}
\begin{split}
\sigma_{\rm insp}	& =  P_{\rm insp}\sigma_{\rm CI}, \\
  			&  \simeq P_{\rm insp} \frac{3{\pi}Gma_{0}}{v_{\infty}^{2}},
\end{split}
\label{eq:cross_section_inspirals_1} 
\end{equation}
where the last equality holds  for the equal mass case.
This factorization is useful in the sense that it separates the contribution coming from the
chaotic RIs from the standard focusing cross section that simply acts as a weight factor.
It is important to notice that $P_{\rm insp}$  depends on the compactness of the initial binary, i.e. its SMA $a_{0}$ and mass $m_{\rm bin}$,
as we will show in Section \ref{subset:ccins}.

\subsection{Phase Space Distribution of Inspirals}

\begin{figure*}[tbp]
\centering
\includegraphics[width=0.48\textwidth]{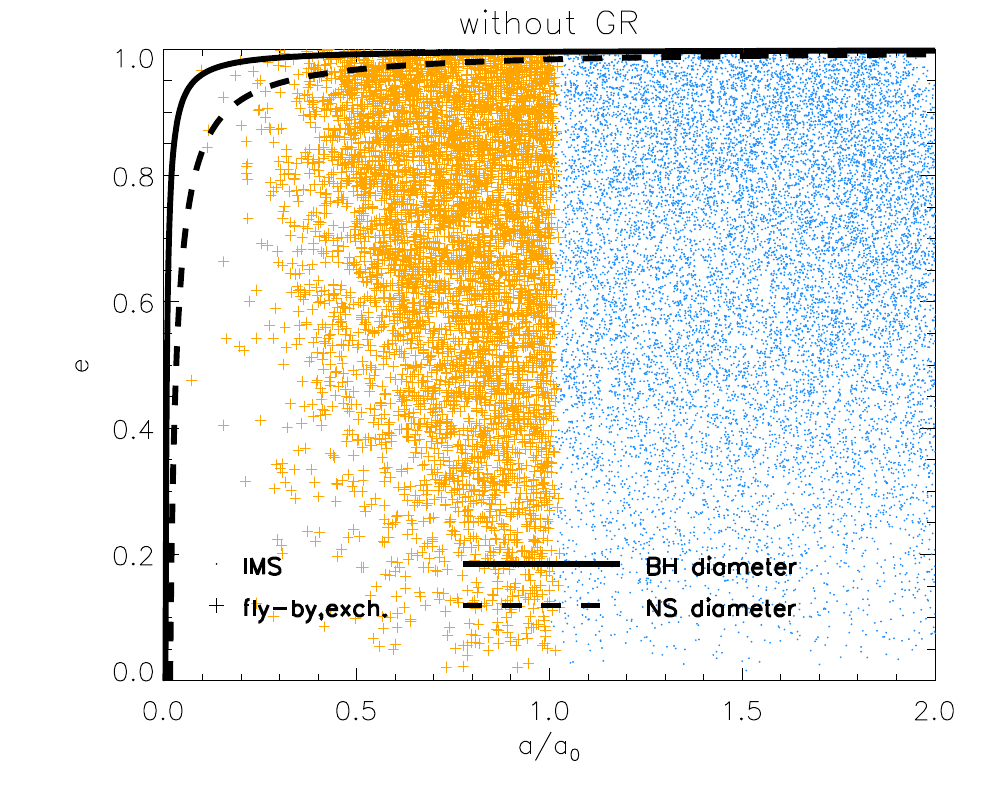}
\hspace{0.3cm}
\includegraphics[width=0.48\textwidth]{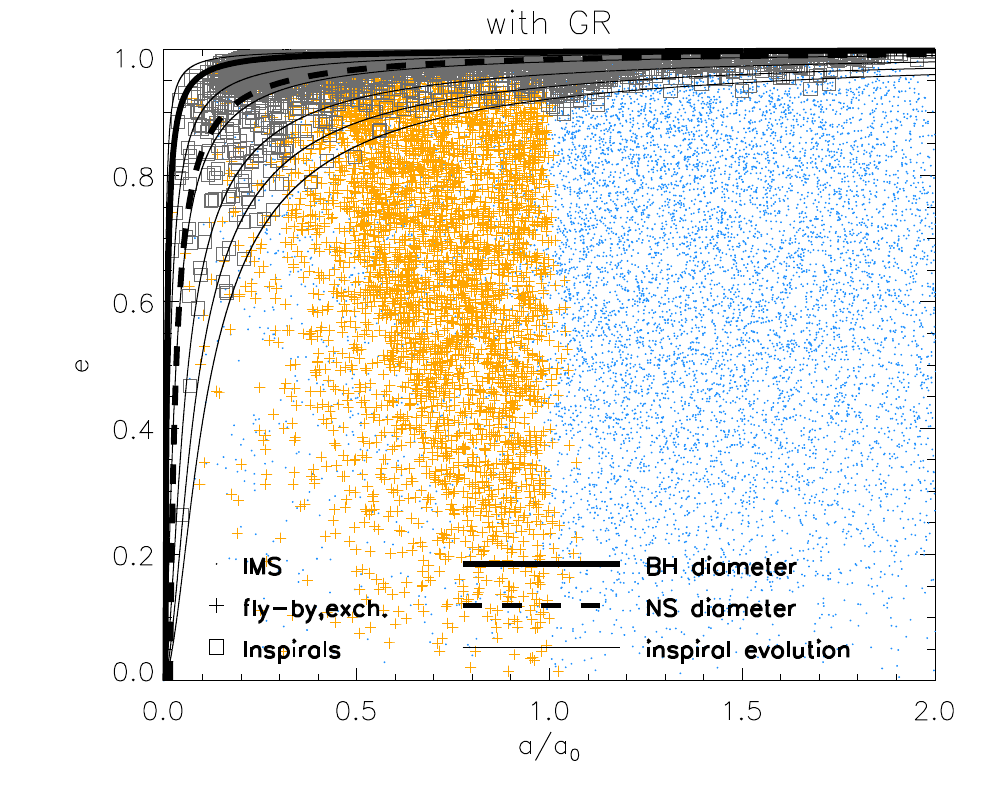}
\caption{Distribution of orbital parameters $(a,e)$ for all identified binaries
from $2\times10^4$ binary-single interactions between equal mass BHs.
The target binaries have a SMA of $a_{0}=10^{-5}{\rm AU}$ and $m_{\rm BH}=1\ M_{\odot}$.
The encounters occur with an incoming velocity of   10 km s$^{-1}$ and  as such are in the extreme HB limit where $v_{\infty}\ll v_{\rm c}$.
{\it  Left:} Without GR. {\it Right:} With GR. The blue points represent  IMS binaries, which are the candidates for {\it inspiral} end states. The sampling of these IMS is nearly homogenous.
The orange  symbols show the endstate binaries from the classical outcomes: {\it exchanges} and {\it fly-bys}.
{\it Inspirals} that arise when GR is included are seen in the right panel as {\it grey squares}. Each identified binary  separation, $a$, is scaled with the initial  $a_{0}$. This is because 
if there is no energy loss and $v_{\infty} \ll v_{\rm c}$, then all intermediate states must have $a/a_{0}>1$ and all final
states with an unbound companion should satisfy $a/a_{0}<1$. This follows directly from conservation of energy $(a\propto1/E)$ and helps
illustrate how binaries tend to harden after a HB interaction. 
If energy is leaking out of the system by GW emission, then the  resultant states shown by the {\it blue-points} can flow across the $a/a_{0}=1$ boundary
as seen on the {\it right} panel. The orbital parameters
for {\it inspirals} are  fast evolving and the grey region is therefore only showing a snapshot of the phase space distribution of these states.
The thin black lines show a few examples of the evolution contours the {\it inspirals} follow  in the $(a,e)$ space.
The black solid line shows the diameter of a $1\ M_{\odot}$ BH, and the dashed shows the
diameter of a  NS with 12 km radius.  In this example, where the interacting objects are three stellar mass  BHs, any formed binary
above the BH limit would lead to a collision. Many of the BH {\it inspirals} would have been collisions instead, if the
objects would have been NSs.
}
\label{fig:example_aet_dist}
\end{figure*}

\begin{figure}[tbp]
\centering
\includegraphics[width=0.97\columnwidth]{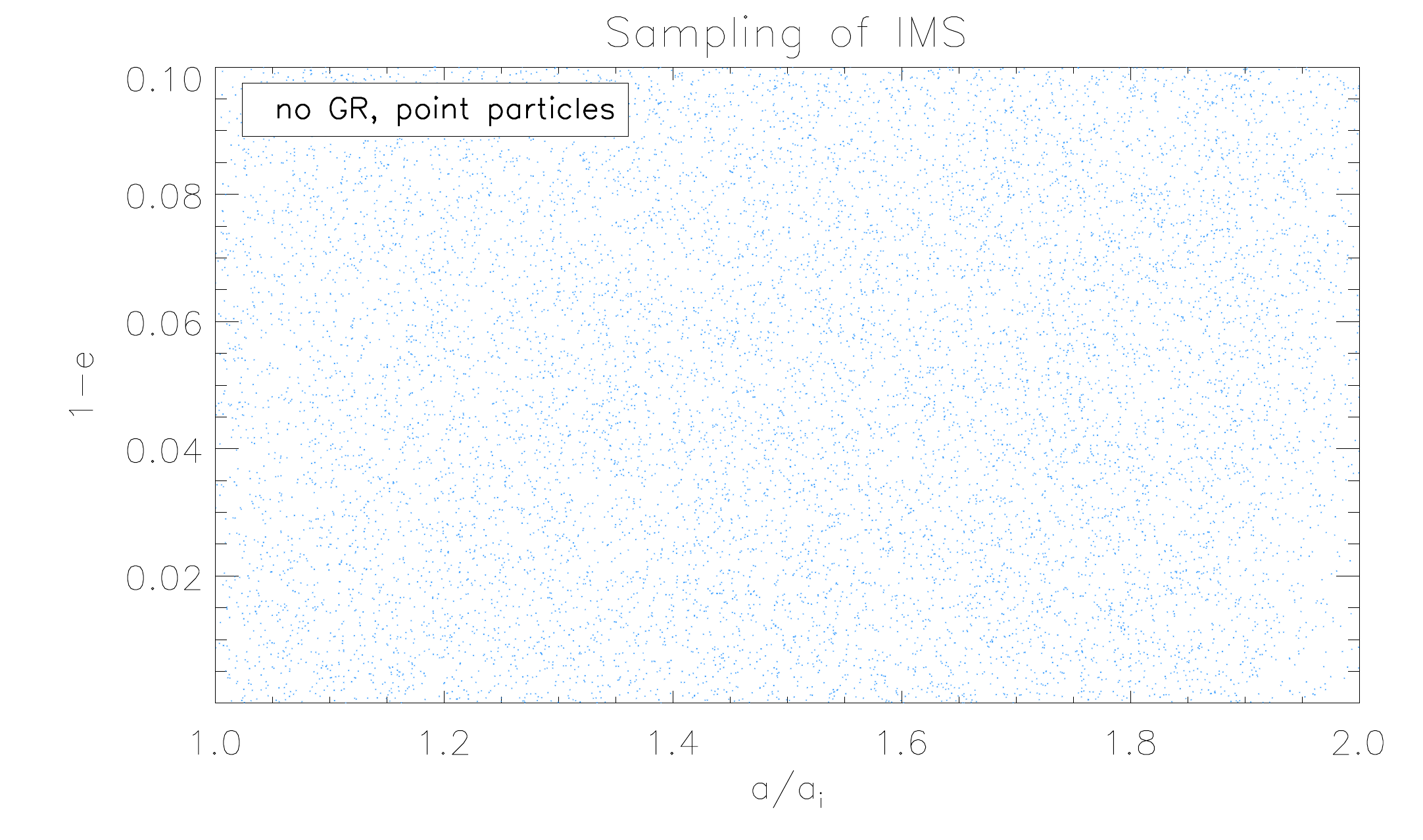}
\includegraphics[width=0.97\columnwidth]{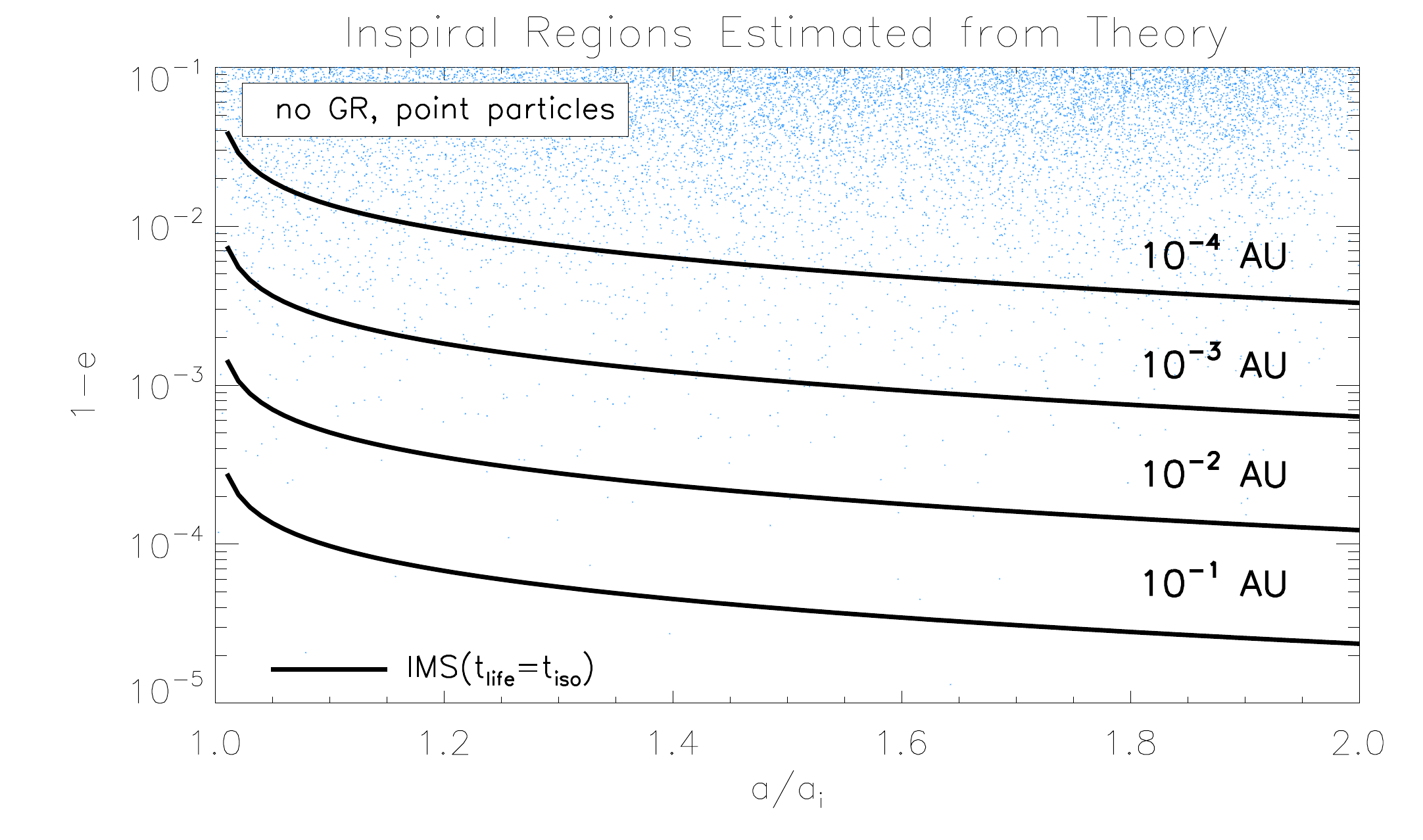}
\includegraphics[width=0.97\columnwidth]{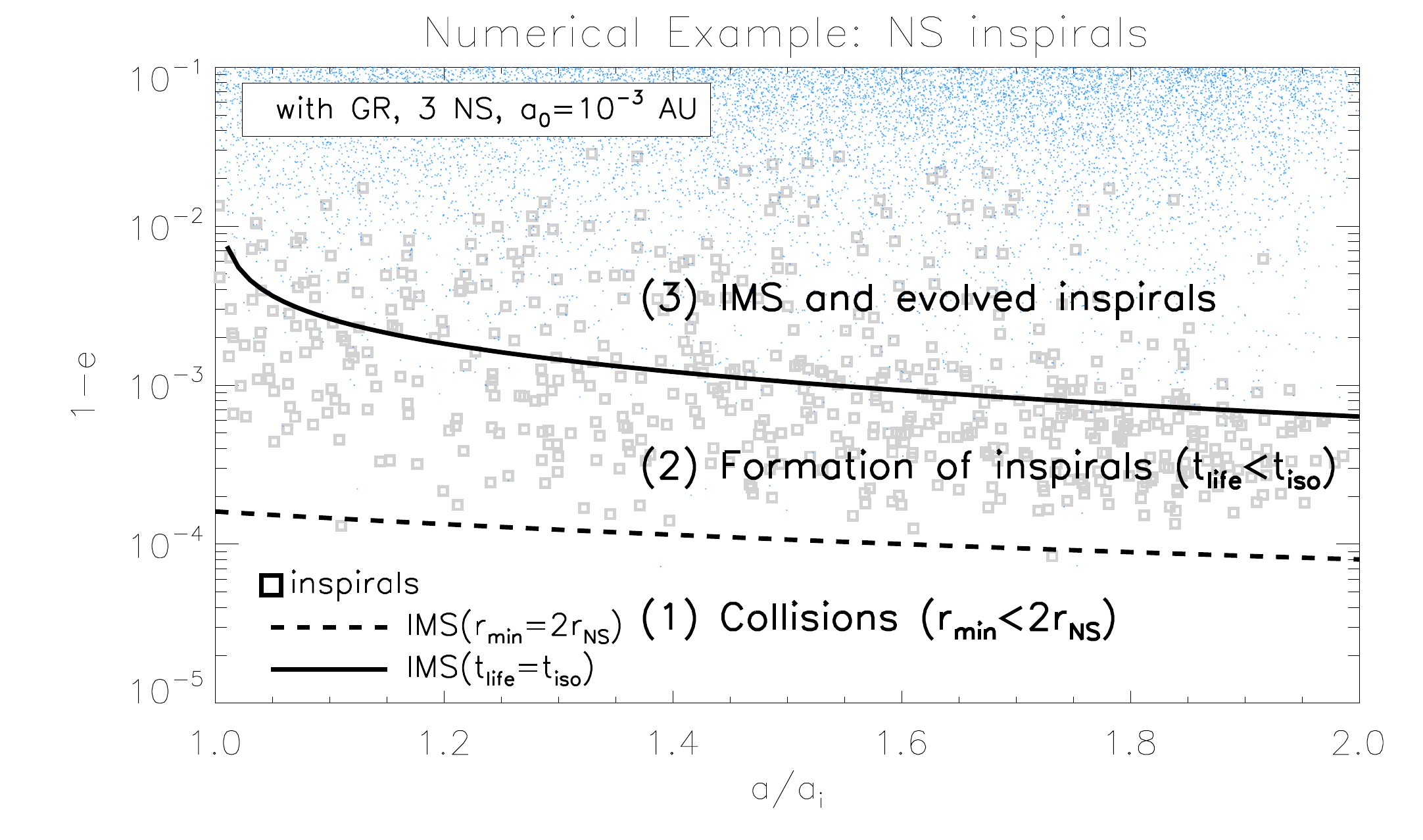}
\caption{Inspirals   in  $(a,1-e)$ space produced  by IMS binaries.
Blue points indicate IMSs whose endstate is not an inspiral.
Grey squares show inspirals at the time of identification. The {\it x}-axis on all three panels show $a/a_{i}=E_{{\rm tot},i}/E_{{\rm bin},i}$ where $E_{{\rm tot},i}$ and $E_{{\rm bin},i}$ are respectively  the total energy of the three-body system and the energy of the IMS binary
at the time of identification: $t_{i}$. 
\textit{Top:} Sampling of IMS near $e\approx 1$. This distribution is relatively  uniform in $(a,1-e)$ space, an observation that makes it
possible to estimate how the number of inspirals produced scales with initial SMA $a_{0}$ (equation \ref{eq:cross_section_inspirals_2}).
\textit{Middle:} Same distribution as in the top panel  but with the y-axis in logarithmic  scale.
To illustrate  which IMS can form inspirals when GR is included, we have plotted using equation \eqref{eq:GRmergingrel} a few lines showing where $t_{\rm life}=t_{\rm iso}$. An inspiral can form when $t_{\rm life}<t_{\rm iso}$.
\textit{Bottom:}  Results from  numerical scattering experiments including GR and  a finite  radius, $r_{\rm NS}$, for the interacting objects. The NS radius introduces a collision boundary given by $r_{\rm min}=2r_{\rm NS}$ where $r_{\rm min}$ is the the pericenter distance of the IMS binary.
As shown in the plot, the $(a,1-e)$ IMS space  divides into three distinct regions:
(1) IMS with $r_{\rm min}<2r_{\rm NS}$  will  produce direct collisions,
(2) IMS with $r_{\rm min}>2r_{\rm NS}$ and  $t_{\rm life}<t_{\rm iso}$ will  form inspirals, and
(3)  IMS with $r_{\rm min}>2r_{\rm NS}$ and  $t_{\rm life}>t_{\rm iso}$ can be  followed by further interactions.
As  inspirals formed in region (2) spiral in, they diffuse into  region (3).
All three panels are based on $2\times10^4$ scatterings between equal $1.4\ M_{\odot}$ objects with $r_{\rm NS}=12\ {\rm km}$.}
\label{fig:ae_states}
\end{figure}

Figure \ref{fig:example_aet_dist} shows
distributions of the orbital parameters $(a,e)$ for all exchange and fly-by binaries ({\it orange}) and intermediate state binaries ({\it blue}) from
$2\times 10^4$ HB binary-single interactions. The division at $a/a_{0}$ indicates energy conservation between the newly formed binary with SMA $a$
and initial binary with SMA $a_{0}$.
The target binary must shrink if the single object becomes unbound, i.e. exchange and fly-by binaries have $a<a_{0}$
while  IMS binaries have $a>a_{0}$.

Inspirals appear in grey in the {\it right-hand} panel in Figure \ref{fig:example_aet_dist} where the 2.5PN term is included in the equation of motion.
These inspirals form from the subset of IMS binaries that merge while the three-body system is still bound and are therefore (mainly) initially created  with $a>a_{0}$.
Since GWs in general carry energy out of the system before an endstate is reached,
then IMS can flow across the initial $a/a_{0}=1$ border line. This means that all outcome distributions are slightly changed when GR is included. Inspiral states  are, however, those that experience the highest energy losses.

Immediately after an inspiralling binary has formed, it evolves according to equation \eqref{eq:a_evolve_from_e_peters}.
Several of these evolutionary trajectories  are shown with thin black lines  in Figure \ref{fig:example_aet_dist}. 
GW emission circularizes the binary as its  SMA is decreased. This migrates binaries from their initial formation region in the right hand side of the $(a,e)$ phase space
to the lower left. Therefore, the exact location of the inspiral event in  Figure \ref{fig:example_aet_dist} 
depends   on when the system was identified in the
code (see Appendix A for a discussion of the selection criteria for states). It is therefore not necessarily representative of the binary's initially assembled position in the formation locus for inspirals. 

The phase space accessible for inspirals depends on $\gamma$ (equation \ref{compactness}). At particular $(a,e)$ combinations with close pericenter approaches, direct collisions can also occur. A direct tradeoff  can the be found between the number of collisions and the number of inspirals. The rates for these particular  end states cannot be independent because they originate from a similar phase space region. Not surprisingly, extended  objects produce relatively fewer inspirals and more collisions than compact ones. The importance  on th object's  size is illustrated   in Figure \ref{fig:example_aet_dist}, in which  we plot the boundaries defined by the BH and NS diameters, respectively. 

\subsection{Analytic Derivation of Inspiral Cross Sections}\label{subset:ccins}

In this section, we  develop an analytical understanding  of what determines the occurrence rate of  inspirals and collisions, including  how the outcomes depend  on the initial SMA and on the mass of the target binary. Each IMS is characterized by three parameters: the
SMA $(a)$ and eccentricity $(e)$ of the IMS binary and the orbital period of the bound
companion, which we denote here as the isolation time $(t_{\rm iso})$.
Since the single object is bound to the binary during  an IMS,  $t_{\rm iso}$ is finite. It then follows 
 that if an IMS binary is formed
with $t_{\rm life}<t_{\rm iso}$, then the binary will inspiral before the return of the bound companion. The lifetime, $t_{\rm life}$, is determined by  equations \eqref{eq:peterssoldadt} and \eqref{eq:peterssoldade} but can be estimated by equations \eqref{tlifee0} and \eqref{tlifee1} in the circular and eccentric limits, respectively. In all of the following calculations, we assume the hard binary limit ($v_{\infty} \ll v_{c}$).

The probability for a particular  outcome to be an inspiral can be estimated by considering the fraction of states during a RI
that satisfies  $t_{\rm life}(a,e)<t_{\rm iso}(a)$.
The isolation time $t_{\rm iso}$ is described by Keplers law
\begin{equation}
t_{\rm iso} = 2\pi\sqrt{\frac{a_{\rm bs}^3}{Gm_{\rm tot}}},
\end{equation}
where $a_{\rm bs}$ is the SMA of the hierarchical triple.
This SMA, $a_{\rm bs}$, can be expressed in terms  of  the initial binary SMA, $a_0$, and the SMA of the IMS binary, $a$, by making use of energy conservation
\begin{equation}
E_{\rm tot} \simeq -\frac{Gm_{1}m_{2}}{2a_{0}} = E_{\rm bin} + E_{\rm bs} = -\frac{Gm_{i}m_{j}}{2a}  -\frac{Gm_{\rm bin}m_{\rm sin}}{2a_{\rm bs}}
\label{eq:encons}
\end{equation}
where `bin' and `sin' respectively refer to the binary and the single bound object in the hierarchical triple.
In the equal mass case, equation (\ref{eq:encons}) reduces to
\begin{equation}
a_{\rm bs} = \frac{2a_0}{1-1/a\prime}
\end{equation}
such that 
\begin{equation}
t_{\rm iso} = \left(\frac{2}{1-1/a\prime}\right)^{3/2}2\pi\sqrt{\frac{a_{0}^3}{Gm_{\rm tot}}},
\label{eq:isolationtime}
\end{equation}
where $a\prime = a/a_{0}$ and the last term in equation (\ref{eq:isolationtime}) is  the orbital time of the initial binary system, $T_{\rm orb,0}$.
Equation \eqref{eq:isolationtime} relates the normalized SMA, $a\prime$, of a given  IMS binary to the time it remains isolated from its bound companion. Since $a\prime>1$ during a resonance, it follows 
that $t_{\rm iso} > T_{\rm orb,0}$. 

We can now compare $t_{\rm iso}$ to $t_{\rm life}$, which, in the high eccentricity limit, is given by equation \eqref{tlifee1}. The ratio  $F_{\rm insp}= {t_{\rm life}}/{t_{\rm iso}}$  describes the lifetime relative to the binary isolation time and can be written as 
\begin{equation}
F_{\rm insp} =  \frac{{C_{\rm F}}c^{5}}{G^{5/2}}\left(\frac{a_0}{m}\right)^{5/2}(1-e^2)^{7/2}a\prime^{5/2}(a\prime-1)^{3/2}
\label{eq:definitionofFt} 
\end{equation}
where $C_{\rm F}=({3\sqrt{3}})/({680\pi\sqrt{2}})\approx 1.7\times 10^{-3}$.  If $F_{\rm insp} < 1$, the binary will inspiral before the third body returns. If, on the other hand, $F_{\rm insp} > 1$ another three-body encounter will take place. 
The boundary defined by $F_{\rm insp}=1$
 produces a clear division in  the $(a\prime,e)$ phase space plane, clearly separating   IMSs that will inspiral to those that 
can be followed by further three-body interactions (Figure \ref{fig:ae_states}).
 
Defining  the allowed phase space region for inspirals as $\Delta_{\rm insp} = 1-e$ and  setting $F_{\rm insp}=1$ in equation \eqref{eq:definitionofFt}, we get
\begin{equation}
\Delta_{\rm insp} \approx \frac{1}{2}\frac{G^{5/7}}{C_{\rm F}^{2/7}c^{10/7}}\left(\frac{m}{a_0}\right)^{5/7}a\prime^{-5/7}(a\prime-1)^{-3/7},
\label{eq:GRmergingrel}
\end{equation}
which implies  $\Delta_{\rm insp} \propto {(m/a_0)}^{5/7}$. Assuming that  the $(a,e)$ sampling of IMSs is relatively 
uniform where $e\sim1$, as observed in Figure \ref{fig:ae_states}, we conclude that  the number of IMSs within the inspiral region is $\propto {(m/a_{0})}^{5/7}$.
This means that the probability for an outcome to be an inspiral given that the interaction is a CI (see \ref{eq:definition_of_Pinsp}) scales as 
\begin{equation}
P_{\rm insp}\propto \left({\frac{m}{a_{0}}}\right)^{5/7} \propto {\gamma}^{5/7}, 
\label{eq:Probability_insp}
\end{equation}
such that
\begin{equation}
\sigma_{\rm insp}=P_{\rm insp}\sigma_{\rm CI}	\propto a_{0}^{2/7}\frac{m^{12/7}}{v_{\infty}^{2}}.
\label{eq:cross_section_inspirals_2} 
\end{equation}
This  illustrates  that the cross section for inspirals is expected to \emph{increase} with the SMA of the target binary. 
The dominant inspiral-producing targets  in a cluster are thus not extremely compact binaries, but instead wide ones.

Collisions occupy a similar phase space region to that populated by   inspirals, with  the size of the interacting objects and the initial SMA of the target binary determining their relative cross sections. 
If an IMS binary is formed with a periapsis $r_{\rm min}=a(1-e)$ that is smaller than twice the radius $r_{\rm obj}$ of the interacting objects, then a collision will occur.
Using  $\Delta_{\rm coll} = 1-e$, the collision boundary is simply given by
\begin{equation}
\Delta_{\rm coll} = (2r_{\rm obj}/a_{0})(a\prime)^{-1},
\label{eq:deltaC}
\end{equation}
which leads to the result that the probability for a collision is $P_{\rm coll} \propto a_{0}^{-1}$. The associated cross section, $\sigma_{\rm coll}$, can  be estimated using equation \eqref{eq:cross_section_inspirals_1}, and it is thus independent of $a_{0}$.

If we compare equations \eqref{eq:GRmergingrel} and \eqref{eq:deltaC}, we can see that the probability for a collision ($\propto a_{0}^{-1}$) decreases faster than the
inspiral probability ($\propto a_{0}^{-5/7}$) as $a_0$ increases.
This means that collisions will occupy a progressively smaller  fraction of the available inspiral phase space as the SMA of the target binary increases.
Inspirals arising from  widely separated binaries are therefore less likely to be depleted by collisions, which in turn 
makes widely separated binaries even better targets for inspiral production.

\subsection{Numerical Determination of the Cross Section}

\begin{figure}
\centering
\includegraphics[width=1\columnwidth]{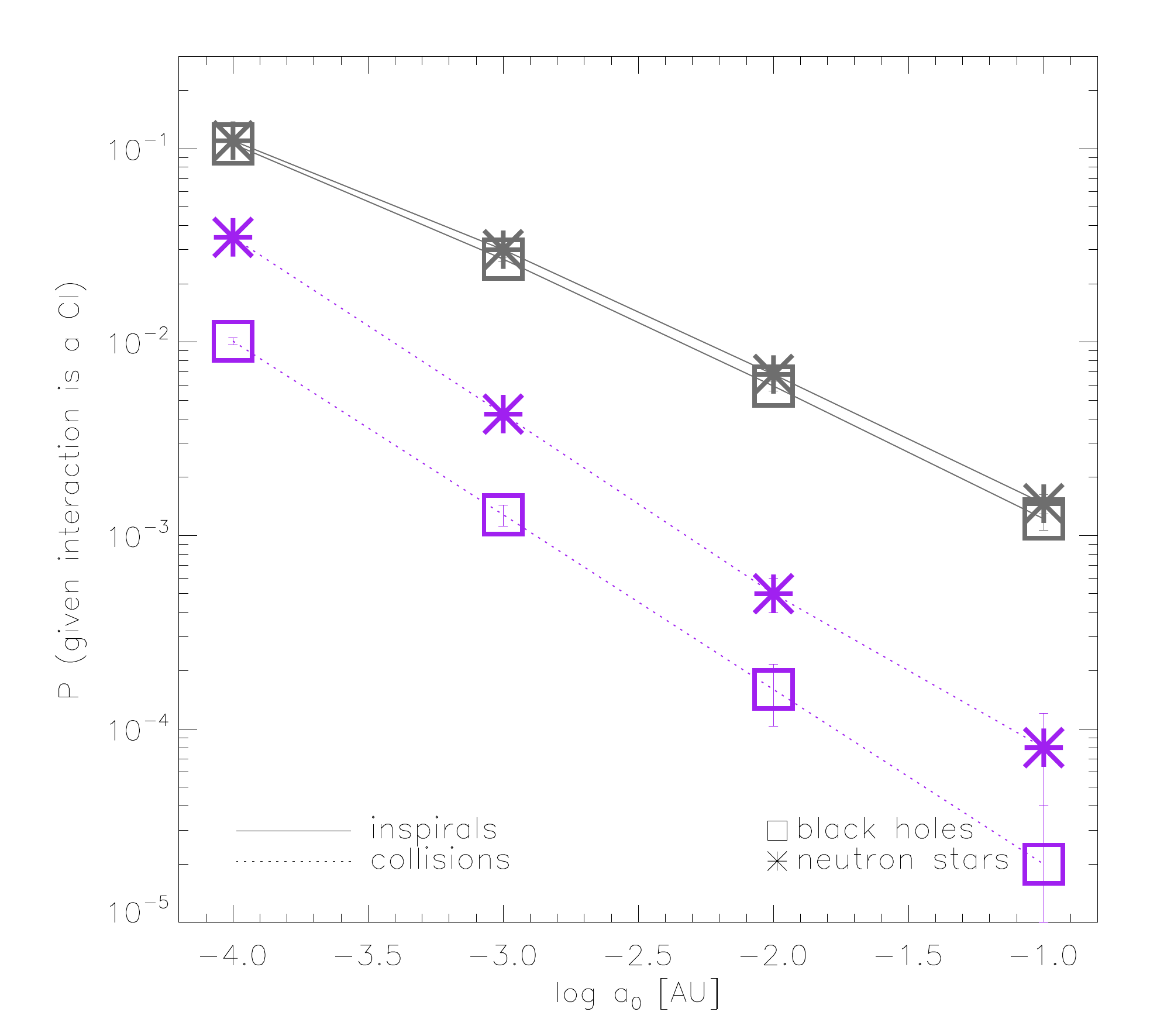}
\includegraphics[width=1\columnwidth]{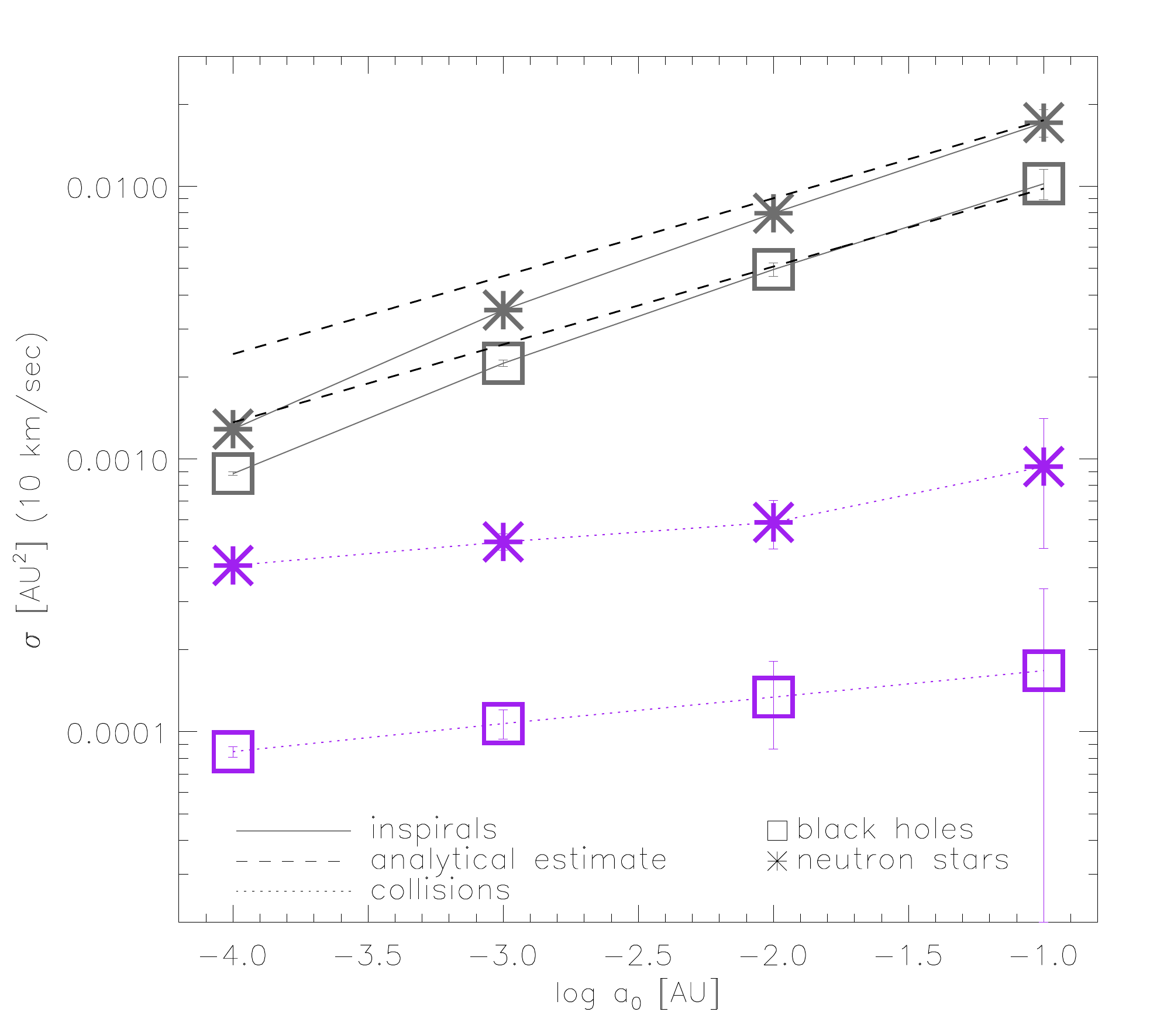}
\caption{Formation of inspirals ({\it grey}) and collisions ({\it purple}) in equal mass binary-single interactions
between either BHs ({\it squares}) or NSs ({\it stars}) as a function of the initial SMA of the target binary.
All BHs have a $1\ M_{\odot}$ mass where the NSs have a $1.4\ M_{\odot}$ mass  and a  $12\ {\rm km}$ radius.
The corresponding analytical estimates, given by equation \eqref{eq:cross_section_inspirals_2}, are shown as  dashed lines.
The general normalization is found by numerical experiments, but as can be seen our analytical  model correctly
separates the cross sections  between NSs and BHs based solely on their mass difference. The reader is refered to the text for a discussion explaining the  slight difference at low SMA between the simple analytical scaling
and the simulations.
\textit{Top:} The probability for an outcome to result either in a collision or an inspiral given a CI.
\textit{Bottom:} The corresponding total integrated cross sections for each outcome.
As expected, the probability for an inspiral  decreases with SMA (equation \ref{eq:Probability_insp}) while  the total cross section increases with SMA (equation \ref{eq:cross_section_inspirals_2}).
Widely separated binaries are thus expected to be the dominant target for producing inspirals. Our numerical results used  $2\times10^5$ scatterings per SMA.}
\label{fig:Prob_outcome_givenCI}
\end{figure}

Figure \ref{fig:Prob_outcome_givenCI} shows the  formation probability and corresponding cross sections   of inspirals and collisions
as a function of initial SMA derived using numerical scattering experiments.  
The symbols show results from our numerical simulations while
the dashed lines show the results from our analytical estimates giving by equation \eqref{eq:cross_section_inspirals_2}.
As discussed in Section \ref{subset:ccins}, the inspiral cross section increases with SMA.
This is  because the gravitational focusing cross section for a CI increases  faster  with SMA ($\propto a_{0}$) than the  probability for an inspiral decreases ($\propto a_{0}^{-5/7}$).

As can be seen in Figure \ref{fig:Prob_outcome_givenCI}, the numerical and analytical scalings are in agreement
in the asymptotic limit but show small differences in slope at low SMA. These differences are caused by having
neglected a series of physical effects in the analytical scaling, such as collisions and GW energy losses before the  interaction has reached its final endstate.
However, these corrections are only important for target binaries in the high compactness  limit. From an astrophysical perspective, these binaries are believed to be  a negligible target population  as these they  are expected to merge before a CI can take place. The reader is refered to Section \ref{sec:discussion} for further discussion.

Since we have now shown that inspirals are a likely outcome even from widely separated binaries, it
is important to compare them with mergers arising from the widely-discussed
single-single GW capture scenario \citep{Hansen:1972il, 2011ApJ...737L...5S, Kocsis:2012ja, East:2012es, 2013PhRvD..87d3004E}.  

\subsection{Comparison to Single-Single Capture}

Inspirals resulting from binary-single interactions and mergers resulting from single-single GW capture can create 
binaries with extremely short merger times and, in some cases,   with very high eccentricity.
Comparing the formation probabilities for eccentric mergers arising from both mechanisms   is thus of great interest.

\begin{figure}[tbp!]
\centering
          \includegraphics[width=1\columnwidth]{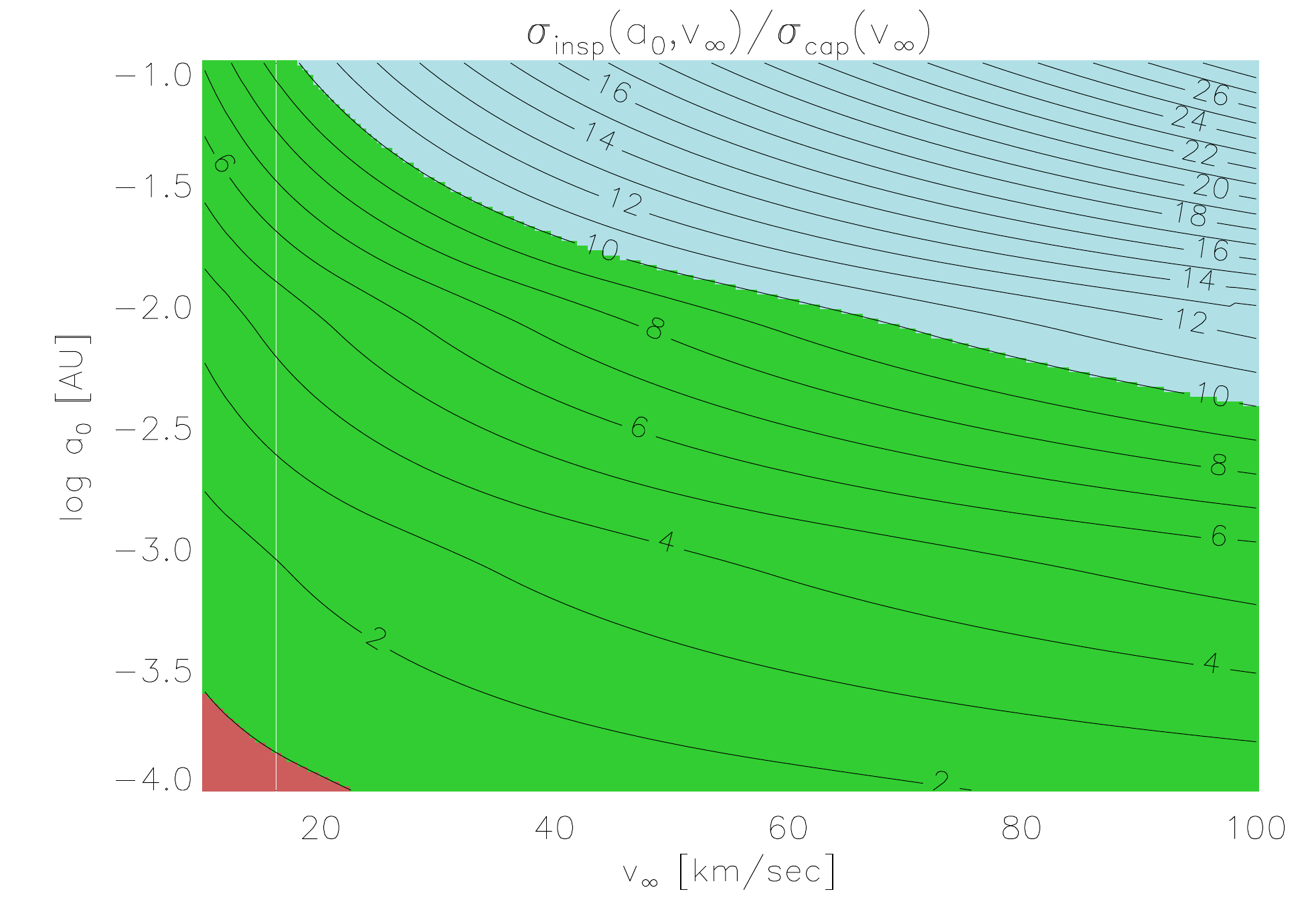}
          \includegraphics[width=1\columnwidth]{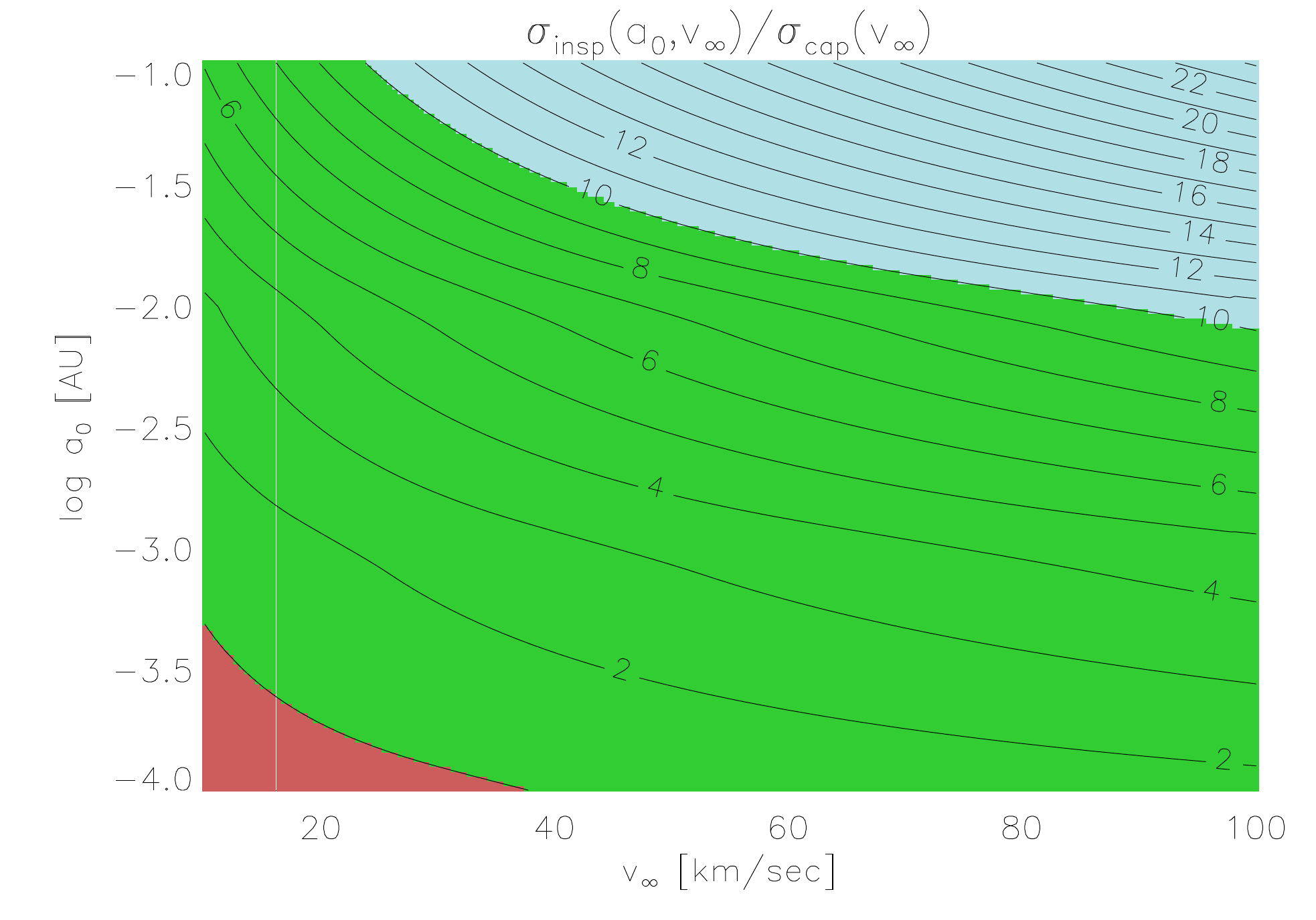}
    	\caption{Ratio between GW single-single capture cross section and binary-single inspirals cross section,
	${\sigma_{\rm insp}}/{\sigma_{\rm cap}}$, as a function of relative velocity at infinity $v_{\infty}$ and SMA, $a_0$, of the target binary.
	All interactions are between equal mass objects.
	{\it Top:} ${\sigma_{\rm insp}}/{\sigma_{\rm cap}}$ for $1\ M_{\odot}$ BHs. {\it Bottom:}  ${\sigma_{\rm insp}}/{\sigma_{\rm cap}}$ for $1.4\ M_{\odot}$ NSs with radius $12\ {\rm km}$.
	The {\it red}, {\it green} and {\it light blue} colors  respectively mark  the regions  where ${\sigma_{\rm insp}}<{\sigma_{\rm cap}}$,
	${\sigma_{\rm cap}<{\sigma_{\rm insp}}<10{\sigma_{\rm cap}}}$ and $10{\sigma_{\rm cap}}<{\sigma_{\rm insp}}$.}
	\label{fig:SSvsBS_crosssections}
\end{figure}

A single-single capture occurs when  two objects pass close enough to each other that the resulting   GW
energy losses 
are larger than their  initial positive energy.
To first order, the energy radiated away during the first passage can be obtained by integrating the GW energy losses
along the initial, \emph{unperturbed} unbound orbit \citep{Hansen:1972il}:
\begin{equation}
{\Delta}E	= -\frac{2}{15}\frac{G^{7/2}}{c^{5}}\frac{m_{1}^{2}m_{2}^{2}(m_{1}+m_{2})^{1/2}}{{r_{\rm min}(a,e)}^{7/2}}h(e),
\label{eq:deltaEunboundSS}
\end{equation}
where $r_{\rm min}=a(1-e)$ is the minimum distance between the two objects in the unbound  orbit and $h(e)$ is a dimensionless constant for which $h(e=1) = 425 \pi/ (8\sqrt{2})$. For capture to occur, we require  ${\Delta}E>(1/2){\mu}v_{\infty}^{2}$, where $\mu$ is the reduced mass.
Combining this with  equation \eqref{eq:deltaEunboundSS},  we find the maximum allowed $r_{\rm min}$ for a capture, which we denote  $r_{\rm cap}$ \citep{Lee:1993dt},
\begin{equation}
r_{\rm cap} = \left(\frac{85\pi}{6\sqrt{2}}\right)^{2/7}\frac{Gm_{1}^{2/7}m_{2}^{2/7}(m_1+m_2)^{3/7}}{c^{10/7}v_{\infty}^{4/7}}.
\end{equation}
All single-single encounters with pericenter distance smaller than $r_{\rm cap}$ become bound.

In analogy with the CI interaction cross section derived in Section \ref{sec:Close Interactions and Their Cross Section},  the  cross section for a single-single interaction with pericenter distance less than  $r_{\rm p,max}$ can be written as 
\begin{equation}
\sigma_{\rm SS}(r_{\rm min}<r_{\rm p,max}) \simeq \frac{2\pi G m_{\rm tot} r_{\rm p,max}}{v_{\infty}^{2}}.
\label{eq:SS_closeencounter_CC}
\end{equation}
The capture cross section can be estimated by inserting $r_{\rm p,max} = r_{\rm cap}$ in  equation \eqref{eq:SS_closeencounter_CC},
\begin{equation}
\sigma_{\rm cap}	= 2{\pi}\left(\frac{85\pi}{6\sqrt{2}}\right)^{2/7}\frac{G^{2}m_{1}^{2/7}m_{2}^{2/7}(m_1+m_2)^{10/7}}{c^{10/7}v_{\infty}^{18/7}}.
\label{eq:ss_capture}
\end{equation}
This cross section can then be compared directly with the cross section for inspirals arising from binary-single encounters. The ratio between the two cross sections can be approximated  using  equation \eqref{eq:cross_section_inspirals_2},
\begin{equation}
\frac{\sigma_{\rm insp}}{\sigma_{\rm cap}}	\propto \left(\frac{a_{0}v_{\infty}^{2}}{m}\right)^{2/7}.
\end{equation}
The number of inspirals relative to single-single captures is then expected to increase with $a_{0}$ and $v_{\infty}$, but
decrease as the mass increases.

Figure  \ref{fig:SSvsBS_crosssections} shows the numerically derived ratio of binary-single inspirals to single-single captures 
based on $8\times 10^5$ binary-single scatterings.
The two mechanisms have similar cross sections for tight binaries and typical cluster velocity dispersions. For binary SMA larger than $10^{-3}$ AU, binary-single inspiral interactions clearly dominate. This implies that inspirals resulting from binary-single interactions may contribute substantially to the inspiraling and eccentric merging binary population in globular clusters. In the next section, we will explore the particularly interesting case of binaries that pass through the {\it LIGO} detector frequency band with high eccentricity.

%
%
\section{Eccentric Inspirals in the {\it LIGO} Band}\label{sec:ecc_insp_in_LIGOband}

Compact merging binaries will be  observed by advanced {\it LIGO} in the near future \citep{Harry:2010hh, Mandel:2010ke, Abadie:2010fn, 2013arXiv1304.0670L}. To detect these inspirals, templates must be convolved with the timeseries data from the interferometer \citep{Abadie:2010fn, 2013arXiv1307.5307T, 2013arXiv1307.1757N, 2013PhRvD..87h2004B}.
The waveforms of relatively high eccentricity differ from those of circular binaries. 
For example, \citet{Huerta:2013wy} find that for eccentricities greater than about $e\approx0.2$, the match to circular templates is degraded by more than 50\%. An understanding of the quantity and origin of eccentric binaries that pass through the {\it LIGO} band is therefore extremely important for future GW searches. 

In the GW inspirals and mergers, one might expect that the majority of binaries will be nearly circular when entering the {\it LIGO} band, since GWs carry away both energy and angular momentum at a rate such that the circularization time is similar to the merging time  \citep{Peters:1964bc, 2004ApJ...616..221G, 2006ApJ...640..156G}. However, as we show in this paper, the dynamical inspiral states formed in binary-single encounters are formed with very high initial eccentricity and rapid merger times. 
As a result,  most of these dynamical formed inspirals will be 
directly observable in the {\it LIGO} band at the time of formation, i.e. when they are still  highly eccentric.
In what follows, we  explore in detail  the fraction of highly eccentric {\it LIGO} sources one expects to come from binary-single interactions as well as making a direct comparison to highly eccentric inspirals formed via single-single interactions.

To quantify the number of eccentric binary mergers in our scattering experiments, we use an approximate form for the gravitational peak frequency \citep{Wen:2003bu},
\begin{equation}
f_{\rm GW}		 = \frac{1}{\pi}\sqrt{\frac{Gm_{\rm tot}}{a^{3}}}\frac{(1+e)^{1.1954}}{(1-e^{2})^{1.5}},
\label{eq:grav_peak_fGW}
\end{equation}
where $\sqrt{{a^{3}}/{Gm_{\rm tot}}}$ is the orbital time, $T_{\rm orb}$.

\subsection{Eccentric Binaries From Binary-Single Interactions}

\begin{figure}
\centering
\includegraphics[width=1\columnwidth]{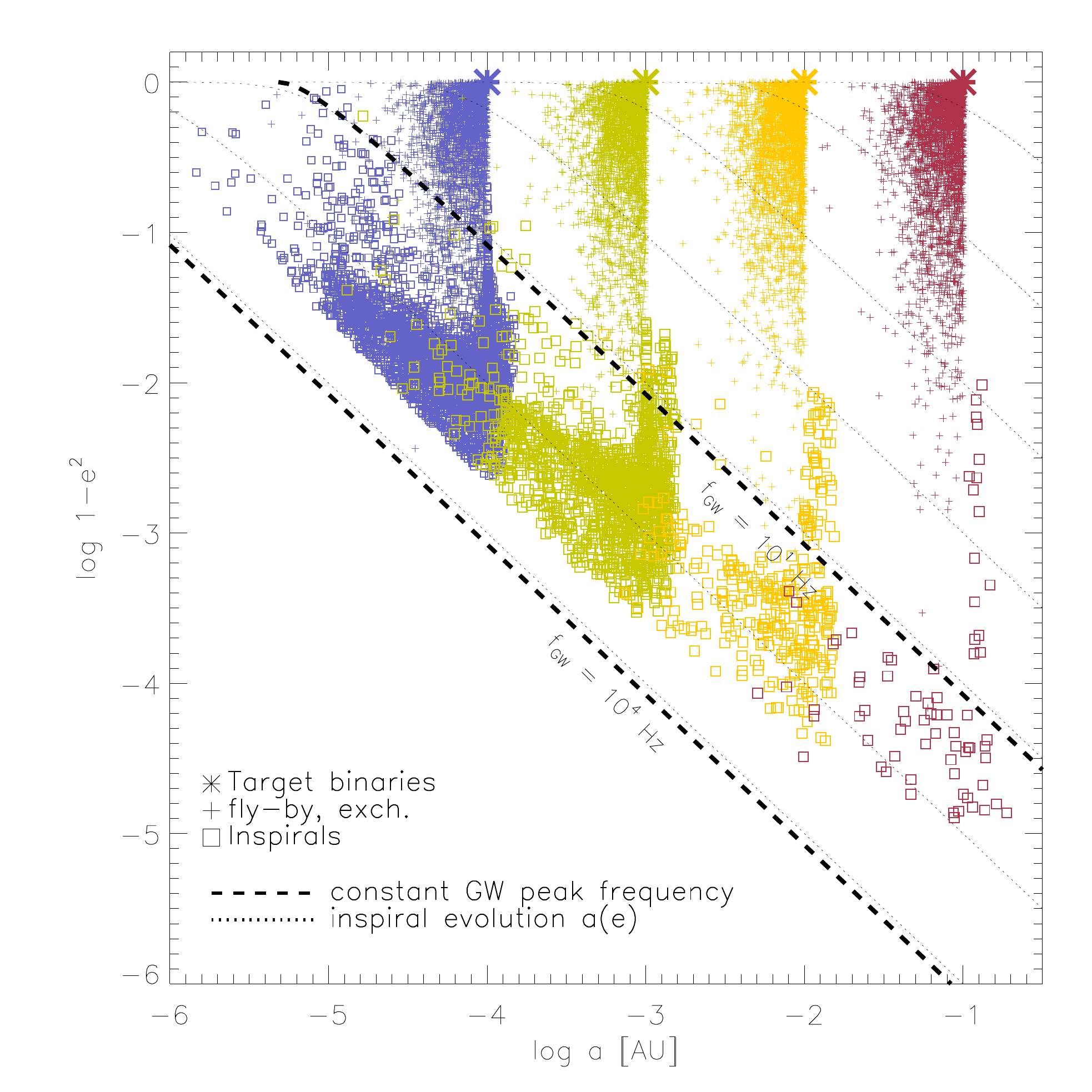}
\includegraphics[width=1\columnwidth]{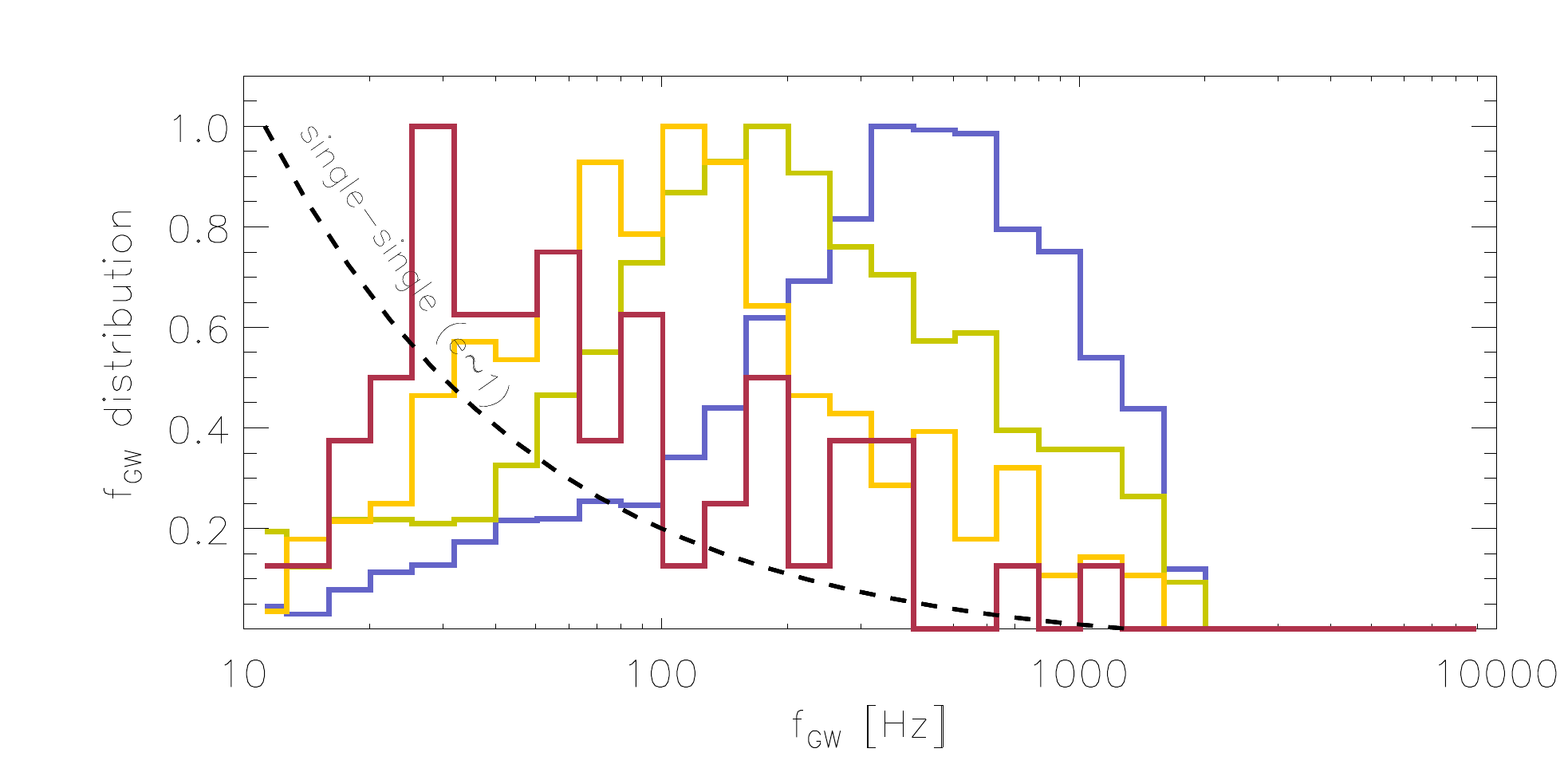}
\caption{Distribution of orbital parameters in the $(a,1-e^{2})$ plane  and the corresponding gravitational peak frequency $f_{\rm GW}(a,e)$
for all endstate binaries resulting from binary-single interactions between NSs with $1.4\ M_{\odot}$ masses and  12 km radii.
The relative velocity between the encounter and the target binary is  $v_{\infty}=10\ {\rm km\; s^{-1}}$.
The plot includes the classical outcomes exchange and fly-by ({\it plus symbols}) and
the GR outcome inspirals ({\it squares}). Different colors denote different initial SMA of the target binary.
\textit{Top:} Orbital parameters at the time of final state identification.
The inspirals fade away as the SMA increases. 
The {\it dashed-black lines} show the GW peak frequencies $10^{1},10^{4}$ Hz that are approximately representative of the  advanced {\it LIGO} window.
The {\it dotted black lines} show a few examples of the inspiral orbital evolution due to GW radiation given by equation \eqref{eq:a_evolve_from_e_peters}.
When $(1-e^{2})\ll 1$, these evolutionary tracks  are parallel to the gravitational peak frequency lines. This 
implies  that if a binary with high eccentricity is \emph{not} formed in the {\it LIGO} band, then it will \emph{never}
evolve into it with high eccentricity. Inspirals are therefore the only states arising from a binary-single interaction
that will have the potential of being observable as high eccentric mergers.
\textit{Bottom:} Distributions of gravitational peak frequencies from all identified  inspirals.
These distributions stay almost unchanged during the inspiral since the binaries evolve with approximately constant GW frequency.
The sensitivity of {\it LIGO} peaks around $\sim200$ Hz. The {\it dashed-black line} shows the eccentricity  distribution expected from merging  binaries
resulting from single-single captures. For illustration purposes, all histograms have been normalized to their peak values.}
\label{fig:BS_LIGO_ae_edist}
\end{figure}

The eccentricity distribution of binaries resulting from binary-single interactions 
includes binaries that \emph{evolve} into the {\it LIGO} band and binaries that are \emph{born} in the {\it LIGO} band.
Figure \ref{fig:BS_LIGO_ae_edist} shows the results from binary-single interactions between NSs with $1.4\ M_{\odot}$ masses  and $12\ {\rm km}$ radius for different initial SMA of the target binary.
The {\it top panel} shows the distribution of all binaries in the $\text{log}(a,1-e^{2})$ plane immediately after final-state identification. Inspirals are shown with {\it large square} symbols.
The distribution of inspirals is not static. Instead, each binary evolves due to GW radiation according to equation \eqref{eq:a_evolve_from_e_peters}. The {\it dotted black} lines show a few of these evolutionary  tracks.   The two {\it dashed-black} lines show constant gravitational peak frequencies $f_{\rm GW}=10^{1}, 10^{4}$ Hz, which 
have been chosen to illustrate the  sensitivity window range for advanced {\it LIGO}  \citep{Harry:2010hh, 2013arXiv1304.0670L}.

By comparing the orbit evolution trajectories  in Figure \ref{fig:BS_LIGO_ae_edist} with the lines of constant $f_{\rm GW}$, we can see that they are parallel for $\text{log}(1-e^{2})\ll0$.
This is because the evolution of $a$ for both scales as $\left(1-e^{2} \right)^{-1}$.
 This implies 
 that high eccentricity mergers that are not born in the {\it LIGO} band cannot evolve into it  with high eccentricity.
 The binaries that are identified inside the {\it LIGO} band are thus the only ones that  are able to be detected  with high eccentricity. 
 This set of binaries is  the dynamically formed inspirals. 
 From the $(a,e)$ distributions shown in the {\it top} panel in Figure \ref{fig:BS_LIGO_ae_edist}
 one can calculate the corresponding $f_{\rm GW}$ distributions by making use of equation \eqref{eq:grav_peak_fGW} ({\it bottom}
 panel in Figure \ref{fig:BS_LIGO_ae_edist}).
 The values of $f_{\rm GW}$  are observed to change only slightly  during inspiral, since the binaries spiral in with almost constant peak frequency.
 As observed in Figure \ref{fig:BS_LIGO_ae_edist}, target binaries with $a \sim 10^{-2}-10^{-3}$ AU  produce inspirals
 with $f_{\rm GW}$ distributions that peak around the most sensitive {\it LIGO} frequency $\approx 200\ {\rm Hz}$.
 The relative normalizations of the distributions shown in the {\it bottom} panel of  Figure \ref{fig:BS_LIGO_ae_edist}   can be derived from  Figure \ref{fig:Prob_outcome_givenCI}.

\subsection{Eccentric Binaries from Single-Single Capture}

Once a binary is formed via single-single GW capture, its subsequent evolution can be  followed in the $(a,e)$ plane according to equation \eqref{eq:a_evolve_from_e_peters}. By analogy with arguments presented above for the binary-single capture case, we can conclude that if binaries formed through single-single capture are not formed with $f_{\rm GW}$ that places them in the {\it LIGO} band, they will circularize before {\it LIGO} can observe them as eccentric binaries. 

\begin{figure}
\centering
\includegraphics[width=0.95\columnwidth]{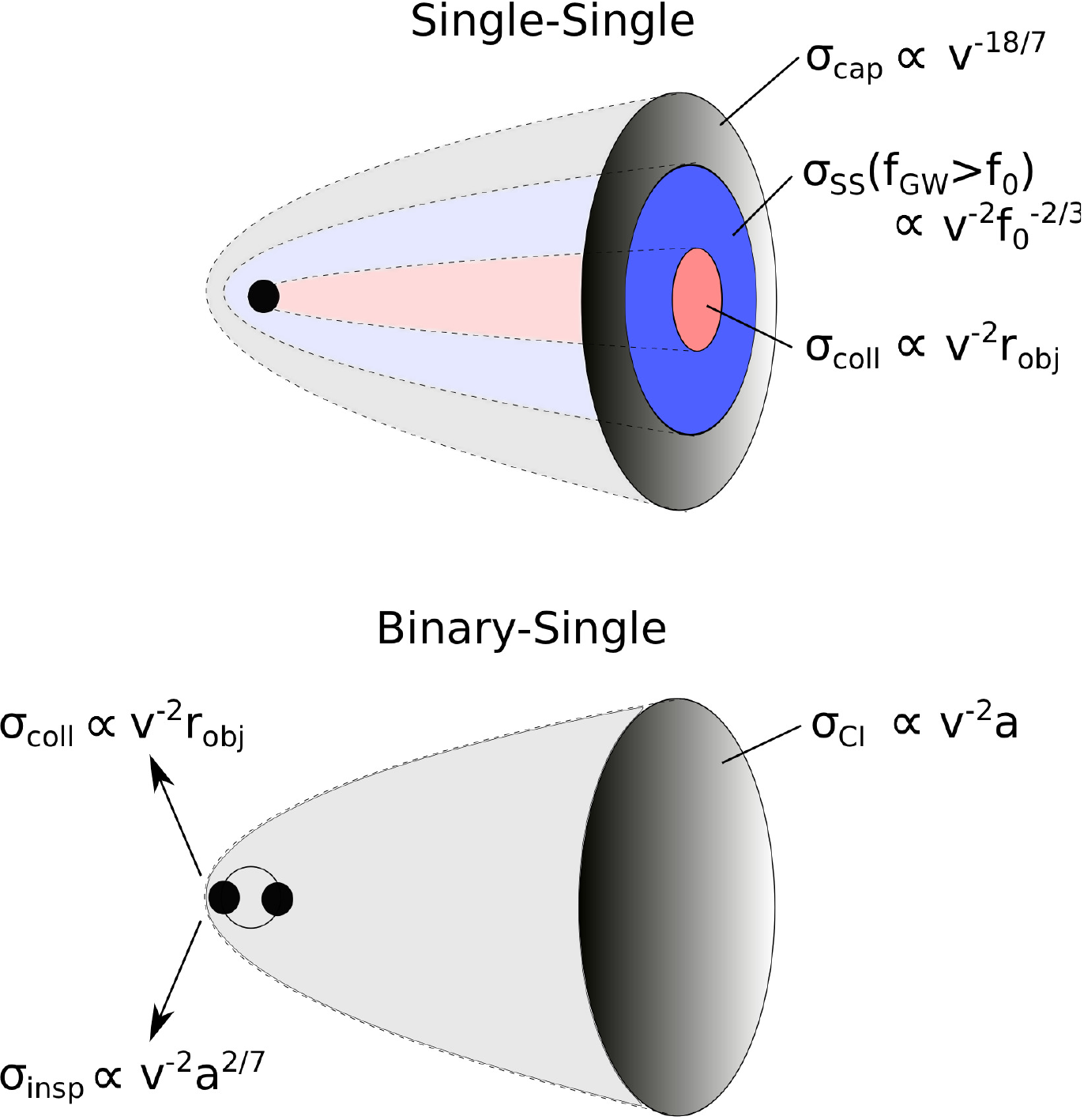}
\caption{Diagram illustrating different outcome cross sections arising from  single-single ({\it top}) and binary-single ({\it bottom}) interactions.
Also  shown are the approximate dependence of the various cross sections on the encounter velocity, $v$, the SMA of the target binary, $a$, the
object radius, $r_{\rm obj}$, and the gravitational peak frequency, $f_{\rm GW}$.
The single-single capture cross section is denoted by $\sigma_{\rm cap}$ (equation \ref{eq:ss_capture}), the high
eccentric single-single capture cross section with $f_{\rm GW}>f_{0}$ by
$\sigma_{\rm SS}(f_{\rm GW}>f_{0})$ (equation  \ref{eq:cross_sec_highecc_ss}), the direct collision cross section 
by $\sigma_{\rm coll}$ (equation \ref{eq:SS_closeencounter_CC}), the CI cross section by $\sigma_{\rm CI}$ (equation \ref{eq:def_of_crosssection}) and
the binary-single inspiral cross section by $\sigma_{\rm insp}$ (equation \ref{eq:cross_section_inspirals_2}).
It is particularly interesting to compare $\sigma_{\rm SS}(f_{\rm GW}>f_{0})$ with $\sigma_{\rm insp}$
because the inspiralling binaries formed in each of these cases give very similar observational signatures. For example,
both channels can form inspirals that enter the {\it LIGO} band with high eccentricity, an event that is not observed when field binaries merge.
Since both channels  scale as $\propto v^{-2}$, their ratio is independent of $v$ as shown in equation \eqref{eq:CSfrac_insp_highecc_ss}.}
\label{fig:illustration_SS_vs_BS}
\end{figure}

To estimate the cross section for highly eccentric {\it LIGO} sources resulting from single-single captures,
we first rewrite equation \eqref{eq:grav_peak_fGW} in the equal mass case and in the high eccentricity limit ($e\sim 1$),
\begin{equation}
\begin{split}
r_{0}			& 	 \simeq \left(\frac{2^{2.3908}}{4\pi^{2}}\frac{Gm}{f_{0}^{2}}\right)^{1/3},
\end{split}
\label{eq:rmin_from_peakfGW}
\end{equation}
where $m$ is the mass of each of the objects, and 
$r_{0}$ is the required pericenter distance for an eccentric binary to have a peak frequency $f_{0}$. It then follows 
that all encounters with pericenter distance $r_{\rm min}<r_{0}$ will have $f_{\rm GW}>f_{0}$. Therefore,
the cross section for a single-single encounter having $f_{\rm GW}>f_{0}$  can be simply calculated 
by setting $r_{\rm p,max}=r_{0}$ in equation  \eqref{eq:SS_closeencounter_CC},
\begin{equation}
\sigma_{\rm SS}(f_{\rm GW}>f_{0}) = \frac{4\pi Gm}{v_{\infty}^{2}}\left(r_{0}-2r_{\rm obj}\right).
\label{eq:cross_sec_highecc_ss}
\end{equation}
To   account for  the object's finite size ($r_{\rm obj}$), we have subtracted  the cross section for direct collisions  in equation \eqref{eq:cross_sec_highecc_ss}.
The velocity dependence ($v_{\infty}^{-2}$) in equation  (\ref{eq:cross_sec_highecc_ss}) implies that the cross section for high eccentricity single-single captures 
scales as the gravitational focusing cross section. 
The single-single capture cross section scales as $v_{\infty}^{-18/7}$ such that  $\sigma_{\rm SS}( f_{\rm GW}>f_0 )/\sigma_{\rm cap}\propto v_{\infty}^{4/7}$.
As the velocity increases, the single-single high eccentricity cross section relative to the capture cross  section will also increase.
The {\it dashed-black} line  in the {\it bottom} panel in  Figure \ref{fig:BS_LIGO_ae_edist} shows the eccentricity distribution given by equation \eqref{eq:cross_sec_highecc_ss} for 
single-single encounters, which  we confirmed using scattering experiments  of single-single objects.
In Figure \ref{fig:illustration_SS_vs_BS}, we show the different cross sections and corresponding scalings for the various outcomes expected from single-single and binary-single encounters.

\subsection{Comparison between Binary-Single and Single-Single}
 
\begin{figure}
\centering
\includegraphics[width=1\columnwidth]{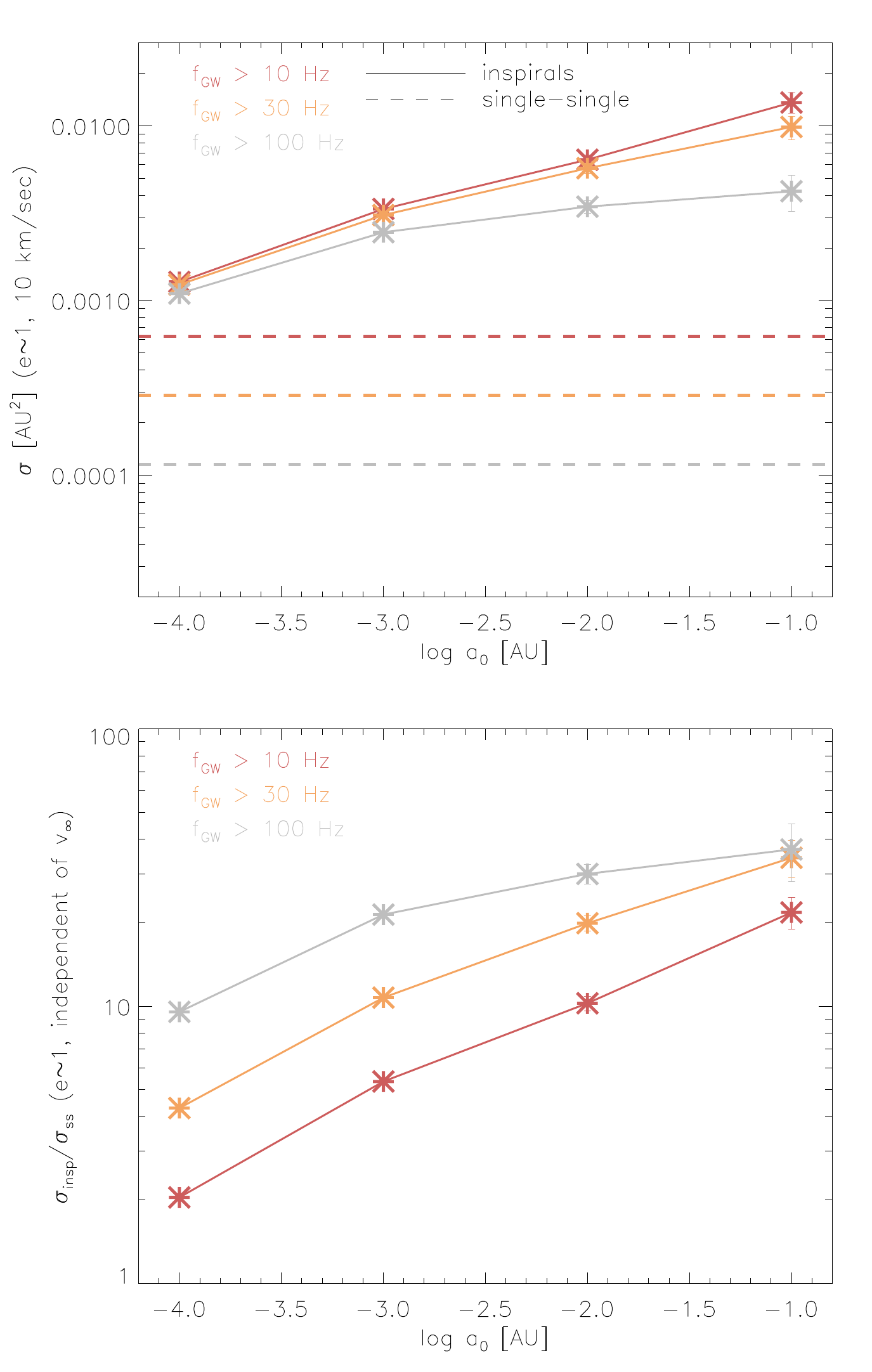}
\caption{Numerically calculated cross sections  for high eccentricity binaries ($e\sim 1$) arising from  binary-single
encounters and their relative importance when compared to those 
produced by  single-single encounters. We divide  the merging  binaries  based on their gravitational peak frequency at formation: $f_{\rm GW}>10,30,100$ Hz. During the inspiral,  the orbital 
parameters $(a,e)$ change according to equation \eqref{eq:a_evolve_from_e_peters} but $f_{\rm GW}$
remains relatively  constant, which means that the results are not altered significantly as the binary evolves.
All results are for  scatterings between NSs with $1.4\ M_{\odot}$ masses  and  $12\ {\rm km}$ radii.
\textit{Top:} Inspiral cross sections. Solid lines show inspirals formed by binary-single interactions
and dashed lines show inspirals formed by single-single captures.
The resultant  high eccentricity binaries formed via  binary-single and single-single
encounters  have different gravitational peak frequencies at formation as shown in Figure \ref{fig:BS_LIGO_ae_edist}.
Each line defined by $f_{\rm GW}$ denotes a cross section that only includes inspirals that are born with a gravitational frequency above the given threshold.
\textit{Bottom:} Ratio between the single-single and binary-single cross sections shown in the {\it top} panel. As described in the text,  both high eccentricity single-single and binary-single
inspirals scale as  $\propto v_{\infty}^{-2}$. This makes the ratio \emph{independent}
of velocity.}
\label{fig:BS_vs_SS_higheccLIGOsources}
\end{figure}

In previous sections, we have computed the scalings for the cross sections of  binary-single interactions and single-single captures;  a summary of our results is  given in Figure \ref{fig:illustration_SS_vs_BS}.
We now turn our attention to the  relative normalization of eccentric inspirals arising from binary-single and single-single capture as a function of binary SMA and GW frequency threshold.
Figure \ref{fig:BS_vs_SS_higheccLIGOsources}
shows the normalization of the numerically computed  inspiral cross sections for interacting NSs given  three frequency thresholds $f_{0} = 10,30,$ and $100\ \text{Hz}$ as a function of the initial binary SMA.
The {\it upper} panel shows the resulting cross sections in AU$^{2}$ for encounters with $v_\infty= 10 \text{ km s}^{-1}$. The {\it lower} panel  shows these cross sections normalized to the corresponding single-single cross sections.
Inspirals become increasingly dominant  relative to the number of single-single eccentric binaries  as the frequency threshold and the SMA increases. 
The ratio between the two cross sections is independent of velocity because both cross sections scale with the gravitational focusing cross section, $v_\infty^{-2}$.
This general behavior can be understood  analytically  by writing out the ratio
\begin{equation}
\frac{\sigma_{\rm insp}}{\sigma_{\rm SS}(f_{\rm GW}>f_{0} )}  \simeq \frac{3}{4}\frac{P_{\rm insp}a_{0}}{r_{0}-2r_{\rm obj}} \propto {a_{0}^{2/7}}{f_{0}^{2/3}}.
\label{eq:CSfrac_insp_highecc_ss}
\end{equation}
The estimation of $P_{\rm insp}$ in this limit is given by equation \eqref{eq:Probability_insp}.
Our numerical and analytical results strongly suggest that the cross section for the formation of eccentric
compact binary inspirals is significantly larger in the binary-single case than in the single-single case even  when  the fraction of compact objects in binaries is relatively modest.

%
%

\section{Discussion}\label{sec:discussion}
\begin{figure}
\centering
\includegraphics[width=1\columnwidth]{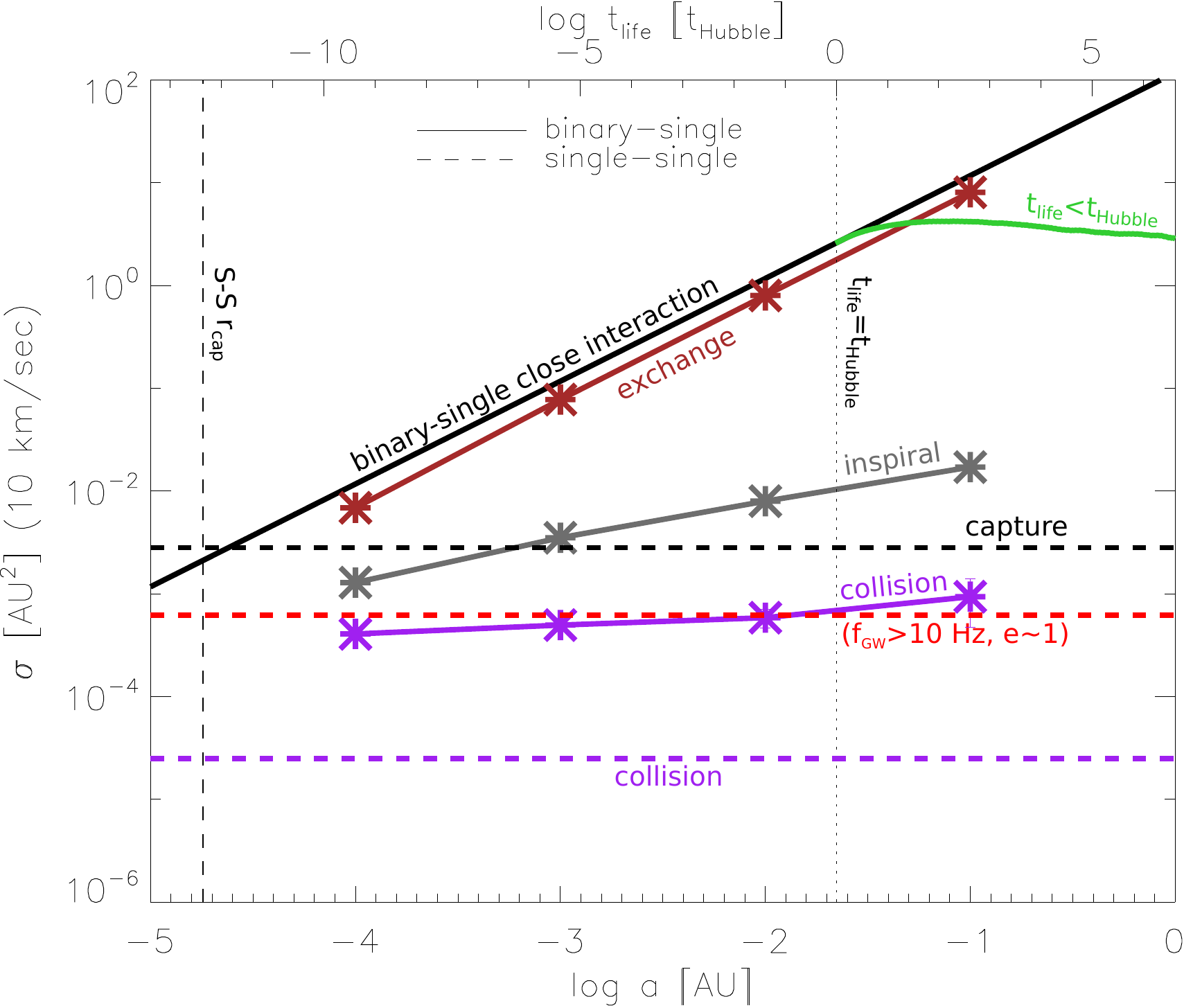}
\caption{Summary of relevant outcome cross sections arising from binary-single and single-single encounters between
equal mass NSs. Each NS has a mass of $1.4\ M_{\odot}$ and radius of 12 km.
The {\it dashed} lines show results from single-single encounters while the {\it solid} lines show results from 
binary-single interactions.
The {\it black} solid line shows the CI crossection, the {\it dark-grey} line  the inspiral cross section and the  {\it purple} and {\it brown} lines the cross sections for   collisions and exchanges, respectively.
The {\it green} line shows the cross section for  binaries that merge in less than a Hubble time.
The {\it black-dashed} shows the single-single capture cross section and the {\it red-dashed} line shows the cross section for single-single high eccentric ($e\sim1$) binary with gravitational peak frequency $f_{\rm GW}>10$ Hz.
The {\it vertical-black-dashed} line shows the single-single pericenter distance for a capture $r_{\rm cap}$. We note that the scaling between lines depends on velocity, here assumed to be 10 km s$^{-1}$.}
\label{fig:summary_all_crosssec_3NS}
\end{figure}

We have discussed the formation of eccentric inspirals in the context of  binary-single interactions and compared them to the more widely discussed single-single capture scenario. 
The expected outcomes for binary-single and single-single interactions of equal mass NSs are shown in  Figure \ref{fig:summary_all_crosssec_3NS}.
The {\it solid-black} line shows the binary-single CI cross section. 
Other outcomes shown are sub-categories of the CI cross section. The {\it solid-red} line shows exchange, the {\it solid-grey} inspirals, and the {\it solid-purple} collisions. The {\it green} line shows binaries with merger lifetimes less than a Hubble time, which  will be  discussed in Section \ref{sec:bin_lifetimes}.
Similarly, the {\it dashed-black} line shows the total cross section for single-single capture, while the {\it dashed-red} line shows only eccentric captures for which $f_{\rm GW} >10$ Hz, and the {\it dashed-purple} line shows the collision cross section. As we emphasized in the previous section, most inspirals occur with $f_{\rm GW} \gtrsim 10$ Hz, so the inspiral cross section may be directly compared to the eccentric component of the single-single cross section. The upper {\it x}-axis label shows the GW inspiral lifetime for binaries separated by a given initial SMA (bottom {\it x-}axis labels). 

Here we turn our attention to the implications of our results and illustrate  how they change with the inclusion of a more extended binary companion by calculating scatterings for WD-NS binaries in Section \ref{sec:wds}. We  discuss the merger lifetime and resulting center-of-mass kicks  in  Sections \ref{sec:bin_lifetimes} and \ref{sec:kicks}, respectively.  We provide a simple estimate of typical event rates in dense stellar systems  in Section \ref{sec:rates}. Finally, we present our  conclusions   in Section \ref{sec:significance}.

\subsection{Target Binaries Containing White Dwarfs}\label{sec:wds}
We have seen that wider binary SMAs lead to an enhancement in the cross section for inspiral outcomes in the case of binaries comprised of NSs and BHs. In widely separated binaries, the binary members need not be compact objects.  In this section, we consider the case where the target binary contains a white dwarf (WD) companion \citep{2009arXiv0912.0009T}. 

WDs have a well defined mass-radius relationship, which takes the following form for lower-mass WDs,
\begin{eqnarray}
r_{\rm WD} &\simeq& \frac{1}{m_{\rm WD}^{1/3}}\frac{(18\pi)^{2/3}}{10}\frac{\hbar^{2}(m_{\rm p}/0.5)^{-5/3}}{Gm_{\rm e}}, \\
&\approx&  2.9 \times 10^9 \left( m_{\rm WD} / M_\sun \right)^{-1/3}\text{cm},
\label{eq:massradius_WD}
\end{eqnarray}
where $m_{\rm e}$ is the electron mass and $m_{\rm p}$ the proton mass \citep{Carroll:1996ut}. 

Another characteristic scale imposed by the size of the WD is the separation at which the WD fills its Roche lobe, 
\begin{equation}
a_{\rm MT}	\simeq r_{\rm WD}\frac{0.6q^{2/3}+\ln(1+q^{1/3})}{0.49q^{2/3}},
\label{eq:mass_transfer_SMA}
\end{equation} 
where $q=m_{\rm WD}/m_{\rm NS}$ \citep{Eggleton:1983fq}. In WD-NS binaries containing moderately massive WDs, the resulting mass transfer is stable, and the binary overcomes the destabilizing effects  produced by GW radiation due to the ongoing mass transfer \citep[e.g.][]{Marsh:2004fv,Paschalidis:2009hk}.

\begin{figure}
\includegraphics[width=1\columnwidth]{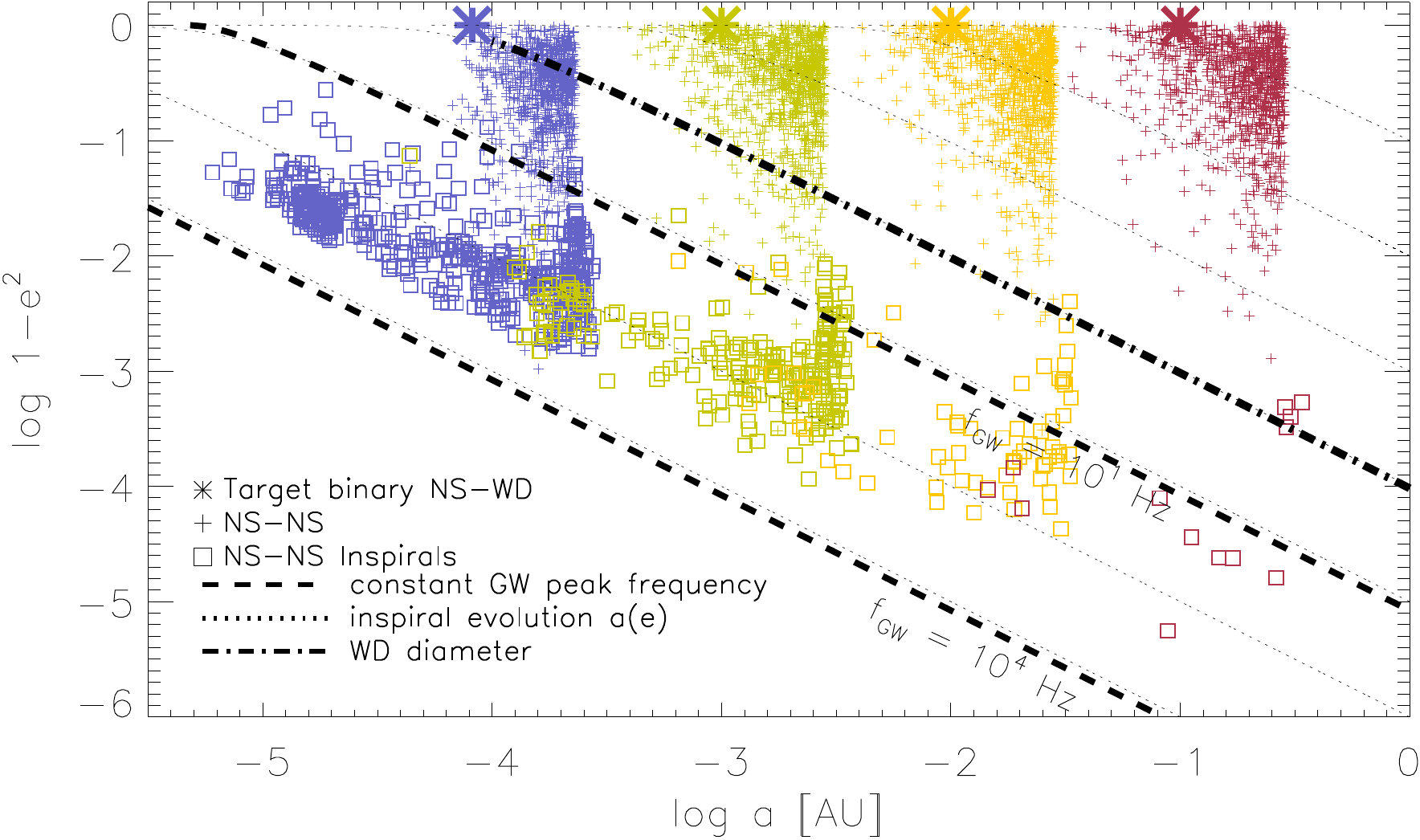}
\includegraphics[width=1\columnwidth]{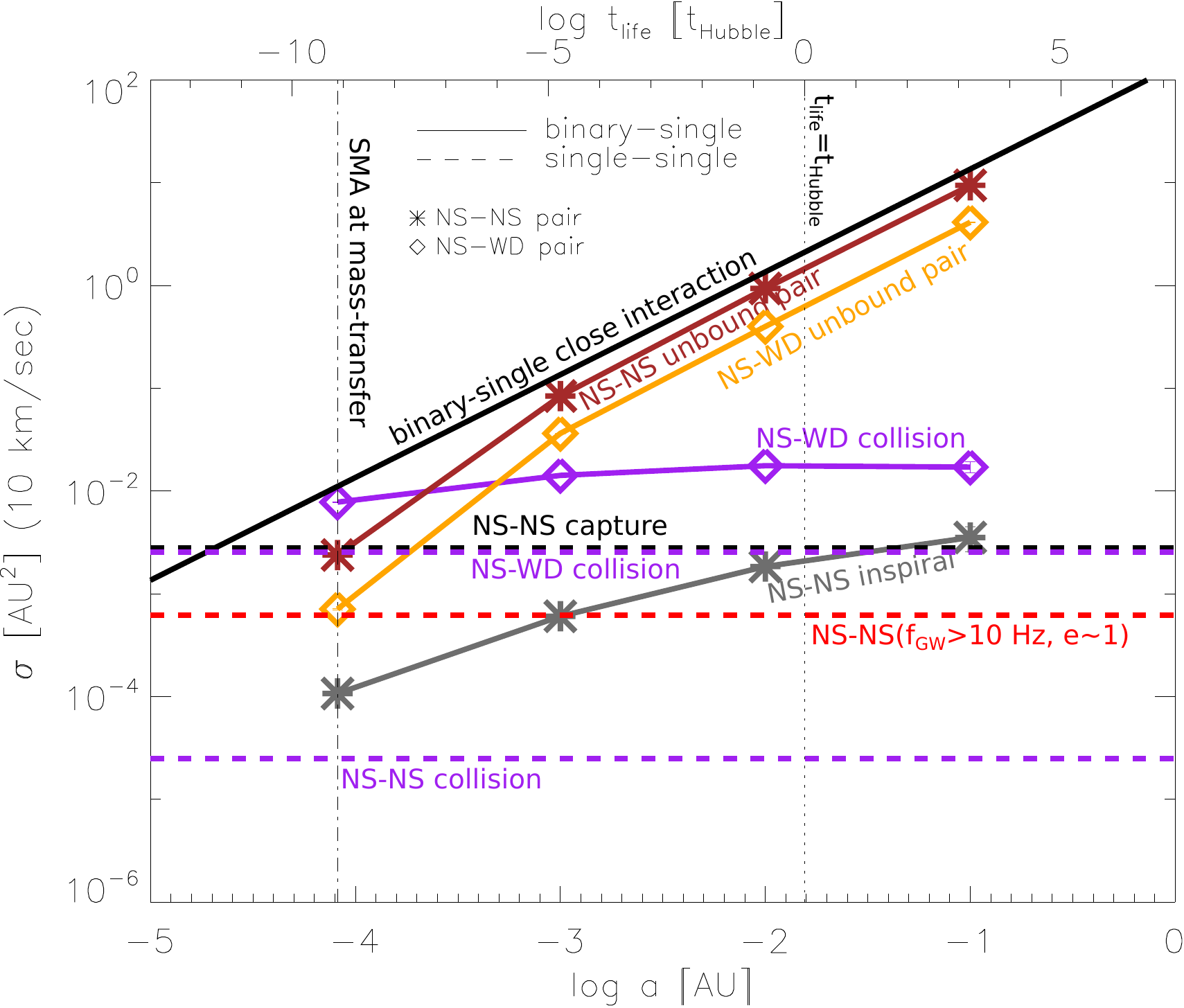}
\caption{Results from scatterings between a
${\rm NS}(1.4\ M_{\odot},{\rm 12\ km})-{\rm WD}(0.5\ M_{\odot})$ binary and a ${\rm NS} (1.4\ M_{\odot},{\rm 12\ km})$ encounter with $v_{\infty}=10\ {\rm km\;s^{-1}}$.
\textit{Top:} Scatter plot of the orbital parameters $(a, 1-e^{2})$ for all endstate NS-NS binaries (similar to Figure \ref{fig:BS_LIGO_ae_edist}). The resulting inspirals are
shown with {\it square} symbols. Each color show results for a given SMA.
The radius of the WD is shown as a {\it dashed-dotted line}. As can be seen, this line is well above the region where inspirals form implying 
that WD inspirals are very unlikely.
\textit{Bottom:} Similar to Figure \ref{fig:summary_all_crosssec_3NS} but for target binaries including a WD companion with $0.5\ M_{\odot}$. The cross section for inspirals is
significantly smaller here than in the equal mass NS case. The three main reasons for this are that only 1 out of 3 endstates can
result in a NS-NS inspiral, collisions with the WD deplete inspiral outcomes, and the relatively small mass of the WD suppresses resonances which could otherwise form inspirals.}
\label{fig:WD05NS_ae_cross_secs}
\end{figure}

The phase space of NS-NS binary outcomes that result from NS scatterings including a companion WD are shown in the upper panel of Figure \ref{fig:WD05NS_ae_cross_secs}, which can  be directly compared to the upper panel of Figure \ref{fig:BS_LIGO_ae_edist}.  These experiments involve a $1.4\ M_\odot$ NS encountering a WD-NS binary containing a  $0.5\ M_{\odot}$ WD and a $1.4\ M_{\odot}$ NS. A  comparison to Figure \ref{fig:summary_all_crosssec_3NS} shows the increased importance of collisions in the WD-NS target case when compared to NS-NS targets. However, we see that inspiral outcomes between  two NSs are still possible, despite the presence of the WD. By contrast, inspirals between the WD and the NS typically do not occur due to the extended radius of the WD \citep[see e.g.][for double WDs seen by {\it LISA}\footnote{http://lisa.nasa.gov/}]{2007ApJ...665L..59W}. 
However, the cross-section for inspirals is reduced somewhat as compared to NS-NS target binaries.  This is partially due to the fact that there is one (rather than three) possible pairwise combination that can result in double NS binaries.  Additionally, in tight binaries with $a\approx a_{\rm MT}$, collisions with the WD play an important role in depleting inspiral outcomes \citep{Lee:2010in}. The hierarchy of masses in the system also likely plays a role by somewhat reducing the typical number of resonances \citep{Sigurdsson:1993jz}.  Despite these effects which tend to deplete the number of inspiral outcomes, we find that NS-NS inspirals have a larger cross section than single-single captures with $f_{\rm GW} > 10$ Hz as long as the binary SMA $a_0 \gtrsim 10^{-3}$ AU. Thus we still expect wide binaries containing WDs to contribute meaningfully to the eccentric inspiral channel, in particular if they dominate the NS-hosting binary population as in \citet{Grindlay:2006ef}. A concern for systems containing extended objects is that tidal dissipation may play in important role in modifying the dynamics \citep[e.g.][]{1986ApJ...306..552M}, an effect we ignore here and  hope to implement in future work.

\subsection{Binary lifetimes}\label{sec:bin_lifetimes}

\begin{figure}
\centering
\includegraphics[width=\columnwidth]{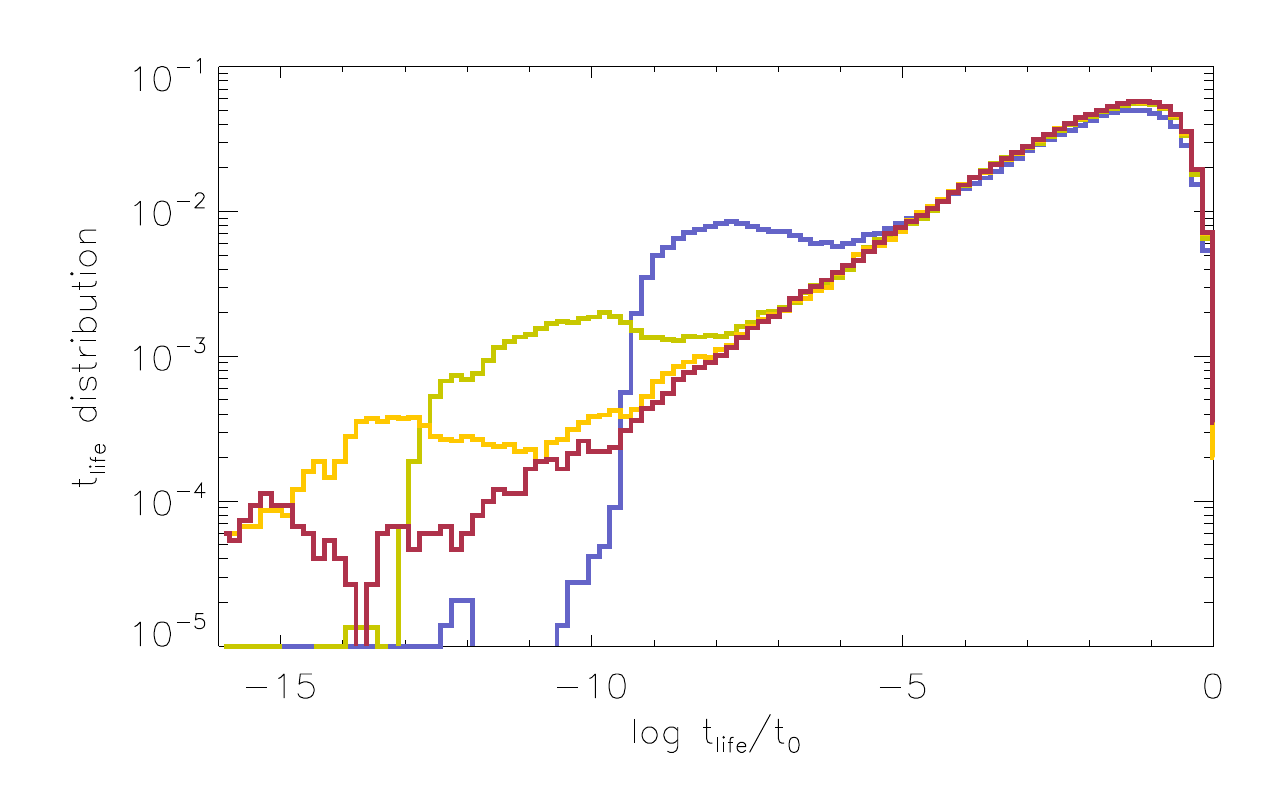}
\caption{Lifetime distributions including all endstate binaries. Colors denote initial binary SMA, $a_0$, from $10^{-4} - 10^{-1}$ AU (blue to red). 
In the Newtonian point-mass case, all initial SMAs would follow the same distribution, but
when collisions and GR are included then the initial SMA $a_{0}$ plays a role in forming the final distribution.
The knee that appears for each distribution is the fast merging inspirals.}
\label{fig:tlifedist}
\end{figure}

Even if the  initial binary lifetime is greater than a Hubble time, $t_{\rm Hubble}$, a fraction of binaries that undergo a  scattering will be either
deposited or exchanged into orbits with very short lifetimes  \citep{Clausen:2012hn}. Thus a fraction of even very widely separated binaries can produce mergers with $t_{\rm life} < t_{\rm Hubble}$. 
Figure \ref{fig:tlifedist} shows the distribution of final binary lifetimes realized following binary-single scatterings with varying binary SMA. In the classical point-mass limit, we
see that an approximate power-law distribution is produced. The inclusion of GW radiation and finite radii introduces two physical scales  that break the self-similarity of the problem. The hard cutoff corresponds to the scales of the objects themselves and depletion by collisions. The inspiral population manifests itself as a knee at scales corresponding to the typical pericenter distances of the rapid inspiral outcomes. 

The cross section for creation of binary products whose lifetime are less than a Hubble time is plotted in Figure \ref{fig:summary_all_crosssec_3NS} for
encounters involving NS.
The key feature of this cross section is that it does not vanish when $a_{0} \gtrsim 10^{-1.7}$ AU, where $t_{0} > t_{\rm Hubble}$.
Instead this cross section remains approximately flat.
The reason for this is that resultant binaries generally have a much smaller pericenter distance than the target binary and therefore also a shorter lifetime
as seen in Figure \ref{fig:tlifedist}.

\subsection{Retention or Ejection of Binary-Single  Outcomes}\label{sec:kicks}

A remaining question is whether final binaries resulting from binary-single interactions are  kicked out, or whether they merge in-situ.  
Kicks relative to the initial center of mass occur when a fraction of the initial binary's binding energy is transferred to the relative motion of the binary and the single \citep{1991Natur.349..220P}.
We denote the resulting binary kick velocity as $v_{\rm kick}$. 
The associated hardening of these binaries leads to a shorter binary lifetime (since $t_{\rm life} \propto a^{4})$
and one therefore expects that a high kick velocity is associated with a short lifetime. A binary that receives a high-velocity kick will therefore not necessarily merge outside of its environment. 

\begin{figure}
\includegraphics[width=\columnwidth]{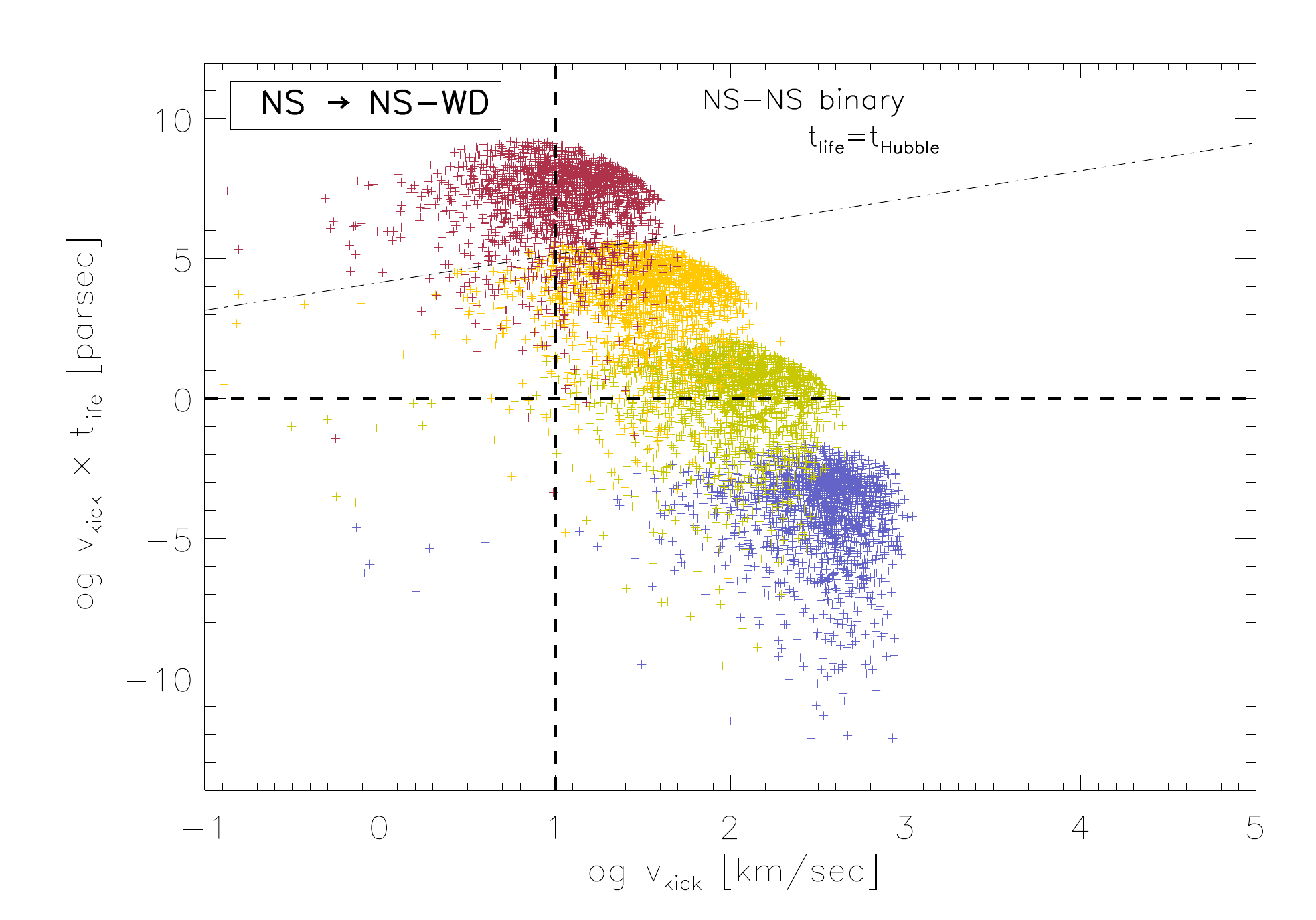}
\includegraphics[width=\columnwidth]{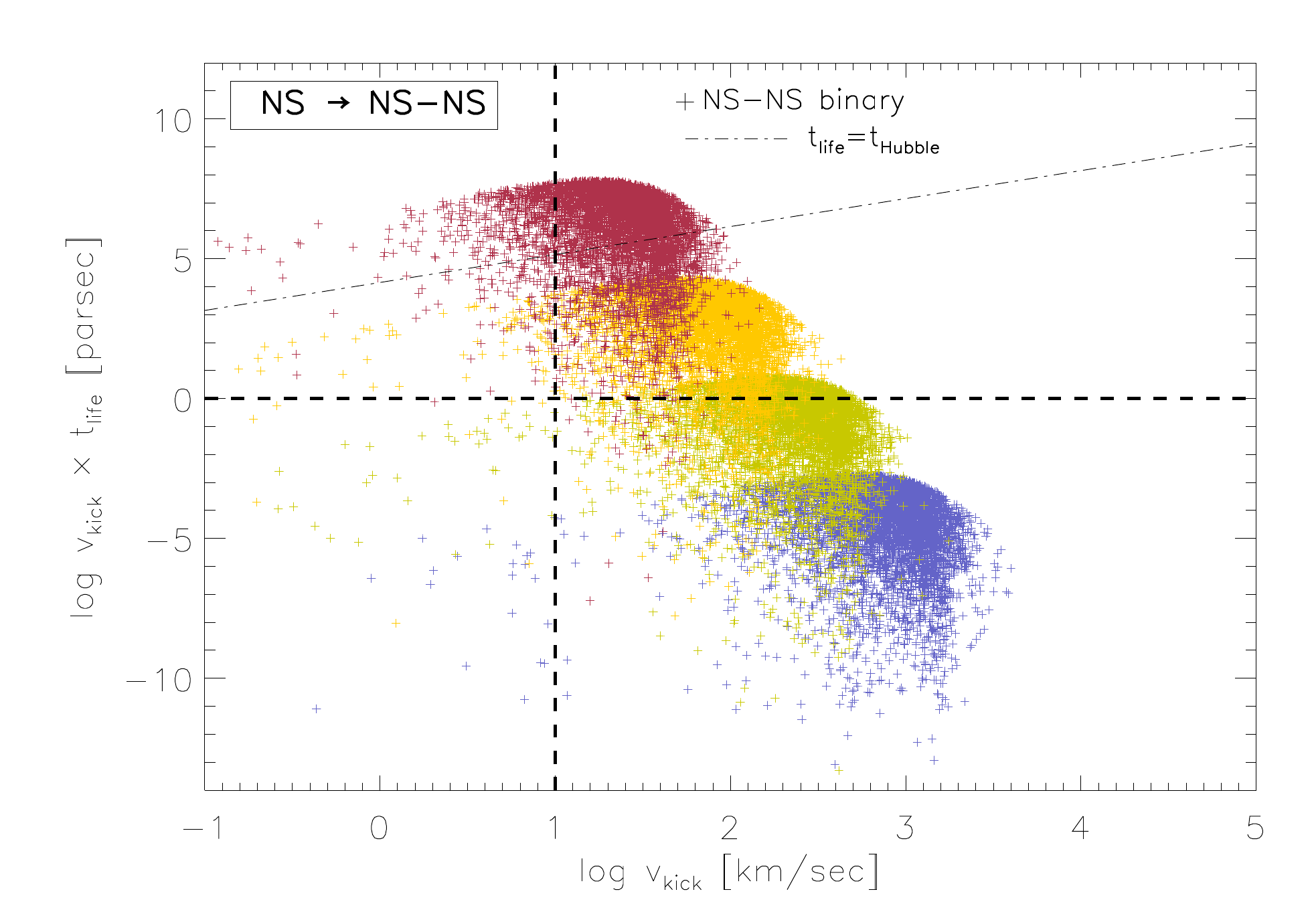}
\caption{Resulting kick velocities $v_{\rm kick}$ and travelled distance before merger defined as $v_{\rm kick} \times t_{\rm life}$
for endstate NS-NS binaries.
The kick velocity is with respect to the initial center of mass of the interaction.
In all scatterings, the encounter comes from infinity with $v_{\infty}=10\ {\rm km\;s^{-1}}$.
\textit{Top:} Results from the scattering ${\rm NS} \rightarrow {\rm NS-WD}~(0.5\ M_{\odot})$.
\textit{Bottom:} Results from the scattering ${\rm NS} \rightarrow {\rm NS-NS}$.
The {\it dashed} lines show characteristic values for a typical globular cluster. In this simple picture
all binaries in quadrants II-IV will merge within the cluster while binaries in the upper right corner will
merge outside. The corresponding single object will be kicked in the opposite direction with a
fraction $m_{\rm bin}/m_{\rm sin}$ of the binary's kick velocity.
The {\it dash-dot} line shows where the binary lifetime is equal to the Hubble time. All binaries below the line will have a lifetime less than a Hubble time. 
Different colors indicate different initial SMA.} 
\label{fig:kicks_vtlife}
\end{figure}

This tradeoff between lifetime and kick velocity is
evident in Figure \ref{fig:kicks_vtlife}. The Figure shows a scatter plot of kick velocity $v_{\rm kick}$ and survival distance,
defined as  $v_{\rm kick}\times t_{\rm life}$ for all endstate NS binaries with respect to the initial center of mass. We use the survival distance to estimate where the binary will merge. Radius and escape velocity for a typical globular cluster are shown with dashed lines.
In this simple calculation only final binaries in the upper right quadrant merge outside the cluster.
If we now assume that binary SMAs are lognormally distributed and we only
consider binaries that merge in less than a Hubble time (below the {\it dash-dotted} line in Figure \ref{fig:kicks_vtlife}), we calculate
that $\sim$ 30\%  (10\%) of all merging binaries arising from NS-NS (NS-WD) targets are kicked out with a median distance of $\sim$ 80 (50) kpc.
While there is little direct evidence that close double neutron star
binaries can form and merge  in globular clusters, the double neutron star system PSR B2127+11C in the Galactic GC M15 \citep{1990Natur.346...42A}
is an example of such a system  and has $t_{\rm life} \approx 2 \times 10^8$ years.

The retention or ejection of binaries has implications for cluster dynamics and merger-induced transients such as e.g.  short gamma-ray bursts \citep{Belczynski:2006dq, Lee:2007em}.
 If binaries are retained, they participate in the continued cluster evolution acting as a heat source or sink depending on their SMA. In some cases the binary distribution may
 reach a steady-state \citep[e.g.][]{2005MNRAS.358..572I}. Merging binaries are expected to show environmental dependance in
 their electromagnetic signatures \citep{2001ApJ...561L.171P,Rosswog:2002fi, Metzger:2012iy, Kelley:2013fl, Rosswog:2013ea}.
   
If a relativistic (short-gamma-ray burst) or a mildly relativistic mass ejection resulted from the merger of two compact objects, the resulting afterglow could then, at least in part, be due to the interaction of the  ejecta with the stellar winds of the red giant cluster members \citep{2012ApJ...751...57D}. Due to the large stellar density in the cluster core, the external shock
would then take place within a more dense medium than the IGM  \citep{Lee:2010in}. In addition, the merger sites of compact binaries will  determine whether we expect the electromagnetic signatures of binary mergers to statistically trace the globular cluster distribution around galaxies \citep{Grindlay:2006ef, Lee:2010in, 2011MNRAS.413.2004C}
or the galactic potential \citep{1999MNRAS.305..763B, Rosswog:2003gj, Belczynski:2006dq, 2007ApJ...665.1220Z, 2009ApJ...705L.186Z, 2010ApJ...708....9F, 2010ApJ...725L..91K, 2013arXiv1307.0819F}.

\subsection{Rates}\label{sec:rates}
Given distributions of target binaries and single encounters, we can  convert the calculated cross sections into event rates.
In this section we present some simple order-of-magnitude  estimates of the rates of dynamical NS-NS inspirals achieved in globular cluster environments.
We denote the total number of NSs by $N_{\rm NS}$, and assume that some fraction $f_{\rm b}$ are in $N_{\rm bin}$ binary systems (target binaries). The remaining
fraction remains single (encounter population), $f_{\rm s} = 1-f_{\rm b}$.
The target binaries are distributed according to their SMA $dN_{\rm bin}/da$, which we assume is lognormal, $dN_{\rm bin}/da \propto a^{-1}$.
The differential rate of inspirals per SMA can then be written
\beq
\frac{d\Gamma_{\rm insp}}{da} = \frac{dN_{\rm bin}}{da}   n_{\rm s} \sigma_{\rm insp} v_{\infty},
\eeq
where $n_{\rm s}$ is the number density of single NSs, $n_{\rm s} = f_{\rm s} N_{\rm NS} / V_{\rm core}$, and $V_{\rm core}$ is the volume of the cluster core over which both single  and binary  objects are distributed.
To obtain the total rate of inspirals, we integrate over the binary distribution,
\beq
\Gamma_{\rm insp} = \int \frac{d\Gamma_{\rm insp}}{da} da.
\eeq
We note here that while we need to evaluate this integral for a given binary distribution and inspiral cross-section as a function of SMA, it
will generally scale as $\Gamma_{\rm insp} \propto N_{\rm NS}^2 f_{\rm b}(1-f_{\rm b}) v_{\infty}^{-1}$.
Below we provide some rate estimates based on simple examples that describe the distribution of NSs in globular clusters. 

In a typical globular cluster, there may be as many as $N_{\rm NS} \sim 10^3$,  for example, as modeled in the case of M15 by \citet{Murphy:2011ej} whose best fit model has $1500$ NSs with a half-mass radius of $0.17$ pc. In what  follows, we take $V_{\rm core} =(0.17\text{pc})^3$, a typical relative velocity $v_\infty =10 \text{km s}^{-1}$, and $N_{\rm NS} = 10^3$.  If 30\% of these NSs are in NS-NS binaries distributed between $10^{-3}$ and $1$ AU in SMA ($f_{\rm b} = 0.3$), the rate of NS inspirals will be
\beq
\Gamma_{\rm insp}^{\rm (NS-NS)} \approx 0.7 \text{ yr}^{-1} \text{ Gpc}^{-3}.
\eeq
To express the above rate in units of volume, we have assumed that the density of galaxies is $n_{\rm gal} = 0.1 \text{ Mpc}^{-3}$ and each galaxy has 100 globular clusters, $N_{\rm GC} = 10^2$, implying 10 GC/Mpc$^{-3}$  \citep{2006ARA&A..44..193B}. 

If we instead assume  NSs are in WD-NS binaries distributed between $10^{-3}$ and $1$ AU in SMA ($f_{\rm b} = 0.3$) and treat our WD-NS scattering cross section as representative for these binaries,  we find 
\beq
\Gamma_{\rm insp}^{\rm (WD-NS)} \approx 0.3 \text{ yr}^{-1} \text{ Gpc}^{-3}.
\eeq
As with the NS-NS case, this numeric result scales $\propto N_{\rm NS}^2 f_{\rm b}(1-f_{\rm b}) v_{\infty}^{-1}$.
This estimate should be treated as an upper limit, because, if, for example the NS is in a binary with a main sequence star, the effects of collisions will be more significant than those with a WD companion. 

These same assumptions imply  a rate of single-single NS captures in globular clusters,
\beq
\Gamma_{\rm cap} = f_{\rm s} N_{\rm NS} n_{\rm s} \sigma_{\rm cap} v_{\infty}  \approx 0.5 \text{ yr}^{-1} \text{ Gpc}^{-3}
\eeq
where we note that the velocity dependence in this case is $v_{\infty}^{-11/7}$.
By the same token, we can calculate  the rate of eccentric binaries in the {\it LIGO} band arising from single-single encounters 
\beq
\Gamma_{\rm SS}(f_{\rm GW}>10\text{\ Hz})  \approx 0.15 \text{ yr}^{-1} \text{ Gpc}^{-3},
\eeq
which has a velocity dependence $v_{\infty}^{-1}$.
Thus, if the binary fraction $f_{\rm b} > 0.18$ (for WD-NS binaries) or $f_{\rm b} > 0.08$ (for NS-NS binaries), the binary-single channel will dominate  the formation of eccentric NS inspirals over the widely discussed single-single channel.

We can also compare to the number of non-eccentric mergers which occur from dynamical interactions. These are defined in our scattering experiments as those binaries arising
from either an exchange or flyby interaction whose lifetime is less than a Hubble time, $t_{\rm life} < t_{\rm Hubble}$. If we take our NS-NS target binary simulations
as representative, non-eccentric merger outcomes have a rate of approximately
\beq
\Gamma_{\rm merge}^{\rm (NS-NS)}  \approx 120 \text{ yr}^{-1} \text{ Gpc}^{-3}.
\label{eq:rates}
\eeq
Binaries with $t_{\rm life} < t_{\rm Hubble}$ are thus more common by a factor of approximately $160$  than inspirals.  \citet{Grindlay:2006ef}, whose rate estimate is in rough agreement with  equation \eqref{eq:rates}, concludes that $\sim 10\%$ of all mergers may be dynamically assembled in globular clusters.  The remainder of mergers are expected to arise from binaries assembled in the field \citep[e.g.][]{Dominik:2012cw,Dominik:2013uj}. However, the exact fraction of mergers in clusters depends sensitively on the distribution of wide binaries containing compact objects which is difficult to constrain observationally.   If this estimate is correct, then the inspiral rate represents a $\sim 1$\%  fraction  of the anticipated total compact object merger rate assembled in cluster.\footnote{This estimate neglects other channels that could lead to eccentric binaries mergers, such
as Kozai resonance in a triple systems \citep{2002ApJ...576..894M, 2011ApJ...741...82T}.} 

Normalized to the rate of eccentric NS mergers from single-single capture for which $f_{\rm GW} > 10$ Hz, We can write a hierarchy of rates as
\beqar
\Gamma_{\rm SS}(f_{\rm GW}>10\text{\ Hz}) : \Gamma_{\rm insp}^{\rm (WD-NS)} : \Gamma_{\rm insp}^{\rm (NS-NS)} : \Gamma_{\rm merge}^{\rm (NS-NS)}  \nonumber  \\
\approx 1:2:5:800.
\eeqar

The expected number, and correspondingly the number density, of BHs in globular clusters remains uncertain. Mass-segregation, for example,  has been  argued to give rise to a BH-dominated subsystem that collapses and dynamically decouples from the remainder of the stellar system \citep{Spitzer:1969jx,Kulkarni:1993jl, 2006ApJ...637..937O}.
In this case, very high BH number densities can be achieved, leading to the formation of a binary population through GW capture. Binary-single and single-single BH interactions are expected to rapidly eject BHs from the cluster after the formation of binaries \citep{Kulkarni:1993jl,Sigurdsson:1993hv}. However, these binary interactions may also produce inspirals and mergers, perhaps even leading to the runaway formation of a massive black hole \citep{2004Natur.428..724P}.
Even if the number of BH binaries is small, the number density of single black holes may be high enough  to produce an inspiral
rate comparable to the NS inspiral rate \citep{2006ApJ...637..937O, 2007PhRvD..76f1504O}.  
However, it is probably unreasonable to expect that a fraction of order unity of globular clusters might  undergoing such an extreme phase at a given time. We therefore expect  NS-NS inspirals rather than BH-BH inspirals  to dominate the inspiral rate. 

\subsection{Significance of Eccentric Inspirals}\label{sec:significance}

We have demonstrated that  binary-single scatterings are likely to dominate the production of  eccentric  binaries. 
In such GW-driven inspirals, the energy change is much more rapid than the angular momentum change, such that the circularization time and inspiral time are similar, $t_{\rm insp} \approx t_{\rm circ}$ \citep{Peters:1964bc}. One consequence of this is that binaries whose peak frequency, equation \eqref{eq:grav_peak_fGW}, is at lower frequency than the {\it LIGO} band will enter the {\it LIGO} band with relatively low eccentricity since these objects tend to circularize as they inspiral. This can be seen most clearly in the trajectories drawn in Figures \ref{fig:BS_LIGO_ae_edist} and \ref{fig:WD05NS_ae_cross_secs}. For a binary to be seen as eccentric in a given waveband, it must have been formed with high eccentricity in that band. 
Eccentric inspirals produce gravitational waveforms which are distinct from those of
circularly inspiraling binaries \citep{Konigsdorffer:2006fj, 2011ApJ...737L...5S, East:2012hg, Gold:2012vt, East:2012es, Gold:2012jg, Huerta:2013wy}.
These may be so distinct that non-circular binaries will go undetected without uniquely created waveform templates \citep{2013PhRvD..87d3004E, Huerta:2013wy},
and the timing between pre-merger GW bursts will contain valuable information about the equation of state.
Close encounters in these systems  can also lead to tidal deformations strong enough to crack the crust of the NS and tap into the $\sim10^{46}$ erg stored in elastic energy,
potentially generating  flaring activity prior to the merger \citep{2012PhRvL.108a1102T, 2013arXiv1307.3554T}.
In contrast to quasi-circular NS-NS mergers, eccentric binary mergers can also result in massive disks even for equal mass binaries \citep{East:2012es}.

Neutron stars that merge with high eccentricity have potentially unique gravitational and post-merger electromagnetic signatures  \citep[e.g.][]{Lee:2010in, East:2012es}. The merger of these binaries may eject copious neutron rich material in tidal tails that will synthesize significantly larger masses of r-process rich material \citep{Lee:2010in, Rosswog:2013ea} 
than the widely discussed, non-eccentric binary mergers \citep{Lattimer:1974ho, Rosswog:1999wz, Rosswog:2003fu, Rosswog:2005dq, Lee:2007em, Metzger:2010hp, Roberts:2011fs, Bauswein:2013wd, Kasen:2013ub, Barnes:2013wg, Tanaka:2013uj, 2013arXiv1307.2943G}.

Multi-messenger astronomy offers tantalizing prospects for probing the nature of
compact objects, their binary assembly, evolution, and eventual merger \citep{Rosswog:2007vp, Rosswog:2007fv, 2009arXiv0902.1527B, Lee:2010in, Rosswog:2012um, Faber:2012tk, Metzger:2012iy,2012PhRvD..86j4035L, Kelley:2013fl, 2013ApJ...767..124N, 2013arXiv1301.7074P, 2013arXiv1307.7372P, 2013arXiv1306.3960B, 2013arXiv1306.4971T,2013CQGra..30l3001B}, in addition to possible
insights into the origin of r-process nucleosynthetic elements and short gamma-ray bursts \citep{Lattimer:1974ho,2002ApJ...577..893L, Rosswog:2003jh, Rosswog:2004vj, 2005ApJ...626L..41M, Roberts:2011fs, Bauswein:2013wd}.  An eccentric GW signal  detection might be one of the most
exciting prospects, as it would provide  a clear signature of the dynamical binary assembly process. In the explicit absence of such  detection, the use of  eccentric waveform template
searches could help exclude a significant dynamically assembled population of  merging compact binaries  in dense stellar systems. 

\acknowledgments{
It is a pleasure to thank J. Goldstein, J. Guillochon, S. H. Hansen, J. Hjorth, D. Kasen, L. Kelley,  W. Lee, L. Lehner,  I.  Mandel,  C. Miller, F. Pretorius, S. Rosswog, and D. Tsang
for helpful discussions. 
M.M. and E.R-R. thank   the DARK cosmology centre for its hospitality.
We acknowledge support from the David and Lucile Packard Foundation, NSF grant: AST-0847563 and  the NSF Graduate Research Fellowship (M.M.).
The Dark Cosmology Centre is funded by the Danish National Research Foundation.}

\begin{appendix}

\section{N-body Integrator with GW energy loss correction}
\begin{figure}[tbp]
\centering
\includegraphics[width=0.65\columnwidth]{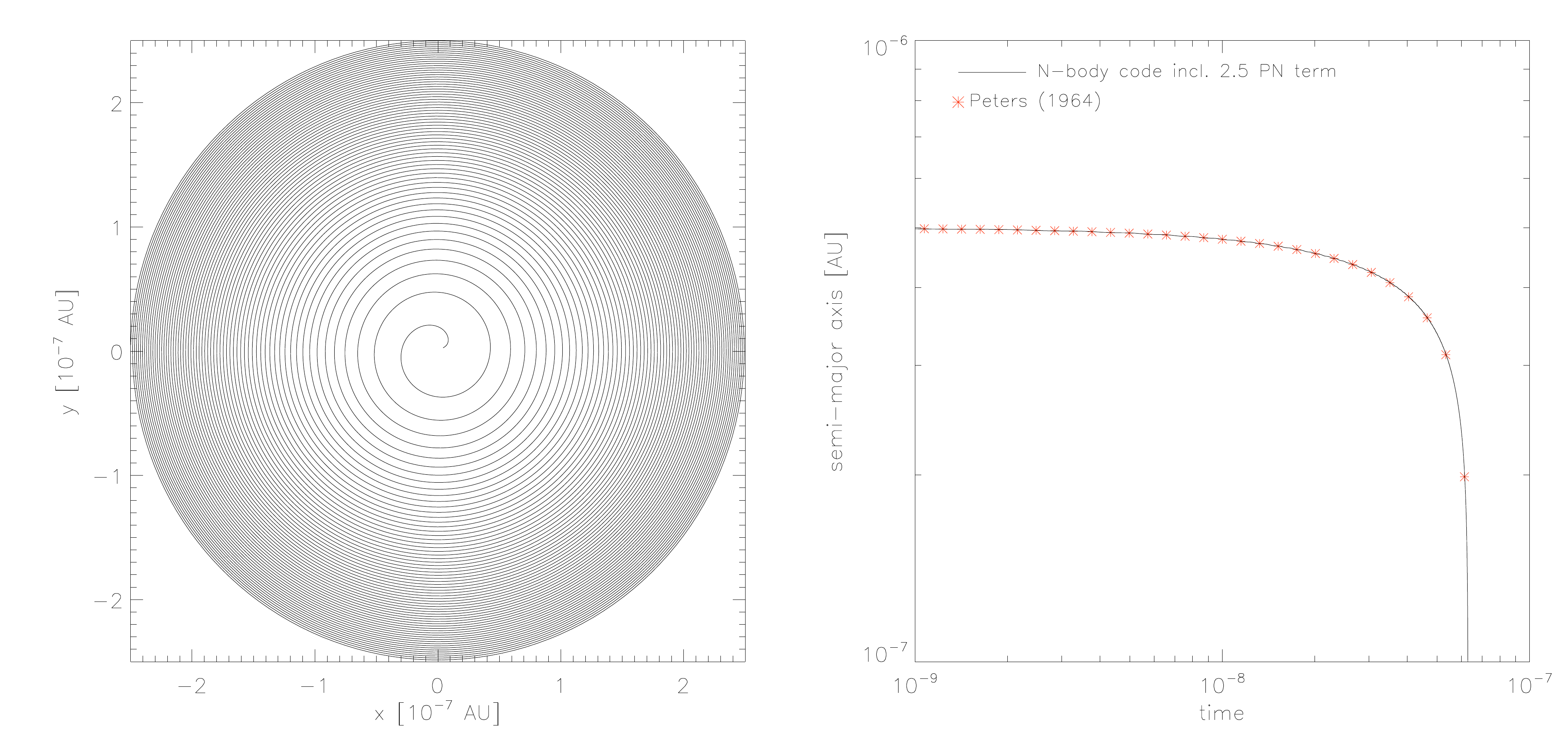}
\includegraphics[width=0.65\columnwidth]{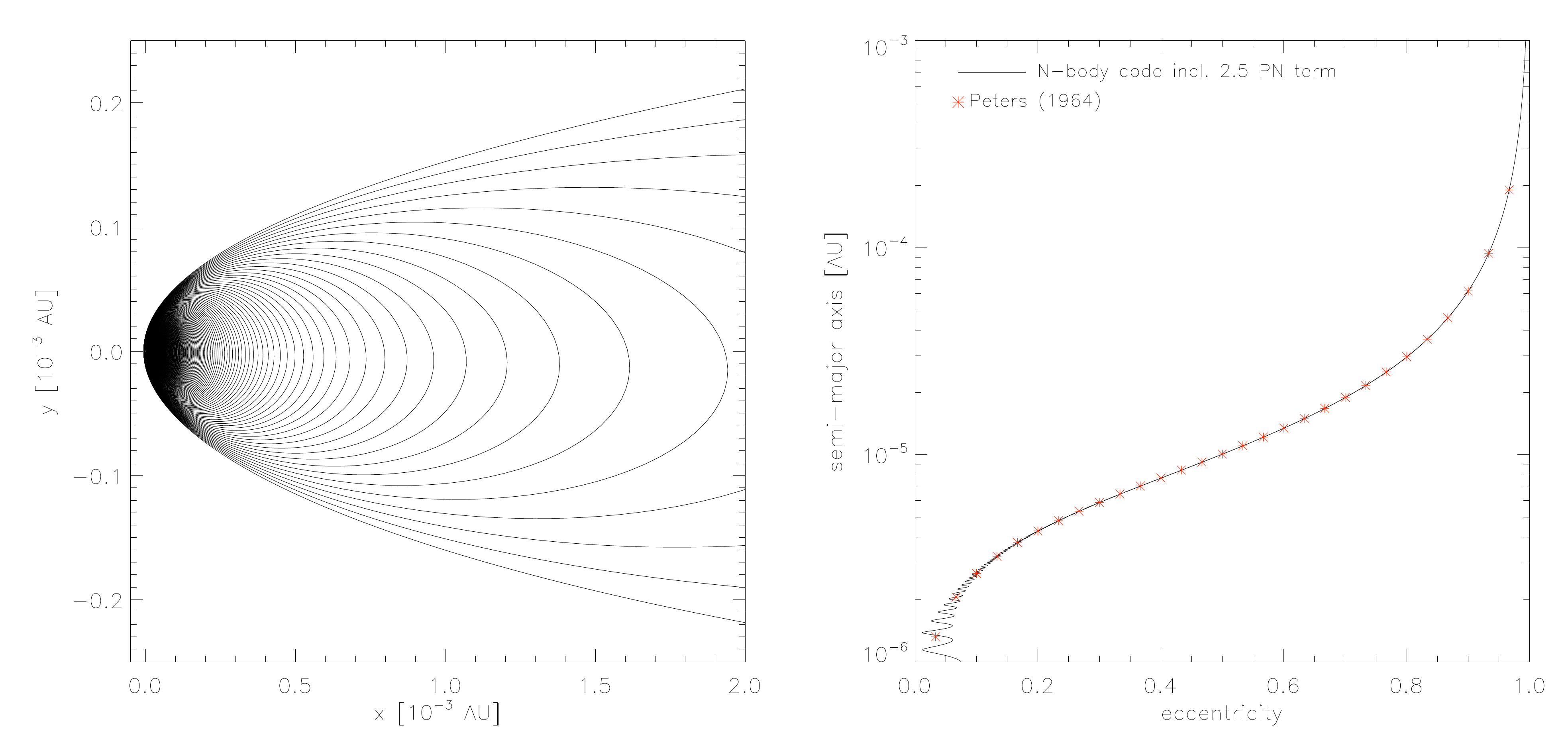}
\caption{Comparison between our N-body code ({\it solid-black lines}) and the analytical solution from \cite{Peters:1964bc} ({\it  red points}).
\textit{Top:} A circular binary that spirals in due to GW radiation. The upper left plot shows the trajectory of one of the objects. The upper right
plot shows how the distance between the two objects decreases with time.
\textit{\it Bottom:} Evolution of an initial highly eccentric binary. The two objects are initially not bound to each other, 
but enough energy is radiated away in terms of GW to make the system bound  after the first orbit. This is an
illustration of a single-single capture. The lower left plot shows the evolution of the incoming single in the rest-frame of the
target object. To the right is shown the evolution in the ($a,e$) plane. The wiggles in the lower left corner (for low $a$ and $e$) illustrate 
the limitation of the integration scheme.
As seen, we find good agreement between our code and the analytical
prediction in both cases.}
\label{fig:GWradtest1}
\end{figure}

We use a Fourth-Order Hermite Integrator with a variable time step to evolve the
N-body system. 
The dynamical effect from GW radiation is included using the Post-Newtonian (PN) formalism \citep{Blanchet:2006kp}
by modifying the Newtonian acceleration term from ${\bf a}_{0}$ to $ {\bf a}_{0}+c^{-5}{\bf a}_{5}$ as described in Section \ref{sec:addingGR}.
This modified PN expansion of the acceleration is strictly valid only for two isolated objects.
However, one can still make use of this approach without introducing significant errors for $N>2$ objects
since the 2.5PN term has a much steeper dependence on the distance $r$ than the Newtonian
acceleration ($r^{-9/2}$ vs $r^{-2}$ for a circular binary). The contribution from the closest pair will therefore
always dominate. Further  justification for  this formalism   can be found  in \cite{2006ApJ...640..156G}.
The 2.5PN term is the first term in the expansion that acts like an energy sink, i.e. carries energy out of the system.
The energy loss from this term is, when orbit averaged, equivalent to the loss calculated
from the quadrupole formalism described in \cite{Peters:1964bc}.  A comparison between the two approaches
 is shown  in Figure \ref{fig:GWradtest1},  which plots the orbital evolution in the $(a,e)$ plane
for a binary that inspirals ({\it top panel}) because of GW radiation and for a single object that captures another single one by emitting GW ({\it bottom panel}).
The {\it black-solid} lines are from our N-body code where the {\it red} dots show the result from
solving for $(a,e)$, using the quadrupole formalism: equations \eqref{eq:peterssoldadt} and \eqref{eq:peterssoldade}. Very good agreement in these tests was found, as can be seen in Figure \ref{fig:GWradtest1}.

To speed up the  binary-single scattering experiments we have propagated  the encounter from infinity to a
distance $r_{\rm proj}$ from the center-of-mass (COM) of the target binary  by modeling
the binary-single system as a two-body system. The distance $r_{\rm proj}$
was chosen to be a fraction of the maximum value of either $r_{\rm bs}$ or $a_{\rm b}$, where $r_{\rm bs}$ is the minimum distance
between the COM of the binary and the interloper in the two-body frame
and $a_{\rm b}$ is the SMA of the binary. This approach ignores the effect from the
binary's dipole gravitational field on the encounter for $r>r_{\rm proj}$,
but the error is insignificant. Further details on the errors related to this strategy
can be  found in \cite{Hut:1983js}.

\section{Identifying States}

\subsection{Binary-single state}

Following \cite{Fregeau:2004fj} we state that the three interacting objects are in a binary-single state
if the binary objects are bound to each other
and the tidal force from the single at the binary's apocenter ($F_{\rm tid}$) is smaller than the relative force at apocenter ($F_{\rm rel}$) by some
fraction $\delta_{\rm tid}$, i.e.  if $F_{\rm tid}/F_{\rm rel}<\delta_{\rm tid}$.
The two force terms are simply given by
\begin{equation}
F_{\rm rel} = \frac{m_{\rm bin,1}m_{\rm bin,2}}{[a(1+e)]^2}
\end{equation}
and
\begin{equation}
F_{\rm tid} \simeq \frac{2(m_{\rm bin,1}+m_{\rm bin,2})m_{\rm s}}{r^3}a(1+e),
\end{equation}
where $m_{{\rm bin},i}$ is the mass of binary object $i$, $m_{\rm s}$ the mass of the single object,
$r$ the distance between the single object and the center-of-mass
of the binary and $a,e$ are the semi-major axis and eccentricity of the binary, respectively.

If a three-body state is identified as a binary-single state and the single object is unbound from the binary, the state
is labeled either as an {\it exchange} or a {\it fly-by}  depending on which objects the binary is composed of.
If the single object is instead bound to the binary, the state is denoted as an {\it intermediate binary-single state} (IMS).
In this case,  the bound single is chosen to have a finite minimum distance to the binary. The chosen threshold, $\delta_{\rm tid}$,
will thus have an influence on the identified number of IMS and the corresponding distribution in $(a,e)$.
There is no dependence on $\delta_{\rm tid}$ if the single is unbound. For this work we use a $\delta_{\rm tid}=0.5$
for identifying IMS and $\delta_{\rm tid}=0.1$ for identifying exchange or a fly-by.

\subsection{Inspirals}

A binary with a bound single companion that inspirals due to GW radiation is denoted an {\it inspiral}.
Since the binaries that inspirals have a bound companion, the inspiral state is a subclass of the IMS discussed above.  In these cases, the $(a,e)$ values for  the orbital parameters of the inspiraling binary are set at initial identification,  when the three-body state is identified as an IMS. The value for this first set of $(a,e)$ depends strongly on the threshold $\delta_{\rm tid}$ since a smaller $\delta_{\rm tid}$  allows more time for the binary to spiral in. However, the total number of inspirals is not affected, and therefore the resulting  cross sections   are also not sensitive to the choice of $\delta_{\rm tid}$.

\subsection{Collisions} 

We assume in all scattering experiments that the objects are rigid spheres with radius $r_{i}$.
We say that object $i$ and $j$ have collided if these spheres ever overlap, $r_{ij}<r_{i}+{r_{j}}$. To distinguish collisions from inspirals we say that collisions are colliding objects that are not in an IMS binary.
This definition is practical, but there is some gray-zone between collisions and inspirals. One can for example have an IMS binary with initial pericenter distance $r_{\rm min}<r_{i}+{r_{j}}$, or a configuration where enough GW energy is radiated away
such that two objects collide before an IMS is identified by the code. In general, this overlap is only important at the very smallest binary SMAs, in which the SMA begins to become comparable to the size of the objects,  of order $10^{-5}$ AU for solar mass compact objects. At larger separations, any sensitivity is lost because the number of inspirals greatly dominates  over the  number of direct impacts.

\end{appendix}

\bibliographystyle{apj}

\begin{thebibliography}{0}
\expandafter\ifx\csname natexlab\endcsname\relax\def\natexlab#1{#1}\fi

\end{thebibliography}


\begin{thebibliography}
\expandafter\ifx\csname natexlab\endcsname\relax\def\natexlab#1{#1}\fi

\bibitem[{Aarseth \& Lecar(1975)}]{Aarseth:1975kf}
Aarseth, S.~J., \& Lecar, M. 1975, \araa, 13, 1

\bibitem[{Abadie {et~al.}(2010)Abadie, Abbott, Abbott, Abernathy, Accadia,
  Acernese, Adams, Adhikari, Ajith, Allen, Allen, Ceron, Amin, Anderson,
  Anderson, Antonucci, Aoudia, Arain, Araya, Aronsson, Arun, Aso, Aston,
  Astone, Atkinson, Aufmuth, Aulbert, Babak, Baker, Ballardin, Ballmer, Barker,
  Barnum, Barone, Barr, Barriga, Barsotti, Barsuglia, Barton, Bartos, Bassiri,
  Bastarrika, Bauchrowitz, Bauer, Behnke, Beker, Benacquista, Bertolini,
  Betzwieser, Beveridge, Beyersdorf, Bigotta, Bilenko, Billingsley, Birch,
  Birindelli, Biswas, Bitossi, Bizouard, Black, Blackburn, Blackburn, Blair,
  Bland, Blom, Blomberg, Boccara, Bock, Bodiya, Bondarescu, Bondu, Bonelli,
  Bork, Born, Bose, Bosi, Boyle, Braccini, Bradaschia, Brady, Braginsky, Brau,
  Breyer, Bridges, Brillet, Brinkmann, Brisson, Britzger, Brooks, Brown,
  Budzy{\'n}ski, Bulik, Bulten, Buonanno, Burguet-Castell, Burmeister,
  Buskulic, Byer, Cadonati, Cagnoli, Calloni, Camp, Campagna, Campsie,
  Cannizzo, Cannon, Canuel, Cao, Capano, Carbognani, Caride, Caudill,
  Cavagli{\`a}, Cavalier, Cavalieri, Cella, Cepeda, Cesarini, Chalermsongsak,
  Chalkley, Charlton, Chassande~Mottin, Chelkowski, Chen, Chincarini,
  Christensen, Chua, Chung, Clark, Clark, Clayton, Cleva, Coccia, Colacino,
  Colas, Colla, Colombini, Conte, Cook, Corbitt, Corda, Cornish, Corsi, Costa,
  Coulon, Coward, Coyne, Creighton, Creighton, Cruise, Culter, Cumming,
  Cunningham, Cuoco, Dahl, Danilishin, Dannenberg, D'Antonio, Danzmann, Dari,
  Das, Dattilo, Daudert, Davier, Davies, Davis, Daw, Day, Dayanga, De~Rosa,
  DeBra, Degallaix, del Prete, Dergachev, DeRosa, DeSalvo, Devanka, Dhurandhar,
  Fiore, Lieto, Palma, Emilio, Virgilio, D{\'\i}az, Dietz, Donovan, Dooley,
  Doomes, Dorsher, Douglas, Drago, Drever, Driggers, Dueck, Dumas, Eberle,
  Edgar, Edwards, Effler, Ehrens, Engel, Etzel, Evans, Evans, Fafone,
  Fairhurst, Fan, Farr, Fazi, Fehrmann, Feldbaum, Ferrante, Fidecaro, Finn,
  Fiori, Flaminio, Flanigan, Flasch, Foley, Forrest, Forsi, Fotopoulos,
  Fournier, Franc, Frasca, Frasconi, Frede, Frei, Frei, Freise, Frey, Fricke,
  Friedrich, Fritschel, Frolov, Fulda, Fyffe, Gammaitoni, Garofoli, Garufi,
  Gemme, Genin, Gennai, Gholami, Ghosh, Giaime, Giampanis, Giardina, Giazotto,
  Gill, Goetz, Goggin, Gonz{\'a}lez, Gorodetsky, Go{\ss}ler, Gouaty, Graef,
  Granata, Grant, Gras, Gray, Greenhalgh, Gretarsson, Greverie, Grosso, Grote,
  Grunewald, Guidi, Gustafson, Gustafson, Hage, Hall, Hallam, Hammer, Hammond,
  Hanks, Hanna, Hanson, Harms, Harry, Harry, Harstad, Haughian, Hayama,
  Heefner, Heitmann, Hello, Heng, Heptonstall, Hewitson, Hild, Hirose, Hoak,
  Hodge, Holt, Hosken, Hough, Howell, Hoyland, Huet, Hughey, Husa, Huttner,
  Huynh-Dinh, Ingram, Inta, Isogai, Ivanov, Jaranowski, Johnson, Jones, Jones,
  \& Jones}]{Abadie:2010fn}
Abadie, J., {et~al.} 2010, Class. Quantum Grav., 27, 173001

\bibitem[{{Anderson} {et~al.}(1990){Anderson}, {Gorham}, {Kulkarni}, {Prince},
  \& {Wolszczan}}]{1990Natur.346...42A}
{Anderson}, S.~B., {Gorham}, P.~W., {Kulkarni}, S.~R., {Prince}, T.~A., \&
  {Wolszczan}, A. 1990, \nat, 346, 42

\bibitem[{Barnes \& Kasen(2013)}]{Barnes:2013wg}
Barnes, J., \& Kasen, D. 2013, eprint arXiv:1303.5787

\bibitem[Bartos et al.(2013)]{2013CQGra..30l3001B} Bartos, I., Brady, P., 
\& M{\'a}rka, S.\ 2013, Classical and Quantum Gravity, 30, 123001 

\bibitem[{Baumgardt {et~al.}(2002)Baumgardt, Hut, \& Heggie}]{Baumgardt:2002eb}
Baumgardt, H., Hut, P., \& Heggie, D.~C. 2002, \mnras, 336, 1069

\bibitem[{Bauswein {et~al.}(2013)Bauswein, Goriely, \& Janka}]{Bauswein:2013wd}
Bauswein, A., Goriely, S., \& Janka, H.~T. 2013, eprint arXiv:1302.6530

\bibitem[{Belczynski {et~al.}(2006)Belczynski, Perna, Bulik, Kalogera, Ivanova,
  \& Lamb}]{Belczynski:2006dq}
Belczynski, K., Perna, R., Bulik, T., Kalogera, V., Ivanova, N., \& Lamb, D.~Q.
  2006, \apj, 648, 1110

\bibitem[{{Berger} {et~al.}(2013){Berger}, {Fong}, \&
  {Chornock}}]{2013arXiv1306.3960B}
{Berger}, E., {Fong}, W., \& {Chornock}, R. 2013, ArXiv e-prints

\bibitem[Blanchet(2006)]{Blanchet:2006kp} Blanchet, L.\ 2006, Living 
Reviews in Relativity, 9, 4 

\bibitem[{{Bloom} {et~al.}(1999){Bloom}, {Sigurdsson}, \&
  {Pols}}]{1999MNRAS.305..763B}
{Bloom}, J.~S., {Sigurdsson}, S., \& {Pols}, O.~R. 1999, \mnras, 305, 763

\bibitem[Bloom et al.(2009)]{2009arXiv0902.1527B} Bloom, J.~S., Holz, 
D.~E., Hughes, S.~A., et al.\ 2009, arXiv:0902.1527 

\bibitem[Brodie 
\& Strader(2006)]{2006ARA&A..44..193B} Brodie, J.~P., \& Strader, J.\ 2006, \araa, 44, 193 

\bibitem[{{Brown} {et~al.}(2013){Brown}, {Kumar}, \&
  {Nitz}}]{2013PhRvD..87h2004B}
{Brown}, D.~A., {Kumar}, P., \& {Nitz}, A.~H. 2013, \prd, 87, 082004

\bibitem[{Carroll \& Ostlie(1996)}]{Carroll:1996ut}
Carroll, B.~W., \& Ostlie, D.~A. 1996, {An Introduction to Modern Astrophysics}

\bibitem[{{Church} {et~al.}(2011){Church}, {Levan}, {Davies}, \&
  {Tanvir}}]{2011MNRAS.413.2004C}
{Church}, R.~P., {Levan}, A.~J., {Davies}, M.~B., \& {Tanvir}, N. 2011, \mnras,
  413, 2004

\bibitem[{Clausen {et~al.}(2012)Clausen, Sigurdsson, \&
  Chernoff}]{Clausen:2012hn}
Clausen, D., Sigurdsson, S., \& Chernoff, D.~F. 2012, \mnras, 428, 3618

\bibitem[{De~Colle {et~al.}(2012)De~Colle, Ramirez-Ruiz, Granot, \&
  Lopez-Camara}]{2012ApJ...751...57D}
De~Colle, F., Ramirez-Ruiz, E., Granot, J., \& Lopez-Camara, D. 2012, \apj,
  751, 57

\bibitem[{Dominik {et~al.}(2012)Dominik, Belczynski, Fryer, Holz, Berti, Bulik,
  Mandel, \& O'Shaughnessy}]{Dominik:2012cw}
Dominik, M., Belczynski, K., Fryer, C., Holz, D.~E., Berti, E., Bulik, T.,
  Mandel, I., \& O'Shaughnessy, R. 2012, \apj, 759, 52

\bibitem[{Dominik {et~al.}(2013)Dominik, Belczynski, Fryer, Holz, Berti, Bulik,
  Mandel, \& O'Shaughnessy}]{Dominik:2013uj}
---. 2013, arXiv:1308.1546

\bibitem[{East {et~al.}(2012)East, Pretorius, \&
  Stephens}]{East:2012hg}
East, W., Pretorius, F., \& Stephens, B. 2012, Phys. Rev. D, 85,
  124009

\bibitem[{{East} {et~al.}(2013){East}, {McWilliams}, {Levin}, \&
  {Pretorius}}]{2013PhRvD..87d3004E}
{East}, W.~E., {McWilliams}, S.~T., {Levin}, J., \& {Pretorius}, F. 2013, \prd,
  87, 043004

\bibitem[{East \& Pretorius(2012)}]{East:2012es}
East, W.~E., \& Pretorius, F. 2012, \apj, 760, L4

\bibitem[{Eggleton(1983)}]{Eggleton:1983fq}
Eggleton, P.~P. 1983, \apj, 268, 368

\bibitem[{Faber \& Rasio(2012)}]{Faber:2012tk}
Faber, J.~A., \& Rasio, F.~A. 2012, Living Reviews in Relativity, 15, 8

\bibitem[{{Fong} {et~al.}(2010){Fong}, {Berger}, \&
  {Fox}}]{2010ApJ...708....9F}
{Fong}, W., {Berger}, E., \& {Fox}, D.~B. 2010, \apj, 708, 9

\bibitem[Fong 
\& Berger(2013)]{2013arXiv1307.0819F} Fong, W.-f., \& Berger, E.\ 2013, arXiv:1307.0819 

\bibitem[{Fregeau(2008)}]{2008ApJ...673L..25F}
Fregeau, J.~M. 2008, \apj, 673, L25

\bibitem[{Fregeau {et~al.}(2004)Fregeau, Cheung, Portegies~Zwart, \&
  Rasio}]{Fregeau:2004fj}
Fregeau, J.~M., Cheung, P., Portegies~Zwart, S.~F., \& Rasio, F.~A. 2004,
  \mnras, 352, 1

\bibitem[{Fregeau {et~al.}(2003)Fregeau, G{\"u}rkan, Joshi, \&
  Rasio}]{2003ApJ...593..772F}
Fregeau, J.~M., G{\"u}rkan, M.~A., Joshi, K.~J., \& Rasio, F.~A. 2003, \apj,
  593, 772

\bibitem[{Fregeau {et~al.}(2009)Fregeau, Ivanova, \&
  Rasio}]{2009ApJ...707.1533F}
Fregeau, J.~M., Ivanova, N., \& Rasio, F.~A. 2009, \apj, 707, 1533

\bibitem[{Fregeau {et~al.}(2002)Fregeau, Joshi, Portegies~Zwart, \&
  Rasio}]{2002ApJ...570..171F}
Fregeau, J.~M., Joshi, K.~J., Portegies~Zwart, S.~F., \& Rasio, F.~A. 2002,
  \apj, 570, 171

\bibitem[{Fregeau \& Rasio(2007)}]{2007ApJ...658.1047F}
Fregeau, J.~M., \& Rasio, F.~A. 2007, \apj, 658, 1047

\bibitem[{Gold {et~al.}(2012)Gold, Bernuzzi, Thierfelder, Br{\"u}gmann, \&
  Pretorius}]{Gold:2012jg}
Gold, R., Bernuzzi, S., Thierfelder, M., Br{\"u}gmann, B., \& Pretorius, F.
  2012, Phys. Rev. D, 86, 121501

\bibitem[{Gold \& Bruegmann(2012)}]{Gold:2012vt}
Gold, R., \& Bruegmann, B. 2012, eprint arXiv:1209.4085

\bibitem[{Goodman \& Hut(1993)}]{Goodman:1993gg}
Goodman, J., \& Hut, P. 1993, \apj, 403, 271

\bibitem[{Grindlay {et~al.}(2006)Grindlay, Zwart, \&
  McMillan}]{Grindlay:2006ef}
Grindlay, J., Zwart, S.~P., \& McMillan, S. 2006, Nat Phys, 2, 116

\bibitem[Grossman et al.(2013)]{2013arXiv1307.2943G} Grossman, D., 
Korobkin, O., Rosswog, S., \& Piran, T.\ 2013, arXiv:1307.2943

\bibitem[{G{\"u}ltekin {et~al.}(2004)G{\"u}ltekin, Miller, \&
  Hamilton}]{2004ApJ...616..221G}
G{\"u}ltekin, K., Miller, M.~C., \& Hamilton, D.~P. 2004, \apj, 616, 221

\bibitem[{G{\"u}ltekin {et~al.}(2006)G{\"u}ltekin, Miller, \&
  Hamilton}]{2006ApJ...640..156G}
---. 2006, \apj, 640, 156

\bibitem[{Hansen(1972)}]{Hansen:1972il}
Hansen, R. 1972, Phys. Rev. D, 5, 1021

\bibitem[{Harry \& {the LIGO Scientific Collaboration}(2010)}]{Harry:2010hh}
Harry, G.~M., \& {the LIGO Scientific Collaboration}. 2010, Class. Quantum
  Grav., 27, 084006

\bibitem[{Heggie(1975)}]{Heggie:1975uy}
Heggie, D.~C. 1975, \mnras, 173, 729

\bibitem[{Heggie \& Hut(1993)}]{Heggie:1993hi}
Heggie, D.~C., \& Hut, P. 1993, \apjs, 85, 347

\bibitem[{Heggie {et~al.}(1996{\natexlab{a}})Heggie, Hut, \&
  McMillan}]{Heggie:1996fza}
Heggie, D.~C., Hut, P., \& McMillan, S. L.~W. 1996{\natexlab{a}}, \apj, 467,
  359

\bibitem[{Heggie {et~al.}(1996{\natexlab{b}})Heggie, Hut, \&
  McMillan}]{Heggie:1996fz}
---. 1996{\natexlab{b}}, \apj, 467, 359

\bibitem[{Hills(1975{\natexlab{a}})}]{Hills:1975io}
Hills, J.~G. 1975{\natexlab{a}}, \aj, 80, 1075

\bibitem[{Hills(1975{\natexlab{b}})}]{Hills:1975jk}
---. 1975{\natexlab{b}}, \aj, 80, 809

\bibitem[{Hills(1976)}]{Hills:1976vi}
---. 1976, \mnras, 175, 1P

\bibitem[{Hills \& Fullerton(1980)}]{Hills:1980bm}
Hills, J.~G., \& Fullerton, L.~W. 1980, \aj, 85, 1281

\bibitem[{Hopman {et~al.}(2006)Hopman, Guetta, Waxman, \&
  Portegies~Zwart}]{Hopman:2006ha}
Hopman, C., Guetta, D., Waxman, E., \& Portegies~Zwart, S. 2006, \apj, 643, L91

\bibitem[{Huerta \& Brown(2013)}]{Huerta:2013wy}
Huerta, E.~A., \& Brown, D.~A. 2013, eprint arXiv:1301.1895

\bibitem[{Hut(1983)}]{Hut:1983by}
Hut, P. 1983, \apj, 268, 342

\bibitem[{Hut(1993)}]{Hut:1993gs}
---. 1993, \apj, 403, 256

\bibitem[{Hut \& Bahcall(1983)}]{Hut:1983js}
Hut, P., \& Bahcall, J.~N. 1983, \apj, 268, 319

\bibitem[{Hut {et~al.}(1992)Hut, McMillan, Goodman, Mateo, Phinney, Pryor,
  Richer, Verbunt, \& Weinberg}]{1992PASP..104..981H}
Hut, P., {et~al.} 1992, Astronomical Society of the Pacific, 104, 981

\bibitem[Ivanova et al.(2003)]{2003astro.ph.12497I} Ivanova, N., 
Belczynski, K., Fregeau, J.~M., 
\& Rasio, F.~A.\ 2003, arXiv:astro-ph/0312497 

\bibitem[{Ivanova {et~al.}(2005{\natexlab{a}})Ivanova, Belczynski, Fregeau, \&
  Rasio}]{2005MNRAS.358..572I}
---. 2005{\natexlab{a}}, \mnras, 358, 572

\bibitem[{Ivanova {et~al.}(2010)Ivanova, Chaichenets, Fregeau, Heinke,
  Lombardi, \& Woods}]{2010ApJ...717..948I}
Ivanova, N., Chaichenets, S., Fregeau, J., Heinke, C.~O., Lombardi, J. C.~J.,
  \& Woods, T.~E. 2010, \apj, 717, 948

\bibitem[{Ivanova {et~al.}(2005{\natexlab{b}})Ivanova, Fregeau, \&
  Rasio}]{2005ASPC..328..231I}
Ivanova, N., Fregeau, J.~M., \& Rasio, F.~A. 2005{\natexlab{b}}, Binary Radio
  Pulsars, 328, 231

\bibitem[{Ivanova {et~al.}(2008)Ivanova, Heinke, Rasio, Belczynski, \&
  Fregeau}]{Ivanova:2008jx}
Ivanova, N., Heinke, C.~O., Rasio, F.~A., Belczynski, K., \& Fregeau, J.~M.
  2008, \mnras, 386, 553

\bibitem[{Ivanova {et~al.}(2006)Ivanova, Heinke, Rasio, Taam, Belczynski, \&
  Fregeau}]{2006MNRAS.372.1043I}
Ivanova, N., Heinke, C.~O., Rasio, F.~A., Taam, R.~E., Belczynski, K., \&
  Fregeau, J. 2006, \mnras, 372, 1043

\bibitem[{Kasen {et~al.}(2013)Kasen, Badnell, \& Barnes}]{Kasen:2013ub}
Kasen, D., Badnell, N.~R., \& Barnes, J. 2013, eprint arXiv:1303.5788

\bibitem[{Kelley {et~al.}(2013)Kelley, Mandel, \& Ramirez-Ruiz}]{Kelley:2013fl}
Kelley, L.~Z., Mandel, I., \& Ramirez-Ruiz, E. 2013, Phys. Rev. D, 87, 123004

\bibitem[{{Kelley} {et~al.}(2010){Kelley}, {Ramirez-Ruiz}, {Zemp}, {Diemand},
  \& {Mandel}}]{2010ApJ...725L..91K}
{Kelley}, L.~Z., {Ramirez-Ruiz}, E., {Zemp}, M., {Diemand}, J., \& {Mandel}, I.
  2010, \apjl, 725, L91

\bibitem[{Kelly \& Shen(2013)}]{Kelly:2013jz}
Kelly, B.~C., \& Shen, Y. 2013, \apj, 764, 45

\bibitem[{Kocsis \& Levin(2012)}]{Kocsis:2012ja}
Kocsis, B., \& Levin, J. 2012, Phys. Rev. D, 85, 123005

\bibitem[{K{\"o}nigsd{\"o}rffer \& Gopakumar(2006)}]{Konigsdorffer:2006fj}
K{\"o}nigsd{\"o}rffer, C., \& Gopakumar, A. 2006, Phys. Rev. D, 73, 124012

\bibitem[{Kulkarni {et~al.}(1993)Kulkarni, Hut, \& McMillan}]{Kulkarni:1993jl}
Kulkarni, S.~R., Hut, P., \& McMillan, S.~J. 1993, \nat, 364, 421

\bibitem[{Lattimer \& Schramm(1974)}]{Lattimer:1974ho}
Lattimer, J.~M., \& Schramm, D.~N. 1974, \apj, 192, L145

\bibitem[{Lee(1993)}]{Lee:1993dt}
Lee, M.~H. 1993, \apj, 418, 147

\bibitem[Lee 
\& Ramirez-Ruiz(2002)]{2002ApJ...577..893L} Lee, W.~H., \& Ramirez-Ruiz, E.\ 2002, \apj, 577, 893 

\bibitem[{Lee \& Ramirez-Ruiz(2007)}]{Lee:2007em}
Lee, W.~H., \& Ramirez-Ruiz, E. 2007, New J. Phys., 9, 17

\bibitem[{Lee {et~al.}(2010)Lee, Ramirez-Ruiz, \& van~de
  Ven}]{Lee:2010in}
Lee, W.~H., Ramirez-Ruiz, E., \& van~de Ven, G. 2010{\natexlab{a}}, \apj, 720,
  953

\bibitem[Lehner et al.(2012)]{2012PhRvD..86j4035L} Lehner, L., Palenzuela, 
C., Liebling, S.~L., Thompson, C., \& Hanna, C.\ 2012, \prd, 86, 104035 

\bibitem[{Lightman \& Shapiro(1978)}]{Lightman:1978go}
Lightman, A., \& Shapiro, S. 1978, Rev. Mod. Phys., 50, 437

\bibitem[LIGO Scientific Collaboration et al.(2013)]{2013arXiv1304.0670L} 
LIGO Scientific Collaboration, Virgo Collaboration, Aasi, J., et al.\ 2013, 
arXiv:1304.0670 

\bibitem[{Mandel \& O'Shaughnessy(2010)}]{Mandel:2010ke}
Mandel, I., \& O'Shaughnessy, R. 2010, Class. Quantum Grav., 27, 114007

\bibitem[{Marsh {et~al.}(2004)Marsh, Nelemans, \& Steeghs}]{Marsh:2004fv}
Marsh, T.~R., Nelemans, G., \& Steeghs, D. 2004, \mnras, 350, 113

\bibitem[{McMillan(1986)}]{1986ApJ...306..552M}
McMillan, S. L.~W. 1986, \apj, 306, 552

\bibitem[{McMillan(1991)}]{1991ASPC...13..324M}
---. 1991, In: The formation and evolution of star clusters (A93-48676 20-90),
  13, 324

\bibitem[{McMillan \& Hut(1996)}]{1996ApJ...467..348M}
McMillan, S. L.~W., \& Hut, P. 1996, Astrophysical Journal v.467, 467, 348

\bibitem[{Metzger \& Berger(2012)}]{Metzger:2012iy}
Metzger, B.~D., \& Berger, E. 2012, \apj, 746, 48

\bibitem[{Metzger {et~al.}(2010)Metzger, Mart{\'\i}nez-Pinedo, Darbha,
  Quataert, Arcones, Kasen, Thomas, Nugent, Panov, \& Zinner}]{Metzger:2010hp}
Metzger, B.~D., {et~al.} 2010, \mnras, 406, 2650

\bibitem[{{Meylan} \& {Heggie}(1997)}]{1997A&ARv...8....1M}
{Meylan}, G., \& {Heggie}, D.~C. 1997, \aapr, 8, 1

\bibitem[{Miller(2005)}]{2005ApJ...626L..41M}
Miller, M.~C. 2005, \apj, 626, L41

\bibitem[{Miller \& Hamilton(2002)}]{2002ApJ...576..894M}
Miller, M.~C., \& Hamilton, D.~P. 2002, \apj, 576, 894

\bibitem[{Murphy {et~al.}(2011)Murphy, Cohn, \& Lugger}]{Murphy:2011ej}
Murphy, B.~W., Cohn, H.~N., \& Lugger, P.~M. 2011, \apj, 732, 67

\bibitem[{{Nissanke} {et~al.}(2013){Nissanke}, {Kasliwal}, \&
  {Georgieva}}]{2013ApJ...767..124N}
{Nissanke}, S., {Kasliwal}, M., \& {Georgieva}, A. 2013, \apj, 767, 124

\bibitem[Nitz et al.(2013)]{2013arXiv1307.1757N} Nitz, A.~H., Lundgren, A., 
Brown, D.~A., et al.\ 2013, arXiv:1307.1757 

\bibitem[{{O'Leary} {et~al.}(2007){O'Leary}, {O'Shaughnessy}, \&
  {Rasio}}]{2007PhRvD..76f1504O}
{O'Leary}, R.~M., {O'Shaughnessy}, R., \& {Rasio}, F.~A. 2007, \prd, 76, 061504

\bibitem[{O'Leary {et~al.}(2006)O'Leary, Rasio, Fregeau, Ivanova, \&
  O'Shaughnessy}]{2006ApJ...637..937O}
O'Leary, R.~M., Rasio, F.~A., Fregeau, J.~M., Ivanova, N., \& O'Shaughnessy, R.
  2006, \apj, 637, 937

\bibitem[Palenzuela et al.(2013{\natexlab{a}})]{2013arXiv1301.7074P} Palenzuela, C., 
Lehner, L., Ponce, M., et al.\ 2013{\natexlab{a}}, arXiv:1301.7074 

\bibitem[Palenzuela et al.(2013{\natexlab{b}})]{2013arXiv1307.7372P} Palenzuela, C., 
Lehner, L., Liebling, S.~L., et al.\ 2013{\natexlab{b}}, arXiv:1307.7372 

\bibitem[Panaitescu et al.(2001)]{2001ApJ...561L.171P} Panaitescu, A., 
Kumar, P., \& Narayan, R.\ 2001, \apjl, 561, L171 

\bibitem[{Paschalidis {et~al.}(2009)Paschalidis, MacLeod, Baumgarte, \&
  Shapiro}]{Paschalidis:2009hk}
Paschalidis, V., MacLeod, M., Baumgarte, T.~W., \& Shapiro, S.~L. 2009, Phys.
  Rev. D, 80, 24006

\bibitem[{Peters(1964)}]{Peters:1964bc}
Peters, P. 1964, Phys. Rev., 136, B1224

\bibitem[{Phinney \& Sigurdsson(1991)}]{1991Natur.349..220P}
Phinney, E.~S., \& Sigurdsson, S. 1991, Nature (ISSN 0028-0836), 349, 220

\bibitem[{Portegies~Zwart {et~al.}(2004)Portegies~Zwart, Baumgardt, Hut,
  Makino, \& McMillan}]{2004Natur.428..724P}
Portegies~Zwart, S.~F., Baumgardt, H., Hut, P., Makino, J., \& McMillan, S.
  L.~W. 2004, \nat, 428, 724

\bibitem[{{Rasio} {et~al.}(2007){Rasio}, {Baumgardt}, {Corongiu}, {D'Antona},
  {Fabbiano}, {Fregeau}, {Gebhardt}, {Heinke}, {Hut}, {Ivanova}, {Maccarone},
  {Ransom}, \& {Webb}}]{2007HiA....14..215R}
{Rasio}, F.~A., {et~al.} 2007, Highlights of Astronomy, 14, 215

\bibitem[{Roberts {et~al.}(2011)Roberts, Kasen, Lee, \&
  Ramirez-Ruiz}]{Roberts:2011fs}
Roberts, L.~F., Kasen, D., Lee, W.~H., \& Ramirez-Ruiz, E. 2011, \apj, 736, L21

\bibitem[{Rosswog(2004)}]{Rosswog:2004vj}
Rosswog, S. 2004, arXiv.org

\bibitem[{Rosswog(2005)}]{Rosswog:2005dq}
---. 2005, \apj, 634, 1202

\bibitem[{Rosswog(2007{\natexlab{a}})}]{Rosswog:2007fv}
---. 2007{\natexlab{a}}, \mnras, 376, L48

\bibitem[{Rosswog(2007{\natexlab{b}})}]{Rosswog:2007vp}
---. 2007{\natexlab{b}}, Triggering Relativistic Jets (Eds. William H. Lee {\&}
  Enrico Ram{\'\i}rez-Ruiz) Revista Mexicana de Astronom{\'\i}a y
  Astrof{\'\i}sica (Serie de Conferencias) Vol. 27, 27, 57


\bibitem[{Rosswog \& Liebend{\"o}rfer(2003)}]{Rosswog:2003fu}
Rosswog, S., \& Liebend{\"o}rfer, M. 2003, \mnras, 342, 673

\bibitem[{Rosswog {et~al.}(1999)Rosswog, Liebend{\"o}rfer, Thielemann, Davies,
  Benz, \& Piran}]{Rosswog:1999wz}
Rosswog, S., Liebend{\"o}rfer, M., Thielemann, F.~K., Davies, M.~B., Benz, W.,
  \& Piran, T. 1999, \aap, 341, 499

\bibitem[{Rosswog {et~al.}(2012)Rosswog, Piran, \& Nakar}]{Rosswog:2012um}
Rosswog, S., Piran, T., \& Nakar, E. 2012, eprint arXiv:1204.6240

\bibitem[{Rosswog {et~al.}(2013)Rosswog, Piran, \& Nakar}]{Rosswog:2013ea}
---. 2013, \mnras, 430, 2585

\bibitem[{Rosswog \& Ramirez-Ruiz(2002)}]{Rosswog:2002fi}
Rosswog, S., \& Ramirez-Ruiz, E. 2002, \mnras, 336, L7

\bibitem[{Rosswog \& Ramirez-Ruiz(2003)}]{Rosswog:2003jh}
---. 2003, \mnras, 343, L36

\bibitem[{Rosswog {et~al.}(2003)Rosswog, Ramirez-Ruiz, \&
  Davies}]{Rosswog:2003gj}
Rosswog, S., Ramirez-Ruiz, E., \& Davies, M.~B. 2003, \mnras, 345, 1077

\bibitem[{Shakeshaft \& Spruch(1979)}]{Shakeshaft:1979gz}
Shakeshaft, R., \& Spruch, L. 1979, Rev. Mod. Phys., 51, 369

\bibitem[{Shapiro \& Teukolsky(1983)}]{Shapiro:1983wz}
Shapiro, S.~L., \& Teukolsky, S.~A. 1983, Research supported by the National
  Science Foundation. New York, Wiley-Interscience, 1983, 663 p., -1

\bibitem[{Sigurdsson \& Hernquist(1993)}]{Sigurdsson:1993hv}
Sigurdsson, S., \& Hernquist, L. 1993, \nat, 364, 423

\bibitem[{Sigurdsson \& Phinney(1993)}]{Sigurdsson:1993jz}
Sigurdsson, S., \& Phinney, E.~S. 1993, \apj, 415, 631

\bibitem[{Sigurdsson \& Phinney(1995)}]{Sigurdsson:1995gh}
---. 1995, \apjs, 99, 609

\bibitem[{Spitzer(1969)}]{Spitzer:1969jx}
Spitzer, L.~J. 1969, \apj, 158, L139

\bibitem[{{Steiner} {et~al.}(2010){Steiner}, {Lattimer}, \&
  {Brown}}]{2010ApJ...722...33S}
{Steiner}, A.~W., {Lattimer}, J.~M., \& {Brown}, E.~F. 2010, \apj, 722, 33

\bibitem[{{Stephens} {et~al.}(2011){Stephens}, {East}, \&
  {Pretorius}}]{2011ApJ...737L...5S}
{Stephens}, B.~C., {East}, W.~E., \& {Pretorius}, F. 2011, \apjl, 737, L5

\bibitem[{Tanaka \& Hotokezaka(2013)}]{Tanaka:2013uj}
Tanaka, M., \& Hotokezaka, K. 2013, eprint arXiv:1306.3742

\bibitem[{{Tanvir} {et~al.}(2013){Tanvir}, {Levan}, {Fruchter}, {Hjorth},
  {Wiersema}, {Tunnicliffe}, \& {de Ugarte Postigo}}]{2013arXiv1306.4971T}
{Tanvir}, N.~R., {Levan}, A.~J., {Fruchter}, A.~S., {Hjorth}, J., {Wiersema},
  K., {Tunnicliffe}, R., \& {de Ugarte Postigo}, A. 2013, ArXiv e-prints

\bibitem[{{The NRAR Collaboration} {et~al.}(2013){The NRAR Collaboration},
  {Hinder}, {Buonanno}, {Boyle}, {Etienne}, {Healy}, {Johnson-McDaniel},
  {Nagar}, {Nakano}, {Pan}, {Pfeiffer}, {P{\"u}rrer}, {Reisswig}, {Scheel},
  {Schnetter}, {Sperhake}, {Szil{\'a}gyi}, {Tichy}, {Wardell}, {Zenginoglu},
  {Alic}, {Bernuzzi}, {Bode}, {Br{\"u}gmann}, {Buchman}, {Campanelli}, {Chu},
  {Damour}, {Grigsby}, {Hannam}, {Haas}, {Hemberger}, {Husa}, {Kidder},
  {Laguna}, {London}, {Lovelace}, {Lousto}, {Marronetti}, {Matzner},
  {M{\"o}sta}, {Mrou{\'e}}, {M{\"u}ller}, {Mundim}, {Nerozzi}, {Paschalidis},
  {Pollney}, {Reifenberger}, {Rezzolla}, {Shapiro}, {Shoemaker}, {Taracchini},
  {Taylor}, {Teukolsky}, {Thierfelder}, {Witek}, \&
  {Zlochower}}]{2013arXiv1307.5307T}
{The NRAR Collaboration} {et~al.} 2013, ArXiv e-prints

\bibitem[{Thompson(2011)}]{2011ApJ...741...82T}
Thompson, T.~A. 2011, \apj, 741, 82

\bibitem[Thompson et al.(2009)]{2009arXiv0912.0009T} Thompson, T.~A., 
Kistler, M.~D., \& Stanek, K.~Z.\ 2009, arXiv:0912.0009 



\bibitem[{{Tsang} {et~al.}(2012){Tsang}, {Read}, {Hinderer}, {Piro}, \&
  {Bondarescu}}]{2012PhRvL.108a1102T}
{Tsang}, D., {Read}, J.~S., {Hinderer}, T., {Piro}, A.~L., \& {Bondarescu}, R.
  2012, Physical Review Letters, 108, 011102
  
  \bibitem[Tsang(2013)]{2013arXiv1307.3554T} Tsang, D.\ 2013, arXiv:1307.3554

\bibitem[{Wen(2003)}]{Wen:2003bu}
Wen, L. 2003, \apj, 598, 419

\bibitem[{Willems {et~al.}(2007)Willems, Kalogera, Vecchio, Ivanova, Rasio,
  Fregeau, \& Belczynski}]{2007ApJ...665L..59W}
Willems, B., Kalogera, V., Vecchio, A., Ivanova, N., Rasio, F.~A., Fregeau,
  J.~M., \& Belczynski, K. 2007, \apj, 665, L59

\bibitem[{{Zemp} {et~al.}(2009){Zemp}, {Ramirez-Ruiz}, \&
  {Diemand}}]{2009ApJ...705L.186Z}
{Zemp}, M., {Ramirez-Ruiz}, E., \& {Diemand}, J. 2009, \apjl, 705, L186

\bibitem[Zheng 
\& Ramirez-Ruiz(2007)]{2007ApJ...665.1220Z} Zheng, Z., \& Ramirez-Ruiz, E.\ 2007, \apj, 665, 1220 

\end{thebibliography}

\end{document}